\documentclass[fleqn,usenatbib,flalign]{mnras}

\usepackage{newtxtext}
\usepackage{lscape}
\usepackage[varg]{newtxmath}

\usepackage[T1]{fontenc}

\usepackage[normalem]{ulem}
\usepackage{graphicx}	
\usepackage{journals}
\usepackage{color}
\usepackage{ae,aecompl}
\usepackage{multicol}

\newcommand{\msun}{{\rm M_\odot}}

\newcommand{\lm}{\log\,(m/{\rm M}_\odot)}

\definecolor{darkgreen}{rgb}{0.0,0.5,0.0}
\definecolor{darkred}{rgb}{0.5,0.0,0.0}
\definecolor{brown}{rgb}{0.65,.16,0.16}
\definecolor{grey}{rgb}{0.4,0.5,0.6}

\definecolor{cgcol}{rgb}{0.0,0.0,0.627}
\definecolor{mnwcol}{rgb}{0.0,0.565,0.0}
\definecolor{bwccol}{rgb}{0.561,0.0,0.561}
\definecolor{mgpcol}{rgb}{0.942,0.,0.}
\definecolor{guocol}{rgb}{0,0.56,0.56}
\newcommand{\CG}{C+G}
\newcommand{\MNW}{MNW}
\newcommand{\BWC}{BWC}
\newcommand{\MCP}{MCP}

\newcommand{\Hen}{Henriques}

\title[The frequency of very young galaxies from SDSS spectra]{The frequency
  of very young galaxies in the local Universe:  II. The view from SDSS
  spectra}  

\author[G. Mamon et al.]
{Gary. A. Mamon$^1$\thanks{E-mail:gam@iap.fr},
Marina Trevisan$^{2,1}$,
Trinh X. Thuan$^{3,1}$,
Anna~Gallazzi$^{4}$,
\newauthor
and
Romeel Dav\'e$^{5,6,7}$
 \\ 
$^1$Institut d'Astrophysique de Paris (UMR 7095: CNRS \& Sorbonne Universit\'e), 98 bis Bd Arago, F-75014 Paris, France\\ 
$^2$Universidade Federal do Rio Grande do Sul -- Departamento de Astronomia -- 91501-970, Porto Alegre-RS, Brazil\\
$^3$Astronomy Department, University of Virginia, P.O. Box 400325, Charlottesville, VA 22904-4325\\ 
$^4$INAF -- Osservatorio Astrofisico di Arcetri, Largo Enrico Fermi 5, I-50125 Firenze, Italy\\
$^5$Institute for Astronomy, Royal Observatory, University of Edinburgh,
 Edinburgh EH9 3HJ, UK\\
 $^6$University of the Western Cape, Bellville, Cape Town 7535, South Africa\\
$^7$South African Astronomical Observatories, Observatory, Cape Town 7925, South Africa
}

\date{Accepted yyyy month dd. Received yyyy month dd; in original form yyyy month dd} 

\pubyear{2018}

\begin{document}

\label{firstpage}
\pagerange{\pageref{firstpage}--\pageref{lastpage}} \pubyear{2017}

\maketitle

\begin{abstract}

Only a handful of galaxies in the local Universe appear to be very young. We
estimate the fraction of very young galaxies (VYGs), defined as those with
more than half their stellar masses formed within the last Gyr. We fit
non-parametric star formation histories (SFHs) to $\sim$280 000 galaxy
spectra from a flux- and volume-limited subsample of the Main Galaxy Sample
(MGS) of the SDSS, which is also complete in mass-to-light ratio, thus
properly accounting for passive galaxies of a given mass. The VYG fractions
decrease with increasing galaxy stellar mass, from $\sim$50~per cent at $m =
10^8\, {\rm M}_{\odot}$ to $\sim$0.1~per cent at $m = 10^{11.5}\,{\rm
  M}_{\odot}$, with differences of up to 1 dex between the different spectral
models used to estimate the SFH and on how we treat aperture effects.
But old stellar 
populations may hide in our VYGs despite our conservative VYG sample built
with galaxies that are globally bluer than within the region viewed by the
SDSS fibre. The VYG fractions versus mass decrease more gradually compared to the Tweed et
al. predictions using analytical and semi-analytical models of galaxy
formation, but agree better with the SIMBA hydrodynamical simulation. These
discrepancies highlight the usefulness of VYGs in constraining the strong
uncertainties in both galaxy formation models and spectral modelling of
galaxy SFHs.
Given the lognormal cosmic SFH, these
mean VYG fractions suggest that galaxies with $m > 10^8\,\rm M_\odot$ 
undergo at most 4 major starbursts on average.

\end{abstract}


\section{Introduction}

Among the wide variety of galaxies in the local Universe, a handful of
low-mass \emph{star-forming galaxies} (SFGs) have drawn attention because of
their extremely low metal abundance, 12 + log O/H $\leq$ 7.3. The
\emph{blue compact dwarf}  (BCD) galaxy  
I~Zw~18, first observed spectroscopically by \cite{Searle&Sargent72},
with oxygen abundances 12 + log O/H $\sim$ 7.17-7.26
\citep{Skillman&Kennicutt93,Izotov&Thuan98} long stood as the lowest
metallicity SFG known.   
The most metal-poor dwarf SFG known to date has 12 +log O/H = 6.98$\pm$0.02 
\citep{Izotov+18}, i.e. 1/50th the Sun's metallicity if 12 + log O/H = 8.7 is adopted for the Sun. In their discussion of I~Zw~18, \cite{Searle&Sargent72}
posed the question: are the most metal-deficient BCDs young galaxies,
i.e. they formed their first stars at a relatively recent time ($\leq$ 1 Gyr
ago) or are they old galaxies which formed most of their stars several Gyr
ago, in which case the present starburst is just the last episode of a series
of such events? For the vast majority of BCDs (more than 99~per cent), the
second case appears to be true: deep CCD images clearly show an
extended low-surface brightness stellar component, indicative of an older
stellar population, whereas the most metal-deficient BCDs,
those with 12 + log O/H $\leq$ 7.3, do not show such an extended low surface
brightness stellar component (e.g. \citealp{Papaderos+02}). This suggests
that the most metal-deficient galaxies in the local Universe have very young
stellar populations, i.e. are
\emph{very young galaxies} (hereafter, VYGs). 

Using \emph{colour-magnitude diagrams} (CMDs) of stars in I~Zw~18, 
resolved through observations with the Hubble Space Telescope
(HST), \cite{Izotov&Thuan04} estimated the bulk of 
the stellar mass to have 
formed within the last 500~Myr. Their age was based on 
the lack of red giant stars. However, the young age of I~Zw~18 was disputed by 
\cite{Aloisi+07} and \cite{ContrerasRamos+11} who obtained deeper 
CMDs with additional HST observations. They found more red giant stars and 
attributed an age of at least 1--2 Gyr to I~Zw~18.  
However, it was not clear from their work whether that older stellar population found constituted only a small fraction of the stellar mass of I~Zw~18, 
or accounted for the
bulk 
of it. In the first case, it could still be called 
a VYG.

In the last few years, there has been an independent development that also suggest young ages, not for low-mass dwarf galaxies like BCDs, but for galaxies 
with considerably larger stellar masses. While the BCDs 
have stellar masses in the range $\sim$ $10^{7-8}\msun$, Dressler et al. (2018)
have found a population of galaxies in the redshift range 0.45$<$ $z$ $<$ 0.75 and with stellar masses $>$ $10^{10} \msun$, 
that formed the majority of their stars within $\sim$2 Gyr of the epoch of
observation, which they called \emph{Late Bloomers} (LBs). These LBs account
for $\sim$20~per cent of $z$\,$\sim$\,0.6 galaxies with Milky Way masses, with a
moderate dependence on mass ($\sim$\,30 per cent at half the Milky Way mass and
5 to 10 per cent at masses greater than that of the Milky Way). According to  
Dressler et al. (2018), major star formation in about 1/5th of the massive galaxies somehow got delayed by billions of years. The LB population declines rapidly at $z$\,$\sim$\,0.3 and is essentially extinct today for $M$\,$>$\,10$^{10}$\,M$_\odot$. 
These LBs are like VYGs except they have higher stellar masses and are at larger redshifts.  
      
These observations have motivated us to look at 
the question of the frequency of VYGs in the
local Universe. \cite{Tweed+18}, hereafter Paper~I, have studied the
predicted frequency of VYGs at $z=0$, defined to have at least half their
stellar mass formed in the last 1~Gyr ($z < 0.08$). They have used various models of
galaxy formation, four analytical ones and 
one \emph{semi-analytical model} (SAM), to predict the VYG fraction as a function of galaxy
stellar mass (hereafter `galaxy mass' or simply `mass').
In Paper~I, we have found that the predicted VYG fraction as a function 
of galaxy mass
can differ between the models, by up to 3 orders of magnitude.
While all models predict low fractions of VYGs at stellar masses
$m>10^{11}\, \msun$, some predict a few percent of VYGs at masses below
$10^{10}\, \msun$, while another found a peak in the VYG fraction at $10^9
\,\msun$. Finally, the SAM predicts only 0.01 per cent
of VYGs at intermediate masses, increasing somewhat at lower masses.

These wildly discrepant model results have led us to ask whether we can
constrain, directly from observations,
the fraction of VYGs in the local Universe. We defined VYGs following Paper~I to have \emph{at least half their stellar
  mass formed within 1 Gyr from the epoch corresponding to their redshift}.

The ideal way to measure a galaxy's \emph{star formation history} (SFH) 
is by combining a high spectral resolution spectrum, covering a wide spectral range, with a colour-magnitude diagram (CMD). Unfortunately, CMDs have been measured in only 186
nearby galaxies.\footnote{According to the NASA Extragalactic
  Database at {\tt http://ned.ipac.caltech.edu/Library/Distances/}}
This sample is much too small to study the mass dependence of the VYG
fraction, especially if the latter reaches peak values of only a few
percent. 
However, one can derive SFHs of galaxies from their \emph{spectral
  energy distributions}
(SEDs). In fact, that was the method used by \cite*{Dressler+18} whose work
was published during the course of our own.  
From SEDs defined by broadband optical and infrared (IR) photometry and optical 
prism observations with $R \approx 30$, those authors 
derived SFHs for 20\,000 galaxies at $0.25 < z < 0.75$ 
and with stellar mass $m>10^{10}\msun$. 

In this article, we make use of the spectral database of the 
\emph{Main Galaxy Sample} (MGS) of the
Sloan Digital Sky Survey (SDSS) \emph{Data Release 12} (DR12),
comparing the SFHs of over 400\,000 galaxies between two SFH codes, using a
total of 7 models of single stellar populations.
This allows us to estimate the fraction of VYGs as a function of galaxy stellar mass
for each spectral model.

We describe our sample in Sect.~\ref{sec:data} and explain our methods in Sect.~\ref{sec:methods}.
In Sect.~\ref{sec:result}, we present the fractions of VYGs inferred 
from the SDSS spectra.
These results are summarized and discussed in Sect.~\ref{sec:discuss}.

\section{Data and Sample Selection}
\label{sec:data}

\subsection{Summary}
\label{sec:datasummary}
We extracted our sample from the MGS of the SDSS data release 12 (DR12), according to the following criteria:
\begin{enumerate}
\itemindent=0pt
    \item \label{flux} flux limit: $r_{\rm Petro}^0 \leq 17.77$;
    \item \label{fibsize} object spectra obtained with original 3 arcsec fibre;
    \item \label{zrange} redshift range: $0.005 < z < 0.12$;
    \item \label{slok} the STARLIGHT \citep{CidFernandes+05} SFH code does not fail when applied to the object;
    \item \label{invespa} object is in the VErsatile SPectral
Analysis database (VESPA)\footnote{The VESPA database is publicly accessible at {\tt http://www-wfau.roe.ac.uk/vespa}.} of SFHs;
    \item \label{nodups} single spectrum for each galaxy;
        \item \label{mulim} surface brightness limit: $\mu_{r,50}^0 \leq 23.0$;
    \item \label{stelmas} stellar mass range: $6 < \log(m/\msun)<12.5$ for
      all spectral models;
        \item \label{mags} reasonable magnitudes: $g_{\rm Petro}>0$, $r_{\rm Petro}>0$, $i_{\rm Petro}>0$;
    \item \label{colors} colours are not extreme:\\ $\hbox{\qquad} -1 < (g-i)_{\rm model}^0 < 2.5$ AND $-1 < (g-i)_{\rm fibre}^0 < 2.5$;
    \item \label{colorgrads} fibre colours are not too different from model colours: \\
    $\hbox{\qquad} |(g-i)_{\rm fibre}^0 - (g-i)_{\rm model}^0|< 1$;
    \item \label{chi2} all 6 spectral fits yield $\chi^2>0$;
    \item \label{zmax} redshift is not too large to fail to see passive galaxies: \\
    $\hbox{\qquad} z < z_{\rm max}(m)$.
    \end{enumerate}
    \newcounter{major}
\setcounter{major}{\value{enumi}}

    In addition, we allowed ourselves to apply the following criteria to the VYG candidates (but not to the parent sample):
    \begin{enumerate}
      \itemindent=0pt
    \addtocounter{enumi}{\value{major}}
       \item \label{noAGN} galaxy does not contain an \emph{Active Galactic Nucleus} (AGN, using the  curve of \citealp{Kauffmann+03} that conservatively separates 
AGN from SFGs 
in the \citealp*{BPT81}, hereafter BPT, diagram);
     \item \label{colorgrads2} fibre colour is not bluer than the model colour:\\
    $\hbox{\qquad} (g-i)_{\rm fibre}^0 > (g-i)_{\rm model}^0$.
\end{enumerate}

\subsection{Basic selection criteria}
\label{sec:basicselection}
According to \cite{Strauss+02}, the MGS sample is defined to be limited in  both
flux ($r_{\rm Petro}\leq 17.77$) and surface brightness
$\mu_{r,50} = r_{\rm Petro} + 2.5\,\log(2\pi\theta_{50}^2) \leq 24.5$,
where $r_{\rm  Petro}$ is the extinction-corrected Petrosian magnitude, while $\theta_{50}$ is the angular
half-light radius.

We first selected galaxies, according to our first 3 criteria, with\\
\vspace{-\baselineskip}
\begin{verbatim}
SELECT (...)
  FROM
    SpecObj as s, 
    PhotoObj as p
  WHERE
    s.bestobjid = p.objid 
    AND s.instrument = 'sdss'
    AND s.Class = 'GALAXY' 
    AND (p.petroMag_r - p.extinction_r) < 17.77 
    AND s.z >= 0.005 & s.z <= 0.12 
\end{verbatim}
This     provided us an initial sample of 424\,506 galaxies. 

The limit in the apparent magnitude (criterion~\ref{flux}), corrected for Galactic extinction, is the limit of the SDSS MGS. 
By selecting the `sdss' instrument, we are avoiding the BOSS spectra that employ smaller fibres (criterion~\ref{fibsize}).
The lower redshift limit ensures a sufficiently accurate distance based on the redshift (the peculiar velocities contribute negligibly to the redshift), while the upper redshift limit ensures that we are not limited to the most luminous galaxies, given the flux limit (criterion~\ref{zrange}).

We added extra requirements based on our stellar population synthesis analysis (see Sect.~\ref{sec:methods}). 
The STARLIGHT SFH code (see Sect.~\ref{sec:methods})
did not provide SFHs (criterion~\ref{slok}) for 842 spectra (2 per cent), leaving us with 423\,664 galaxies.
The VESPA galaxy sample (see Sect.~\ref{sec:methods}) is limited in surface brightness to 
$\mu_{r,50}^0 < 23.0$ (criterion~\ref{invespa}), which leads to a smaller
sample of 406\,005 galaxies.
We then excluded 200 galaxies that have more than one spectroscopic observation per galaxy
(criterion~\ref{nodups}):  
after visually inspection, we kept the spectrum that was closest to the
galaxy centre. This left us with 405\,805 unique galaxies.

The $\mu_{r,50}< 23.$ surface brightness limit
 (criterion~\ref{mulim}) 
of the VESPA galaxy sample had been applied to photometry from the 7th data release (DR7) of the SDSS. Since the SDSS photometry is different in DR12, a few (361) galaxies end up with $\mu_{r,50}^0 > 23$ and are removed, leaving us with 405\,444 galaxies.

We also required the stellar masses to be in the range from $10^6\,\msun$ to $10^{12.5}\,\msun$ for each of our 6 spectral models (criterion~\ref{stelmas}), making us lose another 84 galaxies, leaving us with 405\,360 galaxies. Note that the stellar masses are obtained by extrapolating the masses within the fibre using the difference between fibre and model magnitudes in the $z$ band.

We also demanded that the $g$ and $z$ Petrosian magnitudes be positive (criterion~\ref{mags}), making us lose another 2 galaxies, now leaving us with 405\,358 galaxies. 

\begin{figure}
  \centering
  \includegraphics[width=\hsize,viewport=0 30 550 540]{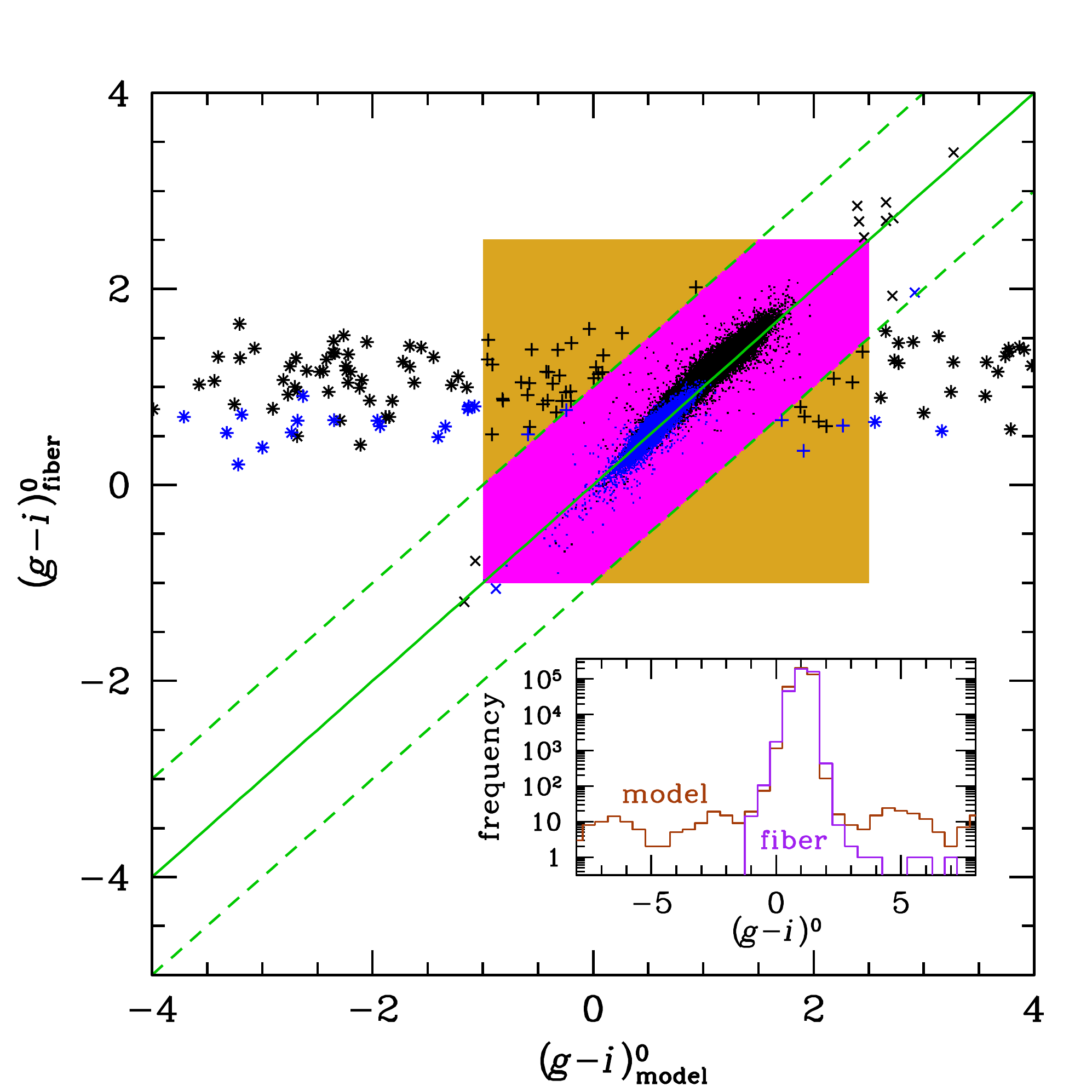}
  \caption{Model versus fibre galaxy colours (both corrected for Galactic extinction) for the sample built with criteria~\ref{flux}-\ref{mags}.
    Our final selection is highlighted by the
    \emph{magenta region}, where  only one random galaxy among 4 is plotted for clarity.
    The \emph{asterisks} show the galaxies whose model colours are bluer than
    --1 or redder than 2.5.
    The \emph{crosses} show the galaxies whose fibre or model colours fail to meet equation~(\ref{gmiselection}).
    The \emph{plus signs} display the galaxies whose model and fibre colours
    are within the limits of equation~(\ref{gmiselection}), but whose difference between model and fibre
    colour is greater than 1 magnitude in absolute value.
    The \emph{blue points} represent VYGs according to the STARLIGHT V15 model (see Table~\ref{tab:sfhmodels}). 
      The \emph{inset} shows the distribution of model (\emph{brown}) and fibre
    (\emph{purple}) colours. 
    \label{fig:colcol}
  } 
\end{figure}

We excluded objects with unreliable photometry, analysing their colours (criteria~\ref{colors} and \ref{colorgrads}) as follows.
Fig.~\ref{fig:colcol} displays the model $(g\!-\!i)^0$ 
colours (corrected for Galactic extinction) as a function of the
corresponding fibre colours, as well as the distributions of model and fibre
colours. Both the model (brown) and fibre (purple) colours are roughly Gaussian-distributed for
\begin{equation}
-1 < (g-i)^0 < 2.5 \ ,
\label{gmiselection}
\end{equation}
but
the fibre colours display much fewer outliers.
We visually inspected all galaxies whose colours that do not
satisfy equation~(\ref{gmiselection})
and found that nearly all were contaminated by neighbouring
saturated stars or large galaxies.
We thus removed from our sample those galaxies whose model or fibre  colours  do not obey equation~(\ref{gmiselection}).
The square shaded box in Fig.~\ref{fig:colcol} indicates our colour-filtered sample.
We then visually inspected all galaxies whose model and fibre colours both satisfied equation~(\ref{gmiselection}), 
but differed by over 1 magnitude in absolute value.
Most of these galaxies were again contaminated by nearby saturated stars or
large galaxies.
We thus made a second colour selection, requesting that the absolute
difference of model and fibre colours be less than unity (magenta region in Fig.~\ref{fig:colcol}).
This led us to discard an additional 336 galaxies, leaving us with a photometrically accurate
sample of 405\,022 galaxies.
We finally excluded 95 galaxies with negative $\chi^2$ values in the spectral
fits in any of the 6 spectral models (criterion~\ref{chi2}), leaving us with
our {\tt clean} sample of 404\,931 galaxies.

\subsection{Handling of selection effects}

Several observational selection effects can influence
 our analysis of VYG fractions.

\subsubsection{Mass-to-light ratio selection effects}
\label{sec:zmax}

When estimating fractions of VYGs, we must also worry that, at a given
stellar mass, galaxies with young stellar populations, which have low
mass-to-light ratios, will be more luminous. We must thus ensure that the
galaxies with old stellar populations are not missed, hence the fraction of
VYGs not overestimated because of this selection effect against high
mass-to-light ratio galaxies.
Since we are measuring VYG fractions as a function of galaxy stellar mass, a
doubly complete survey in volume and luminosity is not complete for the
passive galaxies. A sample complete in stellar mass would be desirable, but
many more galaxies are found by requesting that the sample is complete in
mass-to-light ratio for given mass bins.

\begin{figure}
\centering
\includegraphics[width=\hsize,viewport=10 20 520 450]{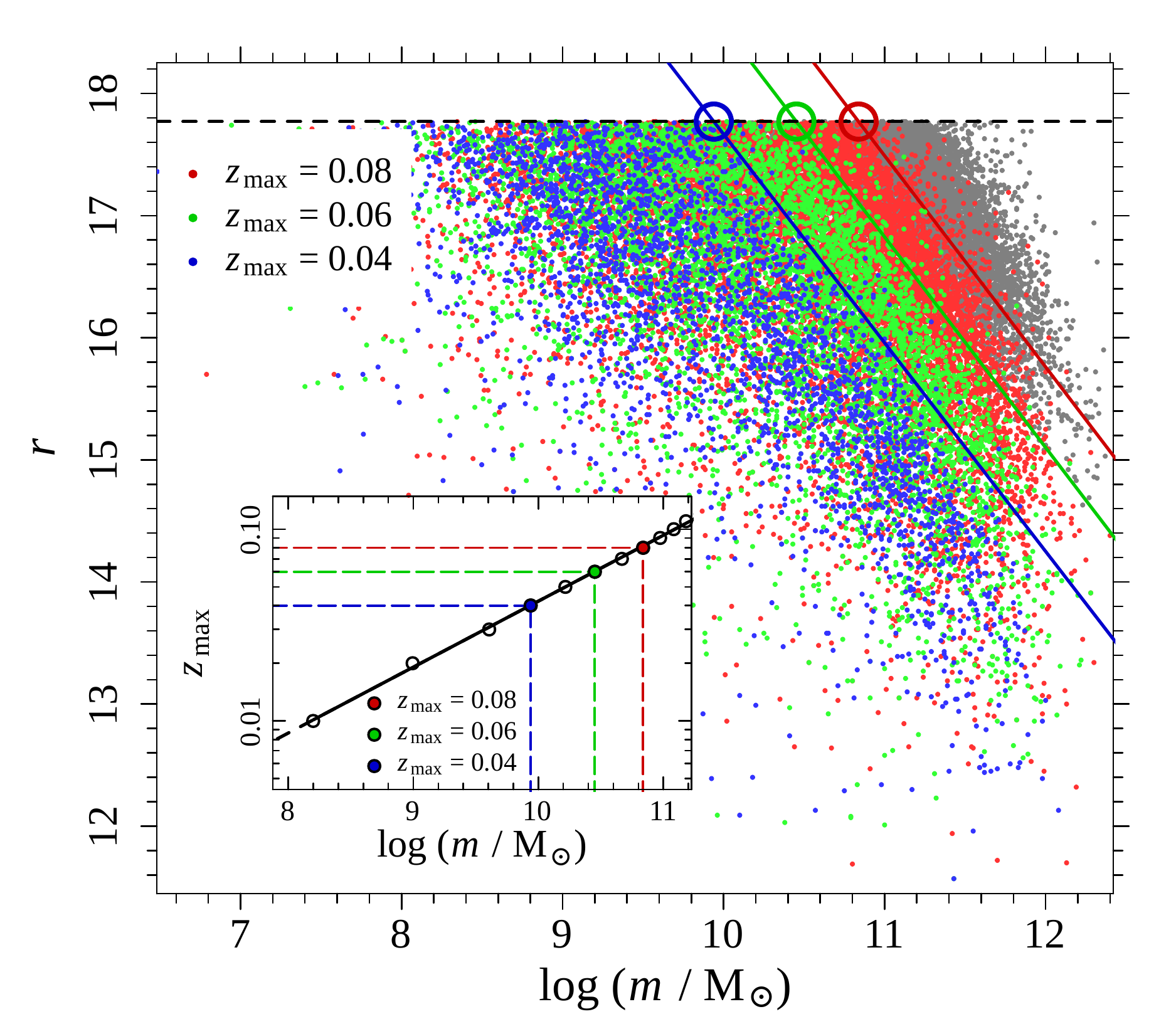} 
\caption{Illustration of the estimation of maximum redshift for completeness
  at a given stellar mass (for the STARLIGHT / V15 model, see Table~\ref{tab:sfhmodels}).
{\bf Main plot}: apparent magnitude vs. stellar mass, for SDSS galaxies in
the {\tt clean} sample,
whose $r=17.77$ flux limit
is shown as the \emph{dashed horizontal line}. The full set of
galaxies is shown in \emph{grey}, with galaxies at 3 representative
redshifts,
$z=0.04$, 0.06, and 0.08,
shown in \emph{blue}, \emph{green}, and \emph{red}, respectively, for
illustrative purposes, with the corresponding
\emph{oblique coloured lines} representing our estimates of 95th percentiles
of mass as a function of apparent magnitude, for that maximum redshift.
The \emph{large coloured circles} indicate the values of $m_{95}(r_{\rm lim})$.
{\bf Inset}: resultant maximum redshift versus stellar mass (with same colour
code), with the 3 representative redshifts and corresponding masses shown in
colour.
\label{fig:compl}
}
\end{figure}

For this,
we estimated the maximum redshift for which the subsample of galaxies
\emph{of a given stellar mass} is complete in luminosity (or in
mass-to-light-ratio) at our flux limit.
This maximum redshift for given mass, $z_{\rm max}(m)$, is estimated in an analogous fashion as \cite*{Garilli+99},
\cite{LaBarbera+10}, and \cite*{Trevisan+17}. 
Here, we considered  a fine grid of maximum redshifts, $z_{ \rm max }$
(in steps of $\Delta z_{ \rm max } = 0.005$), estimated
for each one the 95th percentile in galaxy stellar mass,
$m_{95}(r)$, in narrow bins of apparent magnitude $r$, and deduced
$m_{95}(r_{\rm lim})$ using a linear fit of $m_{95}(r)$ vs. $r$.

\begin{table*}
  \begin{center}
    \caption{Star formation history models
    \label{tab:sfhmodels}}
      \begin{tabular}{llllllllcccc}
        \hline
        \hline
      SFH       & SSP   & \multicolumn{1}{c}{acronym} & \multicolumn{1}{c}{reference} & stellar   &  evol.   & \multicolumn{1}{c}{IMF}    & dust  & age   & metal. & $a_0$ & $a_1$\\
      code      & code  &         & &          library             & tracks   &        & model & bins  & bins \\
      \multicolumn{1}{c}{(1)} &
      \multicolumn{1}{c}{(2)} &
      \multicolumn{1}{c}{(3)} &
      \multicolumn{1}{c}{(4)} &
      \multicolumn{1}{c}{(5)} &
      \multicolumn{1}{c}{(6)} &
      \multicolumn{1}{c}{(7)} &
      \multicolumn{1}{c}{(8)} &
      \multicolumn{1}{c}{(9)} &
      \multicolumn{1}{c}{(10)} &
       \multicolumn{1}{c}{(11)} & 
        \multicolumn{1}{c}{(12)}
      \\
     \hline
      SL & BC03  & S\_BC03 &  Bruzual \& Charlot (2003)  &STELIB   & Padova & Chabrier & screen & 15 & 5 &  --4.677&0.331\\
      SL & V15   & S\_V15 & Vazdekis et al. (2015) & BaSTI & MILES      & Kroupa & screen     & 15 & 6 & --4.814 & 0.344\\
      VESPA     & BC03  & V\_BC03\_d1  & Bruzual \& Charlot (2003)  & STELIB &   Padova & Chabrier & mixed & 16 & 5 & --5.218 & 0.391 \\
      VESPA     & BC03  & V\_BC03\_d2 &  Bruzual \& Charlot (2003) &STELIB& Padova & Chabrier & 2-comp & 16 & 5 & --5.228 & 0.392\\
      VESPA     & M05   & V\_M05\_d1 & Maraston (2005) & Basel & Padova & Kroupa & mixed & 16 & 4 & --5.121 & 0.382\\
      VESPA     & M05   & V\_M05\_d2 & Maraston (2005) &  Basel   &Padova & Kroupa & 2-comp & 16 & 4 & --5.202 & 0.390\\
     \hline 
      \end{tabular} 
  \end{center}
  \parbox{\hsize}{Notes: The columns are:
  (1): star formaiton history code (`SL' stands for STARLIGHT);
  (2): single stellar population code;
  (3): spectral model acronym;
  (4): reference;
  (5): stellar library;
  (6): evolutionary tracks;
  (7): initial mass function (Chabrier 2003 and Kroupa 2003);
  (8): dust model (screen, mixed or mixed with additional component around young stars); 
  (9): number of log age bins;
  (10): number of metallicity bins;
  (11) and (12): parameters of $z-{\rm max}$ vs. $m$ (eq.~[\ref{zmaxvsm}]).
  }
\end{table*}

As seen in the inset of Fig.~\ref{fig:compl}, the relation between maximum redshift and stellar mass is a power law, 
\begin{equation}
    \log z_{\rm max} = a_0 + a_1\,\log\,\left({m\over \msun}\right) \ ,
    \label{zmaxvsm}
\end{equation}
whose coefficients are given in Table~\ref{tab:sfhmodels}.
This constitutes criterion~\ref{zmax}.
This procedure is illustrated in Fig.~\ref{fig:compl} for 3 values of $z_{ \rm
  max }$, 
where the oblique lines
indicate the linear fits to $m_{95}(r)$ vs. $r$ and the big circles show the
linear extrapolation for $m_{95}(r_{\rm lim})$.
The inset of Fig.~\ref{fig:compl} shows
the resulting maximum redshift as a function of stellar mass.
There are usually no solutions $z_{ \rm max } < 0.005$ (below which we do not
trust our stellar masses, given the importance of peculiar velocities that
reduce the accuracy of redshifts as distance indicators) for $\log\,(m/{\rm
  M_\odot}) \leq 8$.
  The fits of equation~(\ref{zmaxvsm}) and Table~\ref{tab:sfhmodels} are thus only adequate for $\log\,(m/\msun) \geq 8$.

\begin{figure}
    \centering
    \includegraphics[width=\hsize,viewport=0 30 550 540]{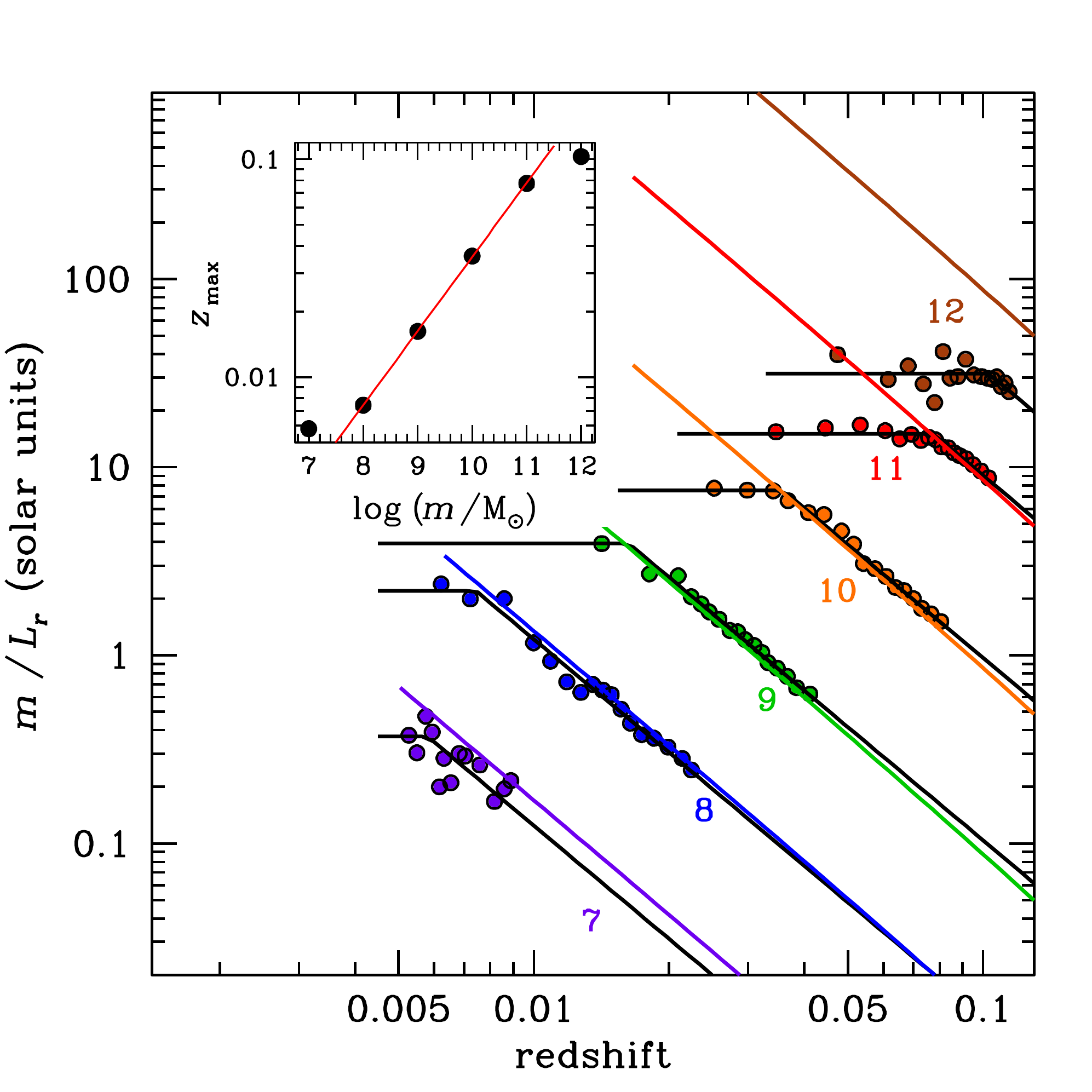}
    \caption{Mass-to-light ratio versus redshift of the {\tt clean} sample in narrow bins of stellar mass (for the STARLIGHT / Vazdekis et al. 2015 model) of $\log(m/\msun)=7$, 8, 9,10, 11, and 12, 
of respective widths 0.25, 0.15, 0.02, 0.01, 0.01, and 0.02 dex.   
     The \emph{symbols}
    indicate the 95th percentiles in equal number bins. The \emph{broken black lines} show the broken power-law fits for low and high redshift  slopes of 0 and --2, respectively, while the \emph{straight coloured lines} indicate the maximum $m/L_r$ given the mass bin maxima and the flux limit.
    The \emph{inset} shows the resultant fit of maximum redshift versus mass, with our adopted linear fit (restricted to $8 \leq \log\,(m/\msun) \leq 11$, \emph{red line}).
    }
    \label{fig:moverlvsz}
\end{figure}

Fig.~\ref{fig:moverlvsz} illustrates this procedure in another fashion. It shows how the variation with redshift of the mass-to-light ratios of galaxies in the {\tt clean} sample are affected by the flux limit. This is shown for different stellar masses (shown as different colours).
The 95th percentiles in equal number bins (circles) show that $m/L_r$ is constant with redshift up to some limiting redshift, then decreases as the flux limit leads to a maximum $m/L_r$ for given mass and redshift (straight coloured lines), which decreases with increasing redshift. 
This behaviour is  seen for all masses except  the lowest mass ($10^7\, \msun$), where constant mass-to-light ratio is not seen at low redshifts, nor at the highest mass in $10^{12}\,\msun$, where the flux limit allows for much higher $m/L_r$ values than found. 

We fitted a broken power-law to each run of $m/L_r$ vs. redshift, with slopes of 0 and $-2$. This fit was obtained with a first estimate of the turning point height using the median $\log(m/L_r)$ of the 25 per cent of the lowest redshifts. Our initial estimate of the turning point redshift used $\hbox{Median}(\log z + 0.5\,\log m/L_r)-\hbox{Median}(\log m/L_r)$, computed on the 25 per cent highest redshifts. We then minimised $\chi^2 = \sum \left [\log (m/L_r)_{\rm data} - \log(m/L_r)_{\rm model}\right]^2$ for the 2 parameters of the broken line break, allowing for $\log m/L_r$ and $\log z$ to lie within 0.3  from the first guess. The best-fit broken lines are displayed in black in Fig.~\ref{fig:moverlvsz}. The inset to the figure shows the best fit values of $z_{\rm max}$, the abscissa of the breaks of the broken lines in the main figure. One notices that $\log z_{\rm max}$ increases linearly with $\log m$, as $\log z_{\rm max} = -4.849 + 0.340\,\log (m/\msun)$ for the S\_V15 model.

Given the similarity between the fits obtained with these two methods,
we adopted the relation of equation~(\ref{zmaxvsm}) with the coefficients shown in Table~\ref{tab:sfhmodels}, which we assume extrapolate to $m < 10^8\,\msun$.

\subsubsection{Discarding AGN}
\label{sec:noAGN}

Although this is not a selection effect, we allow ourselves to discard AGN from our candidate VYGs (criterion~\ref{noAGN}).
For this, we discard VYG candidates that lie above the \cite{Kauffmann+03} curve of the BPT diagram.

\subsubsection{Aperture effects}
\label{sec:aperture}

The finite (3 arcsec) aperture of the SDSS fibres limits the spectra
to the inner regions of many galaxies. 
Fig.~\ref{fig:colcol} shows that the galaxies that are considered VYGs by
one of the spectral models (STARLIGHT V15, see Sect.~\ref{sec:methods}) before filtering by colour and the $\chi^2$ of the spectral fits) are associated with blue fibre colours. But there is very little relation between global (model) colour and galaxy youth (blue symbols) estimated from the spectrum measured within the fibre.
Many (21 per cent of) galaxies in the {\tt clean} sample have bluer nuclei than their global bodies (lying below the solid green curve of Fig.~\ref{fig:colcol}), leading to spectral fits pointing to a younger stellar population than averaged over the galaxy. 
For example, the galaxy  NGC~838 in the HCG~16 compact
group, whose stellar mass is $\approx 5\times 10^{10}\rm\, M_\odot$
(\citealp{O'Sullivan+14a} and references therein), has a nucleus (the SDSS
fibre subtends an angular radius of 400 pc) whose stellar population has an
important very young component (younger than 300 Myr), which accounts for
nearly half of the stellar mass of the nucleus, depending on the SFH and SSP
model
(\citeauthor{O'Sullivan+14a}), while the bulk of the galaxy is older
\citep{Vogt+13}. 
Some of these galaxies with blue nuclei may thus be classified as VYGs even if the bulk of their stellar mass is old.

This aperture effect is potentially serious, since the median fraction of galaxy light
subtended by the fibre is only 26 per cent, and only 3 per cent of our sample have
fibres covering more than half of the galaxy light.
Moreover, building a galaxy sample where the fibres see over, say, 30 per
cent of the galaxy light (39 per cent of the sample), leads to an effective
cut at higher surface brightness than our limit of $\mu_{r,50} = 23$, 
leading us to lose 99.8 per cent of all galaxies
with $21.5 < \mu_{r,50} < 23$.

Instead, we handle the aperture selection effect by restricting our VYGs to
galaxies with \emph{blue colour gradients},  ${\rm d} (g-i)^0/{\rm d} R<0$
(where $R$ is the projected distance to the center),
i.e. with a global (model) colour that is bluer than the fibre colour. This
constitutes criterion~\ref{colorgrads2}.

\subsection{Final samples}
\label{sec:finalsamps}

\begin{table}
  \caption{Galaxy samples}
  \begin{center}
    \tabcolsep 3pt
    \begin{tabular}{lcrrrr}
      \hline
      \hline
      Sample & Ages & \multicolumn{4}{c}{Size} \\
      \cline{3-6}
      & (Gyr) & \multicolumn{1}{c}{S\_V15} & \multicolumn{1}{c}{V\_BC03\_d2} & \multicolumn{1}{c}{V\_M05\_d2} & \multicolumn{1}{c}{Gallazzi} \\
      \hline
      {\tt clean} & all & 404\,931 & 404\,931 & 404\,931 & 98\,624 \\
      {\tt liberal} VYG & $<\,1$ & 29\,775 & 5974 & 5780 & 303 \\
      {\tt conserv.} VYG & $<\,1$ & 16\,212 & 3041 & 3011 & 124 \\
      \hline
      {\tt complete} & all & 264\,847 & 287\,748 & 291\,317 & 74\,836\\
      {\tt liberal} VYG & $<\,1$ & 1461 & 1196 & 1187 & 39\\
      {\tt conserv.} VYG & $<\,1$ & 649 & 729 & 746 & 15\\
      \hline
      \end{tabular}
  \end{center}
    \parbox{\hsize}{Notes: The {\tt clean} galaxy sample is filtered against
      bad measurements, while the {\tt complete} one is also complete in mass-to-light
      ratio. The {\tt conservative} VYG sample is like the {\tt liberal} one,
      except it excludes AGN and galaxies with red colour gradients.
    Only the main spectral models are shown.}
  \label{tab:samples}
\end{table}

The sample extracted with criteria~\ref{flux}-\ref{chi2} is our {\tt clean}
sample (404\,931 galaxies). 
The sample with criteria~\ref{flux}-\ref{zmax} 
is our {\tt complete} sample (typically 280\,000 galaxies, depending on the spectral model).
We consider two samples of VYGs: {\tt liberal} samples (one for each spectral
model) 
and
a {\tt conservative} sample that excludes AGN (criterion~\ref{noAGN}) as well
as galaxies with red colour gradients (criterion~\ref{colorgrads2}).
The two parent samples and the two VYG samples are shown in Table~\ref{tab:samples}.

\section{Ages from galaxy spectra}
\label{sec:methods}
Galaxy SFHs and ages were inferred from the SDSS-DR12 spectra through stellar population synthesis analysis.
These spectra have the advantage of
fairly high spectral resolution ($R \approx 2000$), excellent signal to noise (above $\sim$20), wide
spectral range (from below $3800\,$\AA\ to $9200\,$\AA) and they are flux calibrated.
They allow to 
estimate the SFH of each galaxy, by fitting the observed spectrum with the predicted spectrum
obtained by the combination of single stellar population (SSP) spectra convolved by
the SFH.

We considered three algorithms to estimate the SFH.
We used two non-parametric codes: 
STARLIGHT \citep{CidFernandes+05}, which we ran ourselves, and
the VESPA database \citep{Tojeiro+09}.
Both codes  fit non-parametric SFHs and metal histories with an internal
extinction, 
 considering typically 16 
(roughly equally spaced) bins of log age, as well as
 4 to 6 bins of metallicity, and a single
 extinction (see Table~\ref{tab:sfhmodels}). This leads to a total of
 typically $16\times 5+1 = 81$ free parameters.

We also used the method of \cite{Gallazzi+05}, who analysed the SFHs of the
175\,128 galaxies in the 2nd data release (DR2) of the
SDSS. \citeauthor{Gallazzi+05} adopted a Bayesian approach to derive median
likelihood estimates of stellar population properties such as $r$-band
weighted and mass-weighted mean stellar age, stellar metallicity and stellar
mass. For each galaxy, observed absorption features were compared with the
predictions from a large library of metallicities and SFHs, described by an
exponential declining law on top of which random bursts are added, convolved
with BC03 SSP models. The probability distribution functions (from which the
16, 50, 84 percentiles are measured) of the light- and mass-weighted mean
ages
are
obtained by marginalising over all the other parameters of the models.
We limited our analysis of  the \citeauthor{Gallazzi+05} sample to the
intersection with our {\tt clean} and {\tt complete} galaxy samples
discarding the 18 per cent of galaxies with no
age measurements, with numbers given in Table~\ref{tab:sfhmodels}.

All four algorithms have been run using the
\citeauthor{Bruzual&Charlot03} (\citeyear{Bruzual&Charlot03}, hereafter BC03) 
model to produce synthetic spectra of SSPs to fit the observed SDSS spectra.  
The SSPs were calculated with the Padova 1994 
stellar evolution tracks \citep{Bressan+93,Fagotto+94a,Fagotto+94b,Girardi+96} and
with the \cite{Chabrier03} \emph{initial mass function} (IMF).
The BC03 model employs the STELIB stellar library \citep{LeBorgne+03}.
VESPA has also been run using the SSPs of \citet[][M05]{Maraston05} 
using the Basel stellar library \citep{Lejeune+98} and the \cite{Kroupa01} IMF. We also ran STARLIGHT on 424\,506 SDSS spectra. STARLIGHT uses
the Medium resolution Isaac Newton Telescope Library of Empirical Spectra
(MILES, \citealp{Sanchez-Blazquez+06}), using the updated version 10.0
(\citealp{Vazdekis+15}, hereafter V15) of the code presented in
\cite{Vazdekis+10}.
These MILES models were computed  with the \cite{Kroupa01} IMF, and
stellar evolution tracks from BaSTI 
(Bag of Stellar Tracks and Isochrones, \citealp{Pietrinferni+04,Pietrinferni+06}).
While STARLIGHT was run assuming a screen dust model, VESPA assumed either a mixed
slab interstellar dust model \citep{Charlot&Fall00} or combined it with
extra dust around young stars (also from \citeauthor{Charlot&Fall00}).
\cite{Gallazzi+05} used a screen dust, with the $\lambda^{-0.7}$ extinction
curve used later by VESPA.

The advantage of the STARLIGHT model with the V15 SSP is its recent stellar
library and evolutionary tracks, while the VESPA models involve a more
refined treatment of dust extinction.
The age and metallicity bins of the two SFH codes are quite similar:
The lowest age bins are 25 Myr for STARLIGHT and less than 20 Myr for VESPA,
while the metallicity bins ranged from $Z = 0.0004$ to 0.05 (VESPA BC03) or 0.04
 (VESPA M05) or from $Z = 0.001$ to 0.048 (STARLIGHT).
Note that the lower limit of metallicity of $Z-0.001$ corresponds to the
 limit of ``safe'' spectral modeling with the MILES based models (see
 http://research.iac.es/proyecto/miles/pages/ssp-models/safe-ranges.php). 

For each galaxy and for each spectral (SFH and SSP) model (see
Table~\ref{tab:sfhmodels}),
we compute the \emph{median age}, $\rm age_{50}$,
when the stellar mass was half its final value, by linear interpolation of the
fractional mass as a function of log age, or use the arithmetic mean
mass-weighted age for
the Gallazzi model.
VYGs are defined as galaxies whose median (or mean for Gallazzi) ages are less than 1 Gyr. 

\subsection{Tests of the spectral models}

\label{sec:tests}

We tested the VESPA and STARLIGHT spectral models in several ways.
\begin{figure}
    \centering
    \includegraphics[width=\hsize,viewport=0 30 550 540]{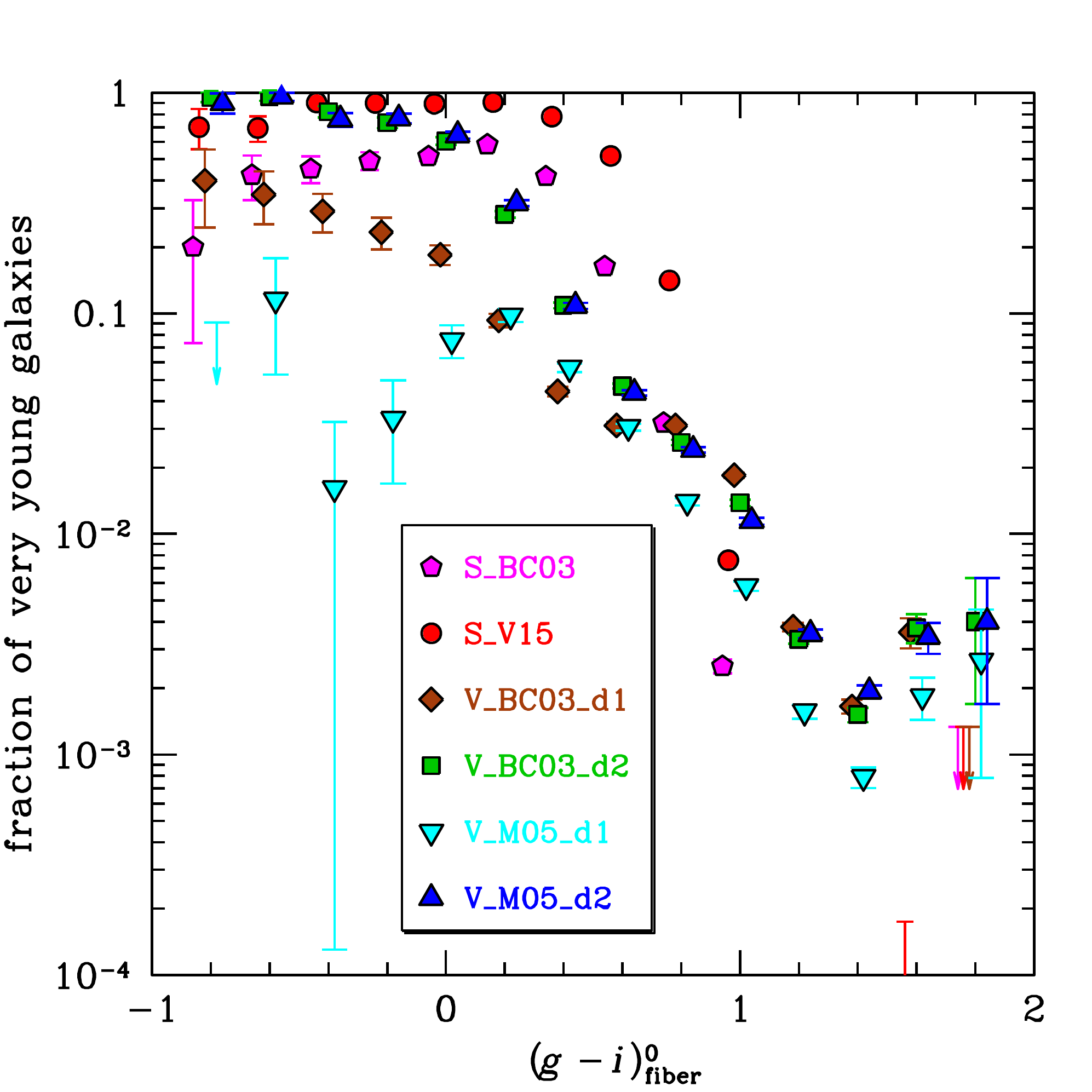}
    \caption{Fraction of very young galaxies (more than half the stellar mass  formed in
  the last Gyr) versus fibre colour, for the
  STARLIGHT and VESPA spectral models (using the acronyms given in
  Table~\ref{tab:sfhmodels}), for  our {\tt clean} galaxy sample.
  The error bars are binomial with Wilson 84 percent confidence ($\sim
  1\,\sigma$) upper limits.
  The abscissa are slightly shifted for clarity.
  }
    \label{fig:fVYGvsfibcol}
\end{figure}
We first check that the fraction of VYGs is very high for the bluest fibre colours.
Fig.~\ref{fig:fVYGvsfibcol} compares the variation of the VYG fraction with fibre colour for the 6 spectral models.
One sees that the VESPA M05 model with 1-component dust fails to achieve a
high fraction of VYGs for the bluest colours.
Only the STARLIGHT V15 model and the VESPA models with 2-component dust reach 
high fractions of VYGs at the bluest fibre colours, i.e. over 82 per cent of galaxies with $(g-i)_{\rm fibre}^0<-0.1$ are VYGs according to these 3 models, while less than 45 per cent of them are VYGs according to each of the other 3 models.

We then  checked that  galaxies  with the largest H$\alpha$ \emph{equivalent widths}
(EWs) have the greatest fractions of VYGs, close to unity.
Indeed, EW measures the ratio of line flux over continuum intensity, which is
proportional to the ratio of ongoing or very recent star formation (SF) rate 
over luminosity, and hence roughly proportional to the ratio of recently formed stellar mass over total stellar mass. 
Therefore,
strong EWs are a sign of high fractions of  recently  formed stars.
This interpretation of the EW is oversimplified, as it depends on the evolution of the stellar mass-to-light ratio, i.e. on the SFH, as well as on the geometrical effects of dust opacity. 
Nevertheless, according to the {\sc STARBURST99} model \citep{Leitherer+99},
assuming a \cite{Salpeter55} IMF extending to star masses of $100\,\rm
M_\odot$ and metallicities of $Z_\odot/20$,
$\rm EW(H\alpha) > 500\,$\AA\ indicates ages lower than 50 Myr for a continuous SFH
(fig.~84 of \citeauthor{Leitherer+99}), and lower than only 5 Myr for an
instantaneous starburst (fig.~83 of \citeauthor{Leitherer+99}). These maximum
ages are reduced by about a quarter when moving to solar metallicity.
For continuous SFH, a burst that started 1 Gyr ago would lead to $\rm EW(H\alpha) = 150\,$\AA.
Of course, if there is an underlying old stellar population at the time of the burst, the continuum would be higher, and the EW  reduced.
On the other hand, the SFHs computed by VESPA and STARLIGHT do not consider the emission lines, hence they provide an independent estimate of galaxy youth.

\begin{figure}
    \centering
    \includegraphics[width=\hsize,viewport=0 30 550 540]{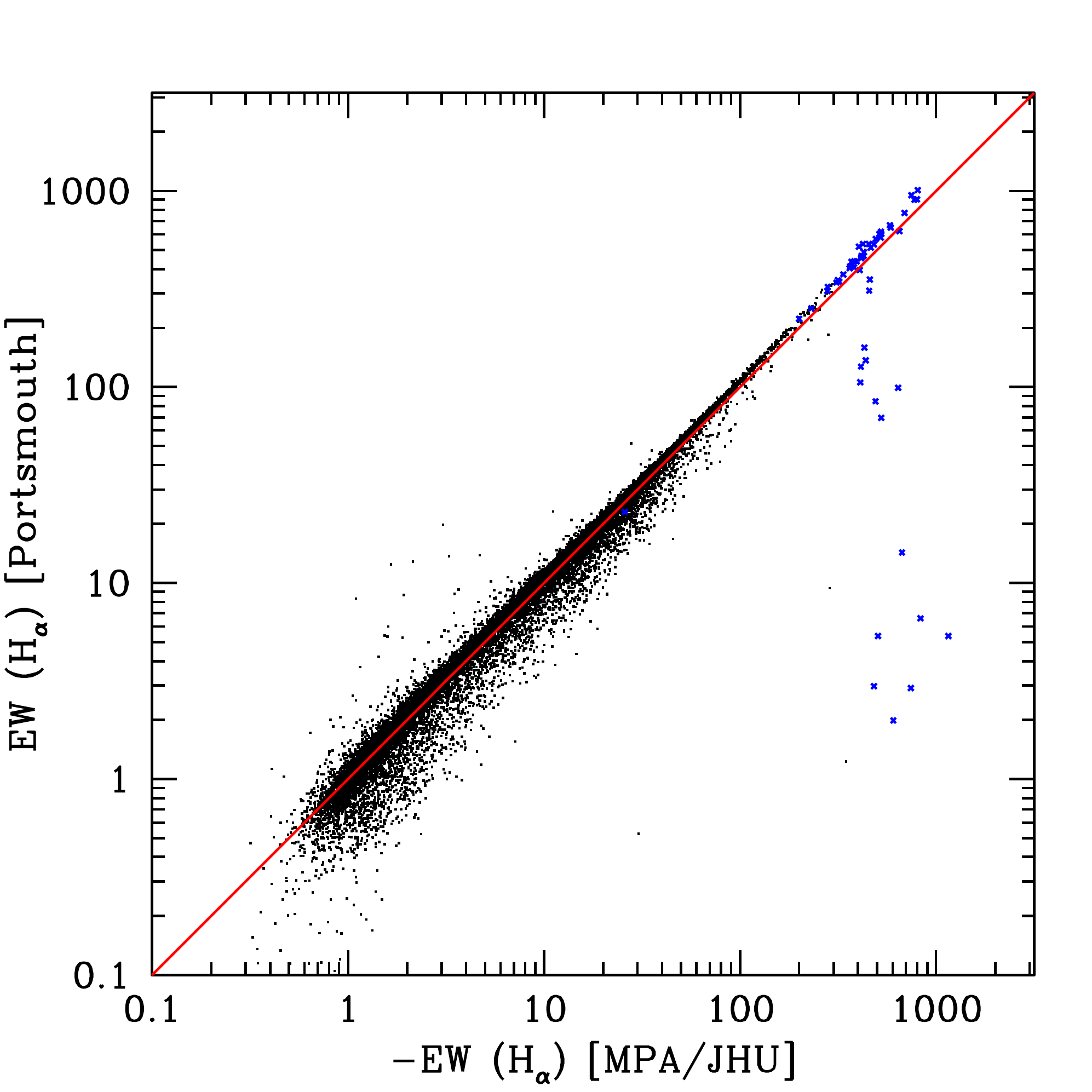}
    \caption{Comparison of the $\rm H\alpha$ equivalent widths between the Portsmouth emission line code (absorption line galaxies have EW = 0) and the MPA/JHU code, which uses the opposite nomenclature (negative EW for emission). Each point is a galaxy (one point in 10 is plotted for clarity) from the {\tt clean} sample, only keeping galaxies whose EW uncertainties are less than one-third of the absolute value of the EW.
    9 galaxies with emission lines according to Portsmouth have absorption lines according to MPA/JHU (thus not shown). The \emph{red line} denotes equality, while the 
    \emph{blue points} show all galaxies with very blue fibre colours $(g-i)^0 < -0.3$.
    }
    \label{fig:compareEW}
\end{figure}

We first compared two measures of SDSS H$\alpha$ line EWs. The first was
taken from the {\tt GalSpecLine} table (MPA/JHU, \citealp{Brinchmann+04,Tremonti+04}), where the continuum is fitted with linear combinations of BC03-MILES simple stellar population models, similarly to STARLIGHT, and the emission lines are then distinguished from the stellar continuum. 
The second meaure of EW was extracted from the
        {\tt emissionLinesPort} table (Portsmouth, \citealp{Thomas+13}), which is similar to MPA/JHU, but uses the single stellar population model of \cite{Maraston&Stromback11}, the Gas and Absorption Line Fitting code, {\sc GANDALF}
        \citep{Sarzi+06},
and Penalized PiXel Fitting, {\sc pPFX} \citep{Cappellari&Emsellem04}.
The Portsmouth EW model, which assigns EW = 0 to absorption line galaxies, fails to assign values for 473 galaxies of our {\tt clean} sample, and assigns astronomically high values of EW (from over $10^4\,$\AA\ to over $10^{30}\,$\AA) for another 120 galaxies.

\begin{figure}
\centering
\includegraphics[width=\hsize,viewport=0 30 550 540]{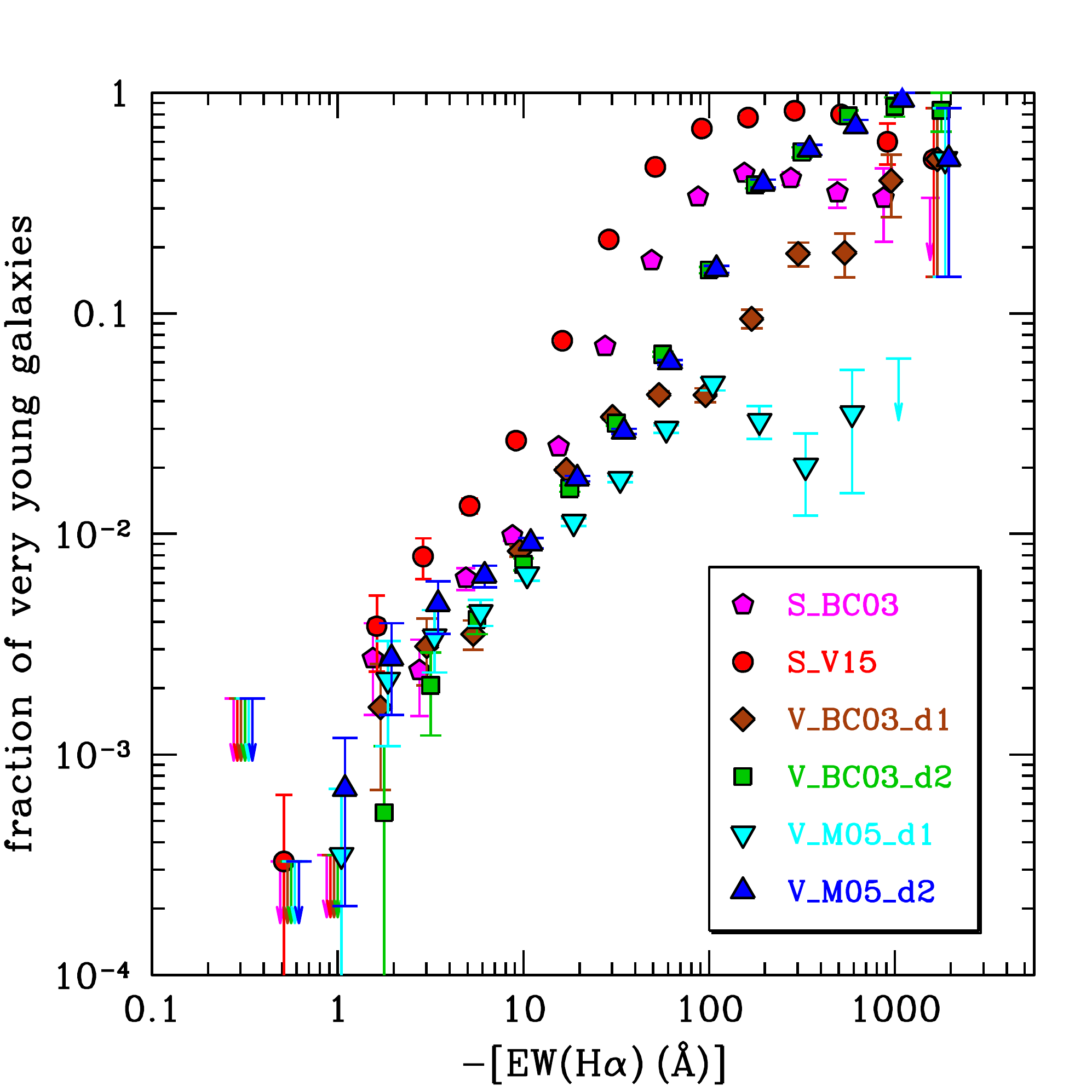}
\caption{Fraction of very young galaxies (more than half the stellar mass  formed in
  the last Gyr) as a function of the H$\alpha$ equivalent width (determined
  from the MPA/JHU code) for the
  STARLIGHT and VESPA spectral models (using the acronyms given in
  Table~\ref{tab:sfhmodels}). We filtered our {\tt clean} galaxy sample against AGN and 
  for the errors in equivalent width to be lower than
  1/3 of the absolute value of the equivalent width.
  The error bars are binomial with Wilson 84 percent confidence ($\sim
  1\,\sigma$) upper limits.
  The abscissa are slightly shifted for clarity.
\label{fig:fVYGvsEW}
}
\end{figure}

\begin{figure*}
\includegraphics[width=0.6\hsize,viewport=0 50 570 770,angle=-90]{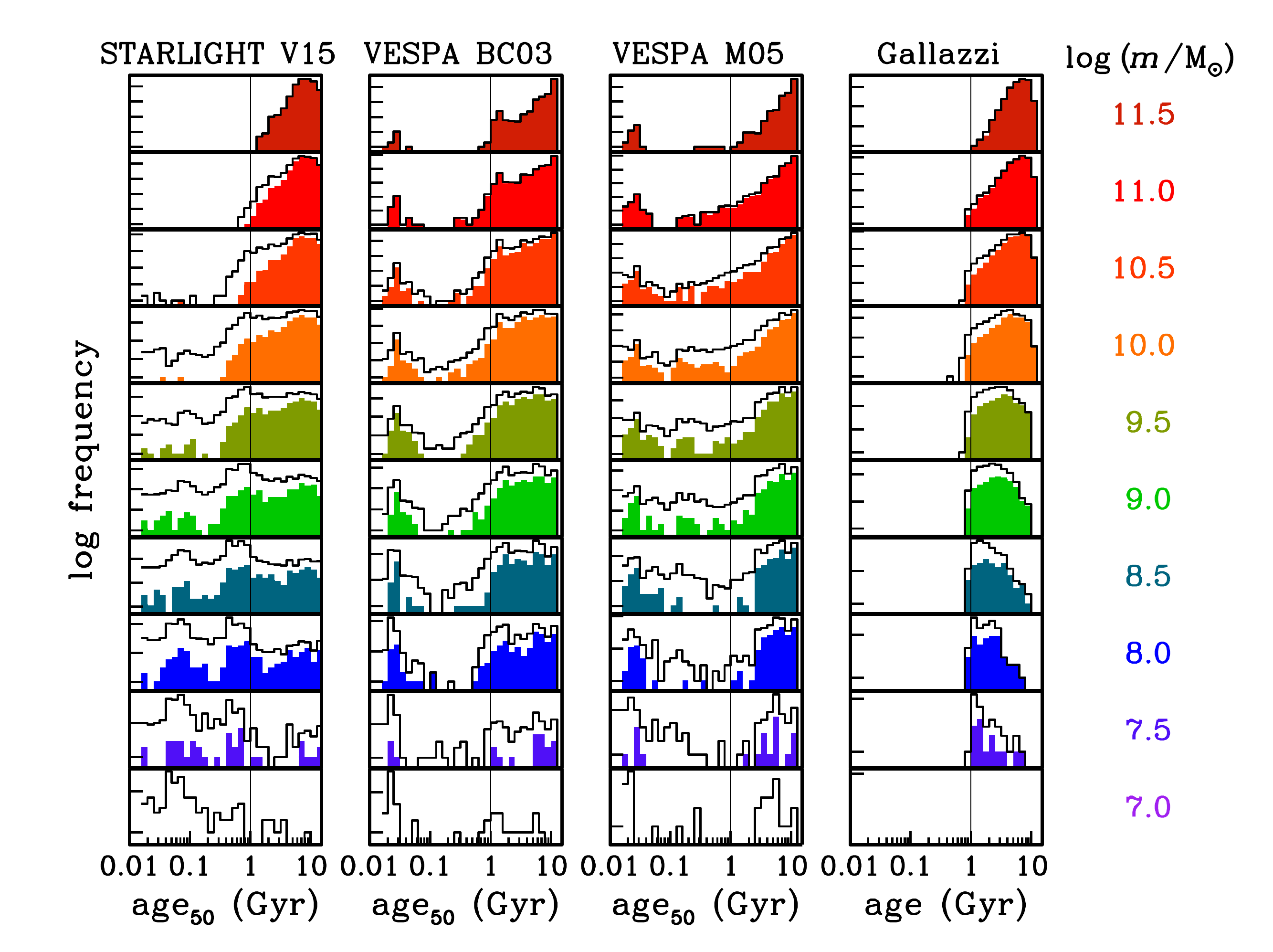}
\caption{Distribution of the median (mass-weighted mean for Gallazzi) ages of galaxies in the {\tt clean} SDSS Main Galaxy  Sample 
in bins of stellar mass, for our four
spectral
models.
The \emph{open} and \emph{filled histograms} respectively indicate the  {\tt clean}
and {\tt complete} (i.e. limited to $z<z_{ \rm max }(m)$, using
the fit given in eq.~[\ref{zmaxvsm}] and Table~\ref{tab:sfhmodels}) samples.
Note that there are no galaxies in the {\tt complete} sample in the lowest mass bin of all 4 spectral models. The intervals between ticks of log frequency are unity (1 dex).
\label{fig:pdfage}
}
\end{figure*}

Fig.~\ref{fig:compareEW} compares the EWs between the MPA/JHU and Portsmouth
for the 304\,418 galaxies for which $0.01\,$\AA\,$< \rm EW(H\alpha) [Portsmouth] < 5000\,$\AA\ and for which the absolute value of the EW is at least 3 times its uncertainty for both models. Note that the MPA values use a nomenclature opposite to that of the Portsmouth values: the MPA EWs are negative for emission lines and positive for absorption lines.
There is good general agreement between the two measures of H$\alpha$ EW. In particular, all galaxies with 
$\rm EW(H\alpha) [Portsmouth] > 100\,$\AA\ have  
$\rm -EW(H\alpha) [MPA/JHU] > 44\,$\AA.
However, the converse is not true: 
among galaxies with $-\rm EW(H\alpha) [MPA/JHU] > 100\,$\AA, 12 have 
$\rm EW(H\alpha) [Portsmouth] < 44\,$\AA, yet most of them have very blue fibre colours, i.e. $(g-i)_{\rm fibre}^0 < -0.3$ (thick blue points), suggesting very recent SF (at least in the nucleus). While most such galaxies with very blue fibre colours lie on
the ridge of equality (taking into account the different sign conventions)
between the two measures, and all of them but one\footnote{One galaxy with 
$-\rm EW(H\alpha)[MPA/JHU] = 26\,$\AA\ has $(g-i)_{\rm fibre}^0 = -0.35$ on top of a galaxy with $(g-i)^0 = 0.36$ (after subtracting the fibre contribution).} have $\rm -EW(H\alpha) [MPA/JHU] > 200\,$\AA, several galaxies with blue fibre colours have 
$\rm EW(H\alpha) [Portsmouth] < 200\,$\AA, and 7 have 
$\rm EW(H\alpha) [Portsmouth]$ lower than $20\,$\AA.
This led us to adopt the
MPA/JHU values for the H$\alpha$ EWs.

Fig.~\ref{fig:fVYGvsEW} displays the VYG fractions as a function of the H$\alpha$ EWs.
It shows that the VYG fractions generally rise with
increasing H$\alpha$ EWs. 
But the M05 model
run with VESPA using a one-component mixed dust model  (cyan) gives only a few per
cent of VYGs for $\rm EW(H\alpha) > 100\,$\AA.
The corresponding VESPA BC03 model (brown) reaches higher values, but not as high as the VESPA models with a 2-component dust extinction (green and blue).
Finally, the BC03
model ran
with STARLIGHT (magenta)
gives intermediate VYG fractions of $\sim 35$ per cent at
these high EWs.
In sum, the figure shows that only 3 spectral models stand out in this regard: STARLIGHT
with V15 (red), and VESPA with 2-component dust (both BC03 [green] and M05 [blue]), as all
three predict that over half the galaxies with $\rm EW(H\alpha)>500\,$\AA\
are VYGs.

The models with the best behaving VYG fractions as a function of fibre colour and H$\alpha$ EW thus appear to be V15 with
STARLIGHT, and the 2-component dust BC03 and M05 models with VESPA.
We therefore adopt these 3 spectral models in our subsequent analysis of VYGs, and will
no longer discuss the other 3 models.
We now  refer to our adopted spectral models as V15 STARLIGHT, VESPA BC03 and
VESPA M05, respectively.

\subsection{Comparison between different spectral models}

\begin{figure}
\includegraphics[width=\hsize,viewport=0 30 550 540]{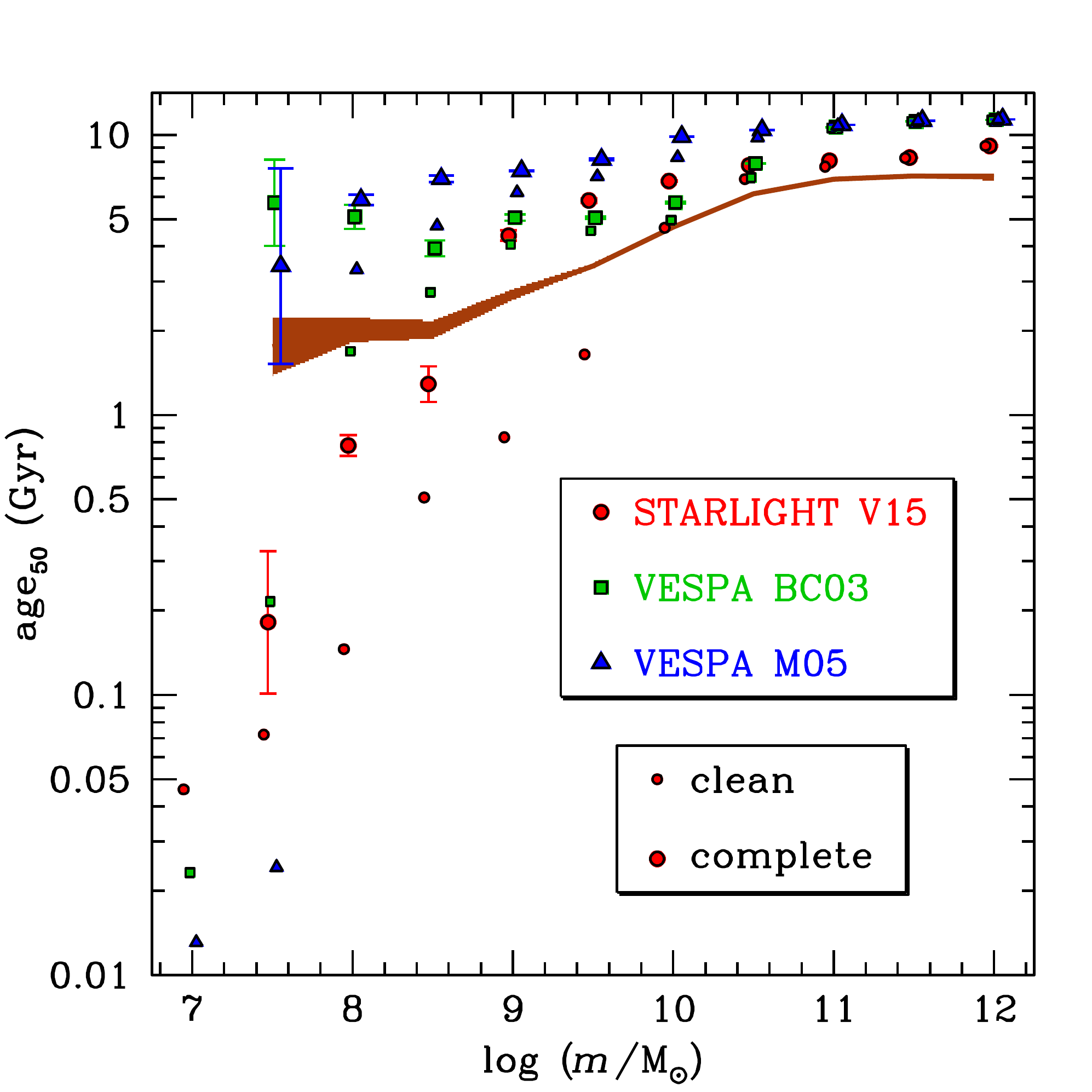} 
\caption{Median (over galaxies within a given mass bin) of `median' stellar ages versus
  stellar mass for galaxies in the {\tt clean} (\emph{small symbols}) and {\tt complete} (\emph{large symbols}) SDSS Main Galaxy 
  Sample according to STARLIGHT with the Vaz15 model
 (\emph{red circles}),
VESPA with BC03 (\emph{green squares}) and VESPA with Maraston et
al. (2005, \emph{blue triangles}). 
The \emph{brown shaded region} shows the mass-weighted arithmetic mean ages
from the intersection of the Gallazzi et al. (2005) sample with the
{\tt complete} sample.
The \emph{error bars} (only shown for the {\tt complete} sample) and the
widths of the shaded regions are the
uncertainties on the medians from 100 bootstraps.
\label{fig:agespecvsm}
}
\end{figure}

Fig.~\ref{fig:pdfage} shows the distribution of median ages of galaxies 
according to our 3 adopted spectral models.
It confirms the well-known \emph{downsizing} trend of
typically lower masses of galaxies with younger stellar populations.
Moreover, the figure shows that the lower age bins are over-represented in
the {\tt clean} sample (open histograms) compared to the {\tt complete}
sample built with $z_{ \rm max }(m)$ as described in Sect.~\ref{sec:zmax}
(filled histograms). This effect is especially evident at low mass, where the
maximum redshift for completeness is considerably lower
(eq.~[\ref{zmaxvsm}], Table~\ref{tab:sfhmodels}, Figs.~\ref{fig:compl} and
\ref{fig:moverlvsz}). 

Comparing the first three panels of Fig.~\ref{fig:pdfage},
it seems odd that the two VESPA SFH models, despite their more realistic treatment of dust
(a mixed slab instead of a simple screen), assign extremely
young ages (less than 80 Myr) to many galaxies at all masses,
yet yield only few galaxies with slightly
older ages (between 80 and 250 Myr), especially with the BC03 SSP. It is difficult to imagine a universe
where the log age distribution of galaxies is so bimodal at all masses. 
This odd behaviour is not seen in the STARLIGHT V15 model, which shows,
within the statistical uncertainties, more continuous distributions of log
age for all galaxy masses.
The Gallazzi model does not show the extension to very low ages
seen in the other models, but this is based on the (arithmetic) 
mass-weighted mean ages, which are higher than the median ages.

Fig.~\ref{fig:agespecvsm} summarises the differences between the SFH/SSP models
by showing the median (over mass bins) of the median age versus galaxy stellar mass.
One notices important differences between the models at low galaxy stellar mass.
In the {\tt clean} sample (small points), the 2 VESPA models show a jump in age$_{50}$ at low masses, while the increase is more gradual with STARLIGHT V15.
On the other hand, the {\tt complete} sample shows high ages at low mass with the VESPA models compared to low ages at low mass for the STARLIGHT model.
In other words, STARLIGHT V15 predicts a strong downsizing
down to low masses, while the downsizing trend is weaker for the VESPA models, with signs of reaching a plateau at low mass (and perhaps even  \emph{upsizing} with VESPA BC03).
In any event, the median ages of galaxies are enhanced when switching from
the {\tt clean} sample to the {\tt complete} sample, especially for the
STARLIGHT V15 model,
but least so for VESPA M05.

Fig.~\ref{fig:agespecvsm} also displays the
mass-weighted arithmetic mean 
ages for the intersection of our complete sample (with S V15) with the sample
extracted from \cite{Gallazzi+05}\footnote{This should not be confused
  with the luminosity-weighted ages reported in \cite{Gallazzi+05}, which are
lower than 
mass-weighted ages, because young stars are so much more luminous for their
mass.}, shown as the brown shaded region. 

\section{Results}
\label{sec:result}

\subsection{Fraction of very young SDSS galaxies versus mass}
\label{sec:fVYG}

\begin{figure}
\centering
\includegraphics[width=\hsize,viewport=0 30 550 540]{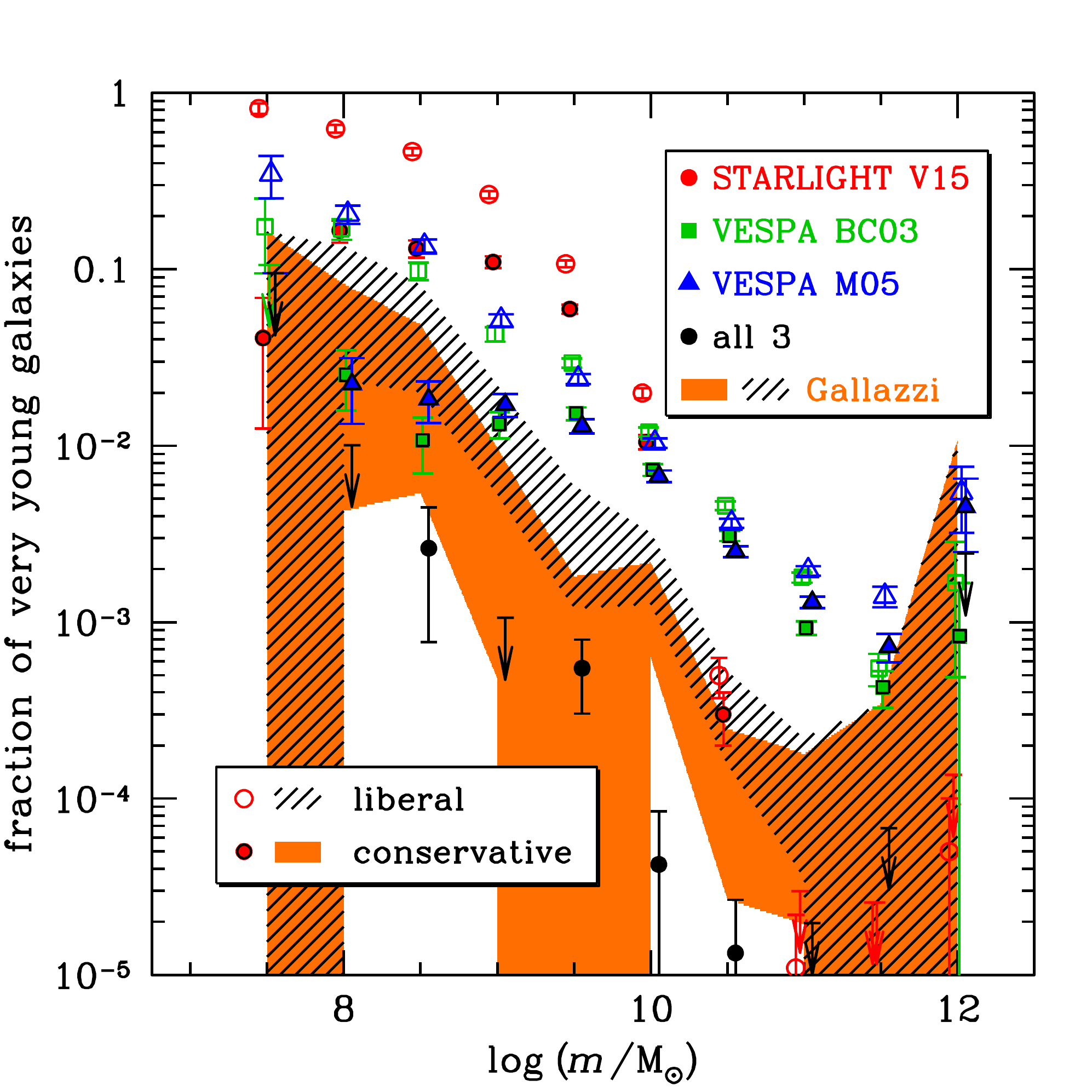}
\caption{Fraction of very young galaxies  (half the stellar mass formed less
  than 1 Gyr before the epoch corresponding to the galaxy's redshift) versus
  stellar mass for the {\tt complete} SDSS sample. The \emph{open symbols}
and \emph{dashed shaded region} indicate the fractions of {\tt liberal} VYGs
for the 3 SFH models and the Gallazzi model, respectively, while the
\emph{solid symbols} and \emph{orange shaded region} show
  the corresponding fractions of galaxies in the {\tt complete} sample that are {\tt
    conservative}  VYGs (not AGN and with bluer global (model) colours than
  the fibre colours).   The
  true fraction of VYGs should lie in between the two sets of symbols. The
  \emph{black filled circles} are the fractions of the galaxies
  that are {\tt liberal} VYGs for all 3 models (the log masses are the means of
  the 3 spectral models with weights 1/2 for STARLIGHT and 1/4 for the two
  VEPSA models). The errors are binomial with Wilson  90 per cent confidence
  upper limits.
  The abscissas are slightly shifted for clarity.
\label{fig:fyoungvsm_sfh}
}
\end{figure}

We now arrive at the focus of our analysis: the fraction of VYGs as a function of galaxy stellar mass.
Fig.~\ref{fig:fyoungvsm_sfh} shows the fraction of VYGs versus galaxy
mass for the {\tt complete} sample of SDSS galaxies, using  the 3 spectral
models, as well as that of \cite{Gallazzi+05}. These models are shown for both
the {\tt liberal} and {\tt conservative} (`not AGN \& $\nabla(g-i)^0<0$') VYG samples.

For all 3 spectral models, the {\tt liberal} VYG  fractions (open
symbols in Fig.~\ref{fig:fyoungvsm_sfh}) decrease with increasing mass. The
STARLIGHT V15 model gives VYG fractions $>$3 per cent at low masses
($m$\,$<$\,$10^{10}\,\msun$) and  50 per cent at the lowest securely complete mass,
$m$\,=\,$10^{8.5}\,\msun$ (see Sect.~\ref{sec:zmax}).
In comparison, the two VESPA models lead to VYG fractions over 1~per cent at low masses ($m$\,$<$\,$10^{10}\,\msun$) and to 12
per cent at $m$=$10^{8.5}\,\msun$.
However at high masses, the VESPA models produce a slowly decreasing VYG fraction with increasing mass, while the STARLIGHT V15 model gives a sharply decreasing VYG fraction for $m$\,$>$\,$10^{10}\,\msun$, 1--2 orders of magnitude lower than the VYG fractions found in VESPA models.

As discussed in Sect.~\ref{sec:aperture}, a serious concern is that the SDSS 3~arcsec diameter fibre is often limited to 
the light from the central
regions of extended galaxies. We may thus mistakenly identify a galaxy as
very young, while it may be an old galaxy with a very young nucleus.
Hence, in Fig.~\ref{fig:fyoungvsm_sfh}, we also plot the fraction of galaxies in the {\tt complete} sample that are {\tt conservative} VYGs, i.e. that are not AGNs, and with global colours that are bluer than their fibre colours.\footnote{Removing AGN from the VYGs turns out to have a negligible  effect on the VYG fraction at low and intermediate masses, while it decreases the VYG fraction by less than 0.2~dex at high masses. However, imposing a blue colour gradient has a very small effect at high masses, but a huge one at low masses, as clearly shown in Fig.~\ref{fig:fyoungvsm_sfh}).}  The {\tt liberal} VYG sample may be contaminated by AGN and by galaxies that are young according to their fibre colours but older globally.
On the other hand, the {\tt conservative} VYG sample may miss galaxies that
are extremely young in the region covered by the SDSS fibre but that are
still very young (but redder and older) overall.
The
true fraction of VYGs should thus lie somewhere between the {\tt liberal} and {\tt conservative} VYG fractions, i.e. they are delimited by the open and filled symbols in Fig.~\ref{fig:fyoungvsm_sfh}. 

While the {\tt conservative} estimates of the VYG fractions
 resemble the {\tt liberal} estimates at high mass, they yield flat trends of VYG fractions at $m<10^{10}\,\msun$ for the VESPA spectral models.
Hence, 
for the 3~adopted spectral models, the difference between the lower and upper limits of the VYG fractions is only important at low mass. 

The black points in Fig.~\ref{fig:fyoungvsm_sfh} represent our most
conservative estimate of the VYG fraction. We obtained these `3-way VYG'
fractions by requesting that all 3 spectral models agree that the median age
of {\tt conservative} VYG candidates is less than 1 Gyr, leading to only 9
3-way VYGs over all masses.
The shaded regions in Fig.~\ref{fig:fyoungvsm_sfh} show that 
the fractions of VYGs derived from the
Gallazzi model (using mass-weighted ages) are typically 10 times lower than
with the
STARLIGHT and VESPA models.

\begin{figure}
  \centering
  \includegraphics[width=\hsize,viewport=0 30 550 540]{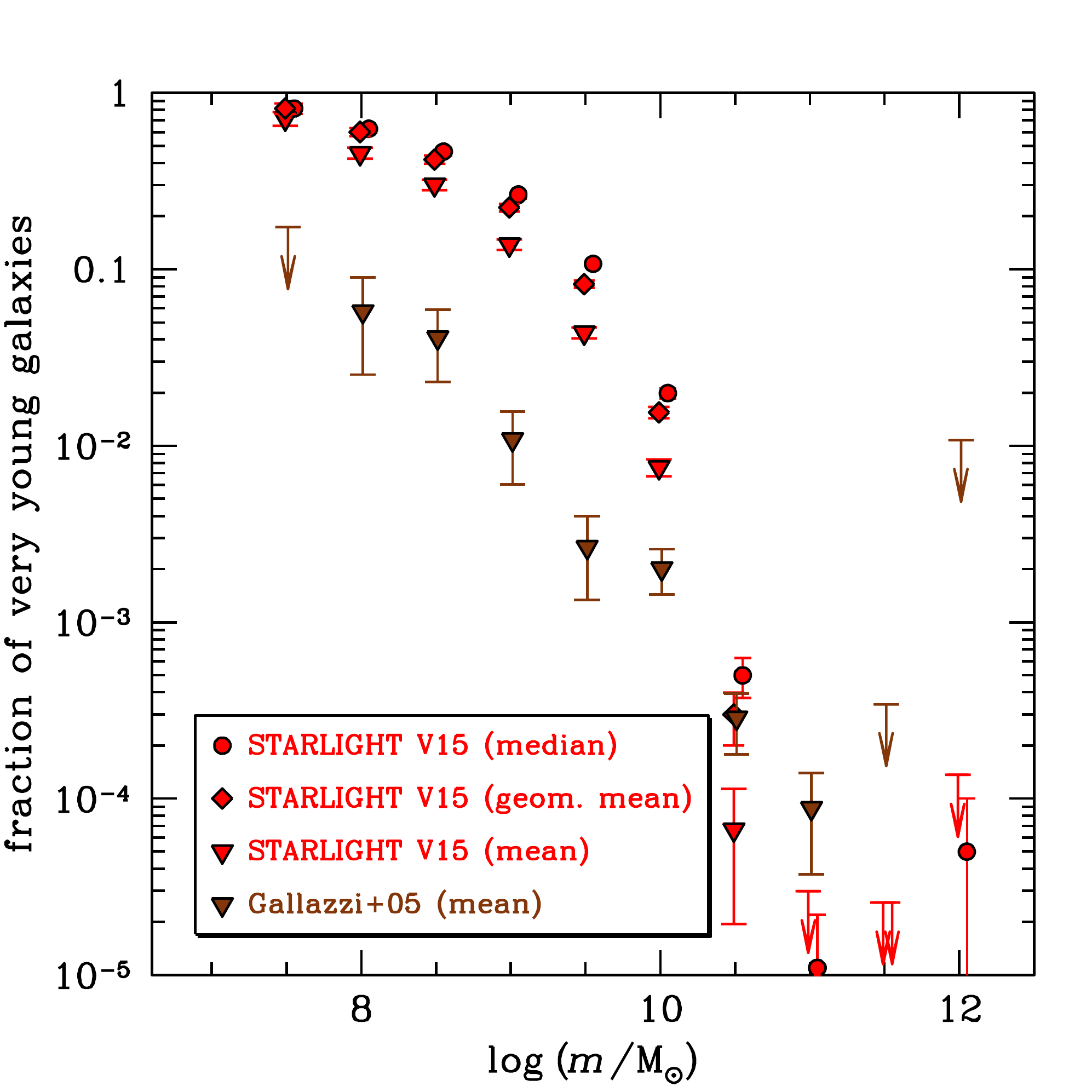} 
  \caption{Effect of the definition of age in the fraction of {\tt liberal} very young
    galaxies in the {\tt complete} sample as a function of stellar mass. The
    galaxy ages are computed as
    medians (\emph{circles}), or mass-weighted with
    geometric means (\emph{diamonds}) or arithmetic means (\emph{triangles}), 
    for STARLIGHT V15 (\emph{red}) and
    the Gallazzi model (\emph{brown}).
    The \emph{error bars} are binomial with 90 per cent Wilson upper
    limits. The abscissa are slighted shifted for clarity.}
  \label{fig:medvsmean}
\end{figure}
Since the SFHs of VYGs are not Gaussian in time but have long high-age tails,
the arithmetic mean mass-weighted age will be greater than the corresponding
median age.
This implies that the VYG fractions defined with arithmetic mean
mass-weighted ages should be smaller than with median ages.
Fig.~\ref{fig:medvsmean} confirms this idea using the STARLIGHT V15 model:
the VYG fractions are indeed lower with arithmetic mean mass-weighted
ages instead of median ages,
in particular at intermediate and high masses for the {\tt complete} sample, where the
decrease reaches a factor 4 (while the VYG fractions obtained with the
mass-weighted geometric  mean ages are
only slightly lower than  those found with the median ages).
Nevertheless, the Gallazzi model (using mean ages) yields even lower VYG fractions than the
STARLIGHT V15 model using similarly defined mean mass-weighted ages.

\begin{figure}
    \centering
    \includegraphics[width=\hsize,viewport=0 30 550 540]{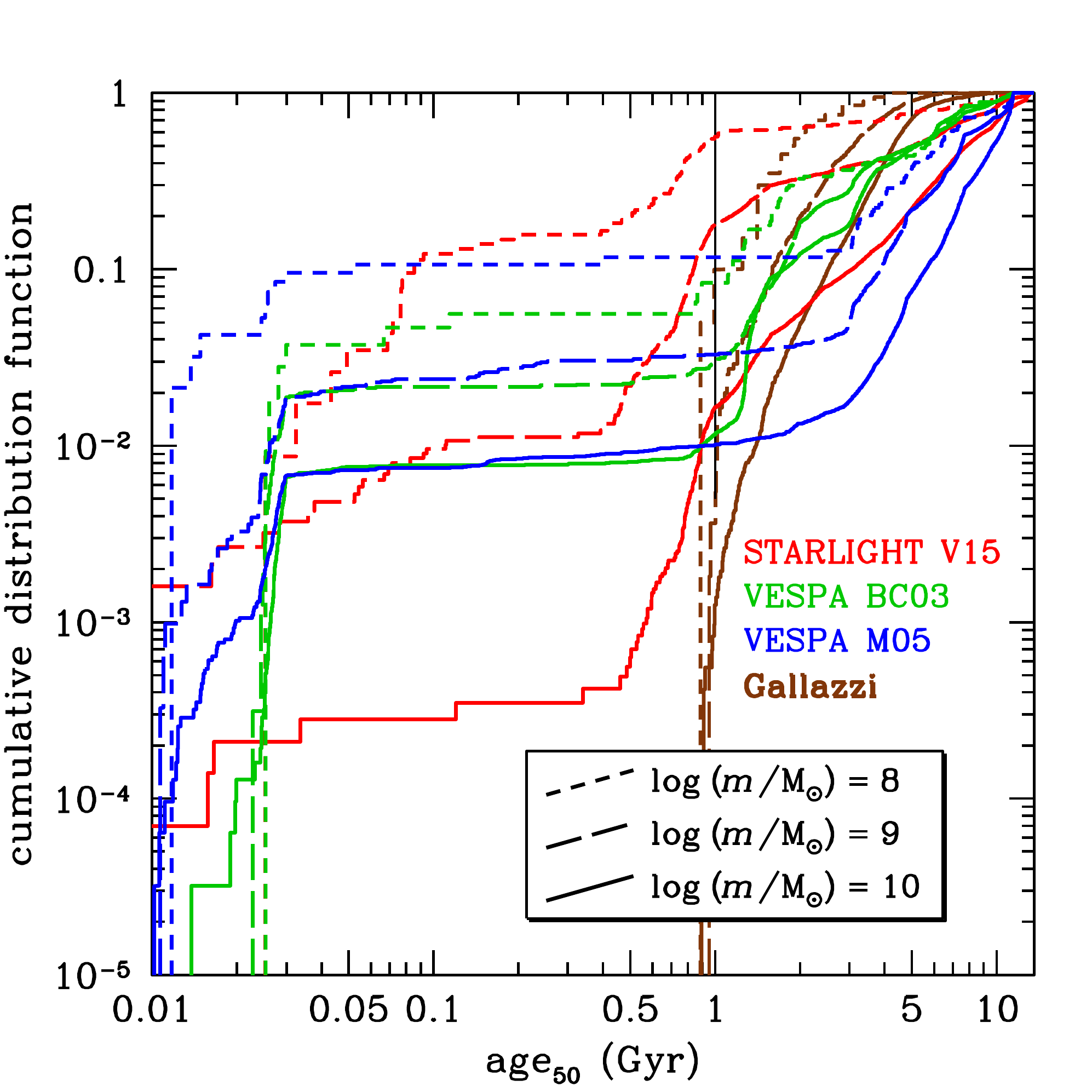}
    \caption{Cumulative distribution functions of median (mass-weighted mean
      for Gallazzi) age for galaxies in
      the {\tt complete} sample, for the four spectral models and for 3 mass bins of width $\pm0.5$ dex. The \emph{vertical line} illustrates our choice of 1 Gyr for defining VYGs.}
    \label{fig:cdfage}
\end{figure}

One question of interest is how the VYG fractions vary when the maximum
median age is changed to values below or above 1 Gyr.  Fig.~\ref{fig:cdfage}
illustrates the distribution of median ages for the 3 spectral models and in
3 bins of stellar mass. This allows us to estimate the fraction of galaxies
younger than any chosen age. For example, with STARLIGHT V15, the fraction of
galaxies younger than 100 Myr is typically one order of magnitude lower than
the fraction of galaxies younger than 1 Gyr (our VYGs). In contrast, the
VESPA models indicate nearly identical fractions of galaxies younger than 100
Myr and younger than 1 Gyr. This difference is related to the wide gap
between 30 Myr and 500 Myr in the distribution of ages seen in
Fig.~\ref{fig:pdfage} for these models (see the filled histograms
corresponding to the {\tt complete} sample).
Finally, whereas the Gallazzi model produces only rare galaxies younger than
1 Gyr, it produces roughly the same fraction of galaxies less than 2 Gyr than
the other 2 SFH codes.  

\section{Discussion}
\label{sec:discuss}

\subsection{Summary of results}

The spectral analyses of SDSS galaxies with
different SFH codes
and different single stellar population models lead to
different age-mass relations (Fig.~\ref{fig:agespecvsm}).
STARLIGHT shows downsizing at all masses down to the limit of $\lm = 8.5$
where we are confident that the data is complete, while VESPA
gives a rough plateau in age at masses below $10^{10}\,\msun$ and Gallazzi is
in between.

The derived VYG fractions fall with increasing mass, with possibly as
many as 50~per cent (STARLIGHT V15, {\tt liberal}) at $\lm = 8.5$ to less
than half a per cent at $\lm = 10.5$.
The  VESPA models show a shallower slope ($\approx\!\!-1.5$) with
stellar mass than STARLIGHT ($\approx\!-3$). Thus, VESPA models predict that VYGs should
occur at high masses, while STARLIGHT V15 does not.
The spectral model of \cite{Gallazzi+05} yields VYG fractions that resemble the lower
VESPA fractions at low mass and the lower STARLIGHT fractions at high mass.

\subsection{Caveats}
\label{sec:caveats}
Estimating the fraction of VYGs from SDSS spectra is difficult for several reasons. 
\begin{enumerate}
    \item  The SDSS spectra are limited to the inner regions of the galaxies, hence do not provide a global view of the SFH of each galaxy. We considered this aperture effect in our {\tt conservative} VYG sample, which discards not only AGN, but also VYG candidates with red colour gradients.
   The true VYG fractions ought to lie in between our {\tt liberal} and {\tt conservative} estimates.
\item Individual SFHs of galaxies suffer from degeneracies caused by their
  different epochs of SF, their metallicities, and their galactic
  extinction in the inner region sampled by the SDSS fibre.
In fact,  old stellar populations may not be 
revealed by the SFH modelling, as we will discuss in
Sect.~\ref{sec:hide}.
\item Our SFHs neglect stellar evolution and the disappearance of high-mass
  stars exploding as type II supernovae.
  Indeed, the VESPA code does not consider these effects,
  and for consistency we have
  chosen to use the same approach with STARLIGHT.
  \begin{figure}
    \centering
    \includegraphics[width=\hsize,viewport=0 30 550 540]{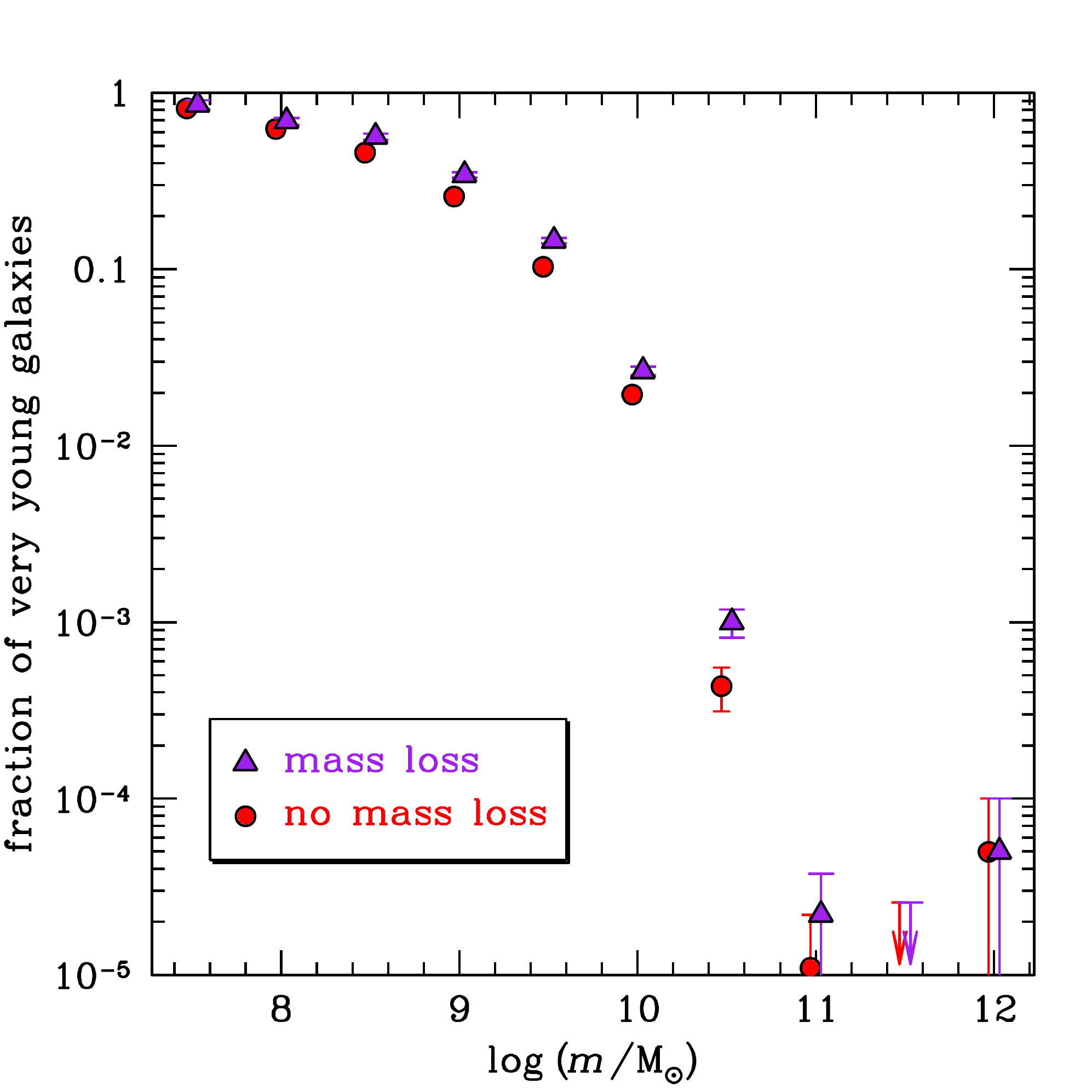} 
    \caption{Effect of considering mass loss in the fraction of VYGs versus
      stellar mass, using the STARLIGHT V15 model on the {\tt complete}
      sample.
      The \emph{error bars} are binomial, with 90 per cent Wilson upper
      limits.
The abscissa are shifted for clarity.
    }
    \label{fig:massloss}
  \end{figure}
  Supernova explosions affect massive stars that are short-lived, thus
  depleting more the stars formed earlier. Similarly, mass loss from stellar
  evolution
    takes time, thus decreases more the stars formed earlier. So both
    type II supernovae and stellar evolution should result in
    enhanced VYG fractions.
    Fig.~\ref{fig:massloss} confirms that mass loss does raise the VYG
    fractions, but never more than by 0.2 dex for low and intermediate galaxy
    stellar masses ($\lm \leq 10$).
  
\item There is disagreement between different spectral models using the same
  SFH code (e.g. BC03 vs. M05 with VESPA) and using different SFH codes
  (e.g. STARLIGHT vs. VESPA).

\item The spectral models may lead to worse fits for low-mass galaxies, which
  show the highest fractions of VYGs. 
  \begin{figure}
    \centering
    \includegraphics[width=0.7\hsize,angle=-90,viewport=30 0 572 729]{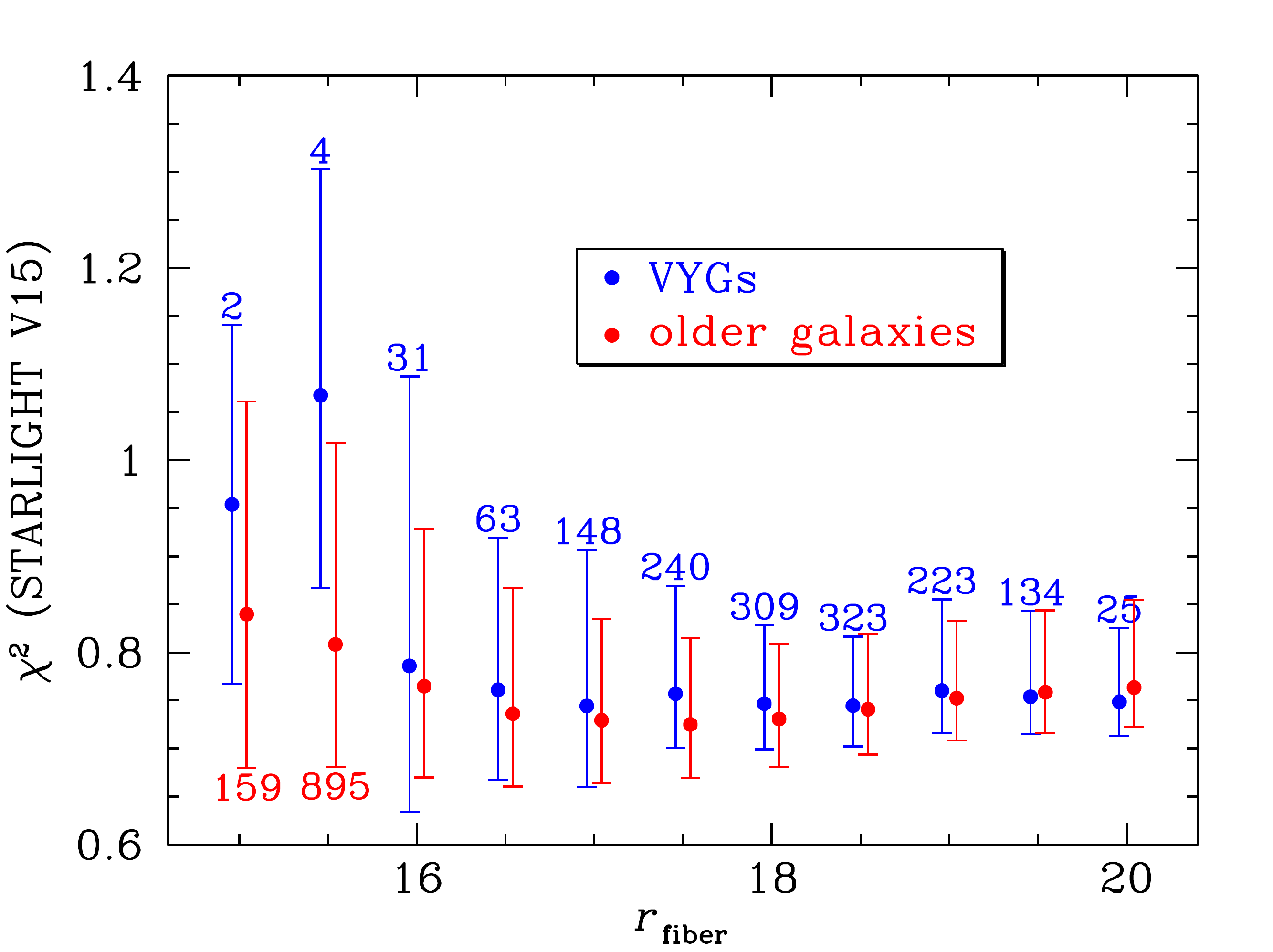}
    \includegraphics[width=0.7\hsize,angle=-90,viewport=30 0 572 729]{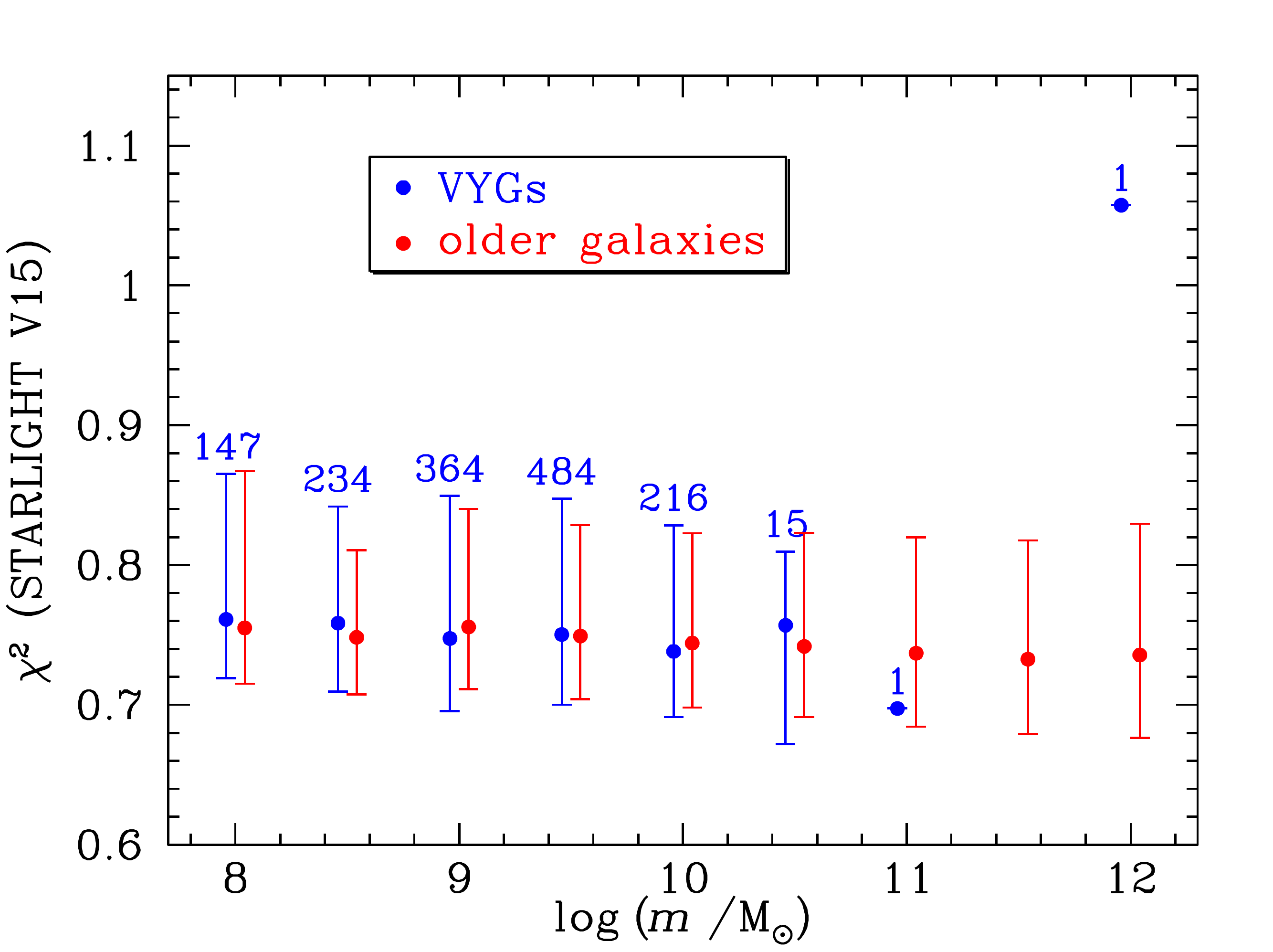}
    \caption{Spectral fit quality (reduced $\chi^2$) for the STARLIGHT V15 fits on the {\tt
        complete} sample, as a function of the fiber
      magnitude (proxy for the signal to noise of the spectrum, \emph{top})
      and of the galaxy stellar mass (\emph{bottom}), for both the VYGs
      (\emph{blue} and the older 
      galaxies (\emph{red}). The \emph{symbols} denote medians, while the
      \emph{error bars} represent 16th and 84th percentiles.
      The numbers above the error bars indicate the number of VYGs per bin.
    }
    \label{fig:chi2vsrfib}
  \end{figure}
  The top panel of
  Fig.~\ref{fig:chi2vsrfib} shows how the reduced $\chi^2$ of
  the spectral fit varies with fiber magnitude. Since all
  observations are in the same angular aperture of 3 arcsec, fiber magnitude
  is directly linked to the mean surface brightness in the fibre, which is an
  excellent proxy for the signal to noise ratio of the continuum.
  The figure indicates that for $r_{\rm fiber} > 15.75$, the spectral fits of
  the VYGs are only slightly worse than those of the older galaxies. This is
  likely due to the enhanced emission lines in the spectra of VYGs (see
  Sect.~\ref{sec:emline} and Fig.~\ref{fig:agevsEW} below).
  However, at very bright fiber magnitudes, the spectral fits are
marginally
  worse for the VYGs (as the numbers on the figure attest,
  this only concerns 6 galaxies, of which 4 have $\chi^2>1$).\footnote{The
    median reduced $\chi^2$ of the STARLIGHT V15 fits is 0.74 for the
    {\tt complete} sample instead of unity, suggesting that the
    uncertainties of the fit are overestimated.}
  However, the older very bright galaxies also show a stronger dispersion in
  $\chi^2$ than less bright galaxies of the same age range, and their high numbers attest that this higher dispersion is
  statistically significant. 
    The bottom panel of Fig.~\ref{fig:chi2vsrfib} shows that the spectral
    fits of VYGs and older galaxies differ very little when $\chi^2$ is
    plotted versus galaxy stellar mass.

    We also ran STARLIGHT V15 for
    I~Zw~18,\footnote{I Zw 18 is too close to satisfy our criterion
      \ref{zrange}.}
    whose stellar mass
is only $10^7\,{\rm M}_{\odot}$ \citep{Papaderos&Ostlin12,Izotov+18}.
We found a median age of only 30 Myr (our lower allowed age), with 100 per
cent of the stars younger than 100 Myr. Theses ages are much lower than those
derived from CMDs (less than 500 Myr according to \citealp{Izotov&Thuan04}
and over 1 Gyr according to \citealp{Aloisi+07}).
This galaxy is known to have strong nebular emission, not only in lines
(EW(H$\alpha) = 410\,\rm \AA$), but also in the continuum
\citep{Papaderos&Ostlin12}.
However, in the spectral range of SDSS, the contribution of the nebular
emission to the continuum rises with wavelength, reaching half of the
continuum at 8200$\,\rm \AA$ (fig.~9b of \citealp*{Izotov+11}).
Since we do not subtract the nebular contribution to the continuum, the
galaxy appears redder, leading the spectral modeling of stellar populations
to overestimate the ages.
Therefore, according to STARLIGHT V15,
the median age of I~Zw~18 should be even lower than 30 Myr.

\item Regardless of the quality of the fits, it is very difficult to extract
  SFHs of galaxies with non-negligible
  young stellar populations, since young stars have so high luminosity/mass
  ratios compared to older ones.
\end{enumerate}

\subsection{Comparison to the literature}

The only other study known to us in estimating fractions of very young
galaxies per bin of stellar mass is that of \cite{Dressler+18}.
These authors
derived SFHs from galaxy SEDs defined by $R=20-50$ prism observations in the optical band, combined with $ugrizJK$ broadband photometry,
following the prescriptions described in \cite{Dressler+16}. They used single stellar population models from \cite*{Conroy+09}, based on the BaSeL 3.1 library \citep{Lejeune+98}, the Padova isochrones \citep{Marigo&Girardi07,Marigo+08}, and an IMF that is virtually identical to that of \cite{Chabrier03}.

\begin{figure}
    \centering
    \includegraphics[width=\hsize,viewport=0 30 550 530]{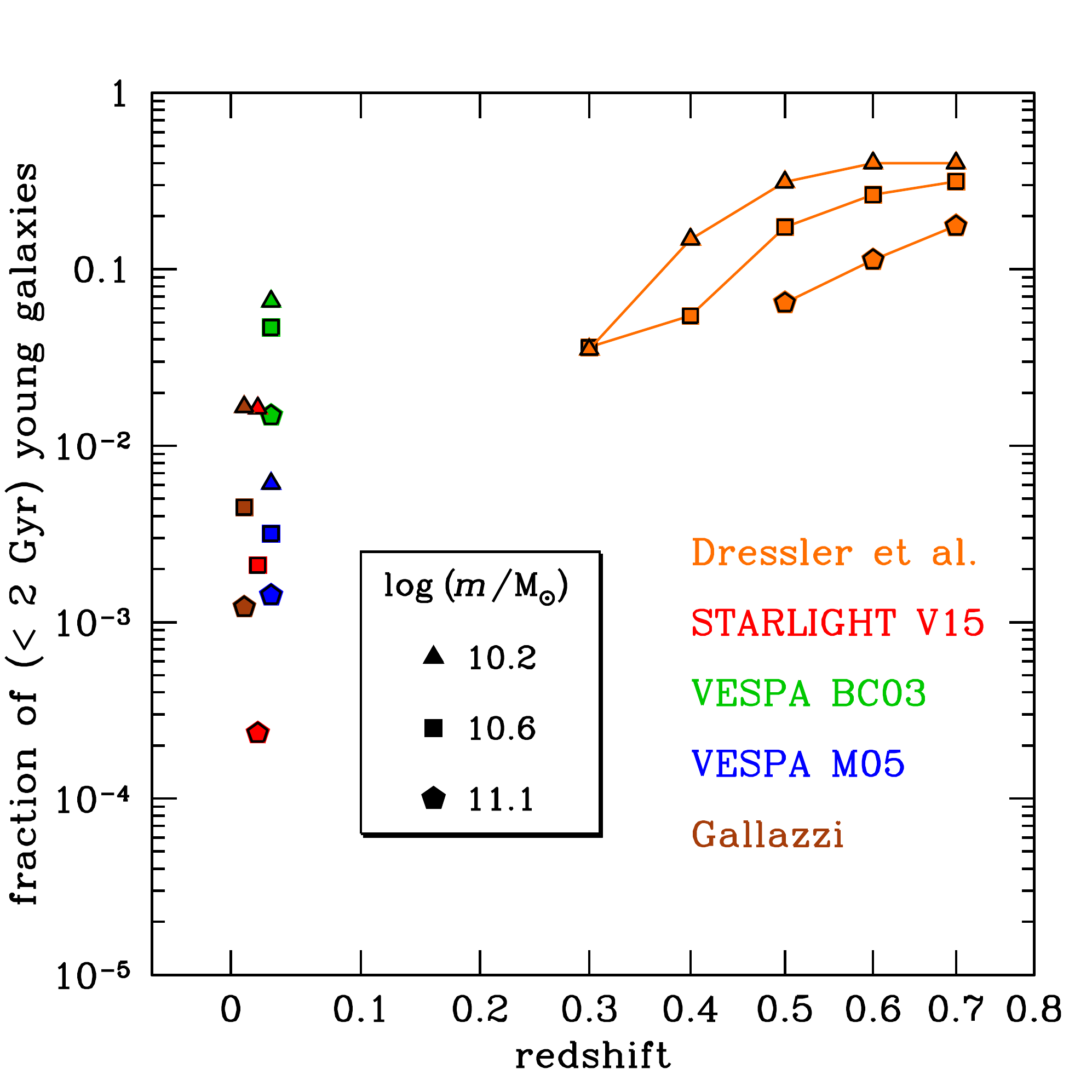}
    \caption{
      Comparison of our analysis of the SDSS Main Galaxy Sample with the
      analysis of Dressler et al. (2018) of a higher redshift sample of the
      mass dependence of the fraction of `young' galaxies, defined here to
      have over half their stellar mass formed within 2 Gyr before the epoch
      of observation, versus redshift in 3 bins of stellar mass.
The SDSS fractions (i.e. low redshift) are for the {\tt conservative} definition
      using the {\tt complete} sample, with slightly shifted abscissas for
      clarity. } 
    \label{fig:Dresslervsz}
\end{figure}
 Note that \citeauthor{Dressler+18} selected their
galaxies at $3.6\,\rm \mu m$, i.e at typically $2.2\,\rm \mu m$ in the rest
frame. 
While galaxies evolve more slowly in the near-IR than in the optical
$r$ band used in SDSS, there is still significant evolution of the
mass-to-light ratio in the $K$ band between 1 and 13 Gyr for single stellar
populations, by a similar fraction as in the $r$ band (see. fig. 5 of \citealp{Charlot&Bruzual91}), but there is no evolution of $M/L_K$ after 4 Gyr, contrary to $M/L_r$. Therefore, the sample of \cite{Dressler+18} may be incomplete for high mass-to-light ratio galaxies. However, we noted in Sect.~\ref{sec:zmax} that this incompleteness is smaller at the high stellar masses probed by \cite{Dressler+18}.   

As seen in Fig.~\ref{fig:Dresslervsz}, the young galaxy fractions measured by
\citeauthor{Dressler+18} decrease with decreasing redshift, as expected,
since the redshifts considered are past the cosmic noon  ($z\approx2$),
the epoch
when the cosmic star SF rate density is highest.
The comparison of their fractions with ours involves an extrapolation of
their fractions to the low redshifts probed by the {\tt complete} subsamples
of the SDSS/MGS ($\left\langle z\right\rangle\approx 0.07$).
The high young galaxy fractions found with the VESPA BC03 spectral model
(also seen in Fig.~\ref{fig:cdfage} for these high masses) appear
inconsistent with these extrapolations.
However, there
is no obvious conflict between the young galaxy fractions found by
\citeauthor{Dressler+18} and those deduced from our other three spectral
models.

\subsection{Star formation histories}

\begin{figure}
\centering
  \includegraphics[width=\hsize,viewport=0 30 550 710]{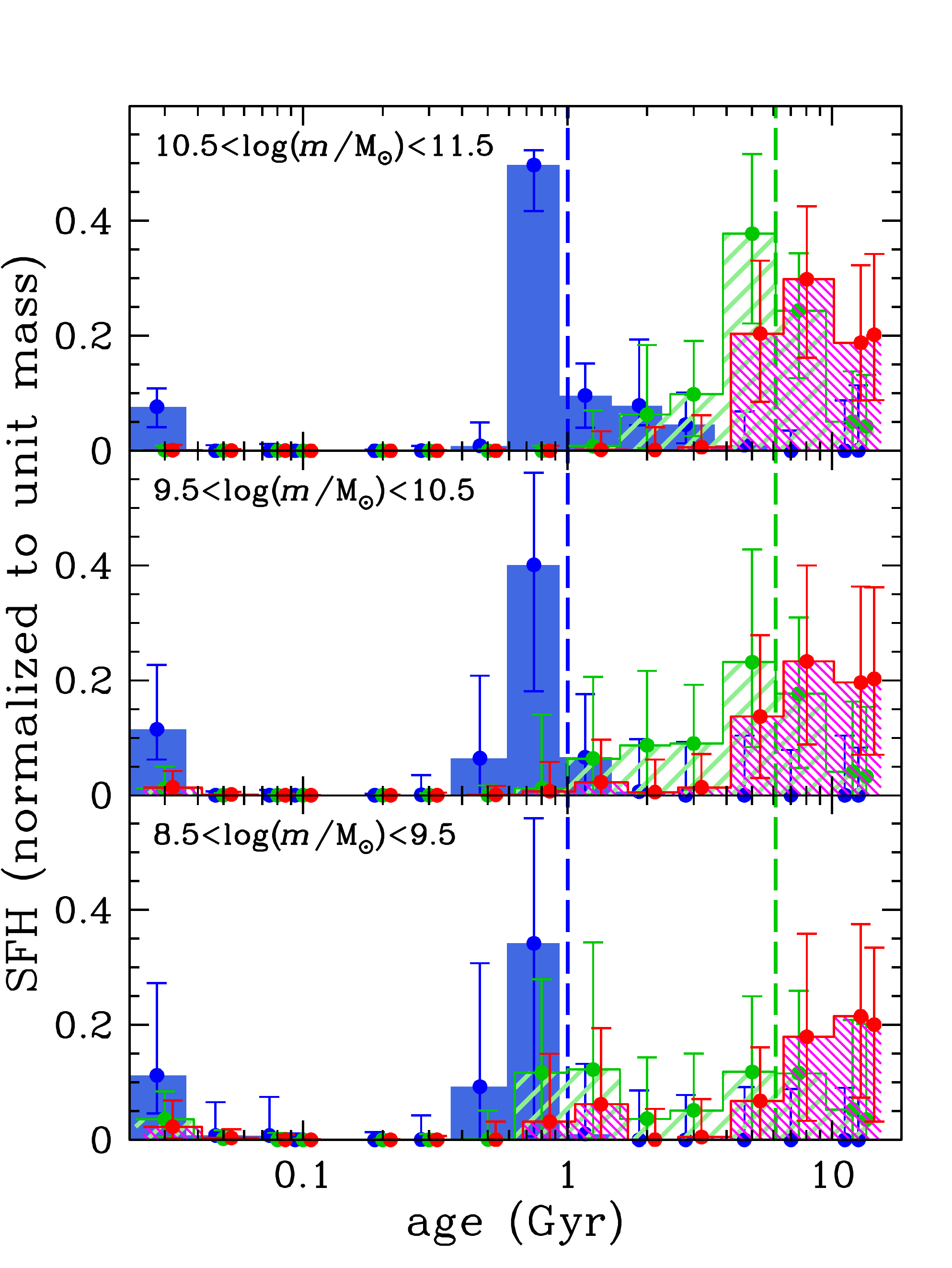} 
\caption{Star formation histories for VYGs (\emph{blue}), 
  intermediate age galaxies ($1 < \rm age_{50} < 6.1\,Gyr$, \emph{green})
  and old galaxies ($\rm age_{50} > 6.1\,Gyr$, \emph{red}) for 3 bins of final stellar
  mass, for the STARLIGHT V15 
  model, using the {\tt complete} sample.
  These star formation histories account for stellar evolution and the
disappearance of massive stars by supernova explosions.
The \emph{points} indicate the median star formation history, while the \emph{error bars}
indicate 16th and 84th percentiles. The \emph{vertical dashed lines} delimit the
maximum allowed median ages for the very young and intermediate galaxy classes.
}
\label{fig:sfh3}
\end{figure}

One can go further and ask whether the SFHs of VYGs  are such that the
distribution of stellar ages has a narrow peak
slightly below 1 Gyr or is more spread out. We also explore
how these SFHs vary with final galaxy
stellar mass, and how varied are the SFHs of VYGs of given
stellar mass.

Fig.~\ref{fig:sfh3} displays, for 3 bins of $z$=0 stellar mass,
the SFHs of VYGs (blue symbols) as well as of
intermediate-age galaxies, which we define as having median ages between 1 and 6.1 Gyr, and
of old galaxies (older than 6.1 Gyr).
The figure shows the STARLIGHT SFHs, accounting for stellar evolution and the disappearance of
massive stars from type II supernova explosions.

At high mass, the VYGs display an intermediate-age (1-4 Gyr) stellar population,
and virtually no population younger than 600 Myr, except for an extremely
young population of 30 Myr. This is probably an unavoidable artefact
of the over-sensitivity of SFH codes to very young stellar populations.
At lower masses, the VYGs no longer display the 1-4 Gyr
population, but progressively enhance their younger 0.4-0.6 Gyr population.
In other words, the SFHs of VYGs, expressed in log age, are positively skewed at high masses and
negatively skewed at low masses.
One also notices an increase in the diversity of the SFHs of VYGs towards lower galaxy stellar masses.
Similar trends are seen for intermediate-age galaxies (green symbols),
indicating a general trend of higher age diversity
at lower masses (see Fig.~\ref{fig:pdfage}).

\subsection{Comparison to predictions from models of galaxy formation}

\begin{figure}
\includegraphics[width=\hsize,viewport=10 30 570 515]{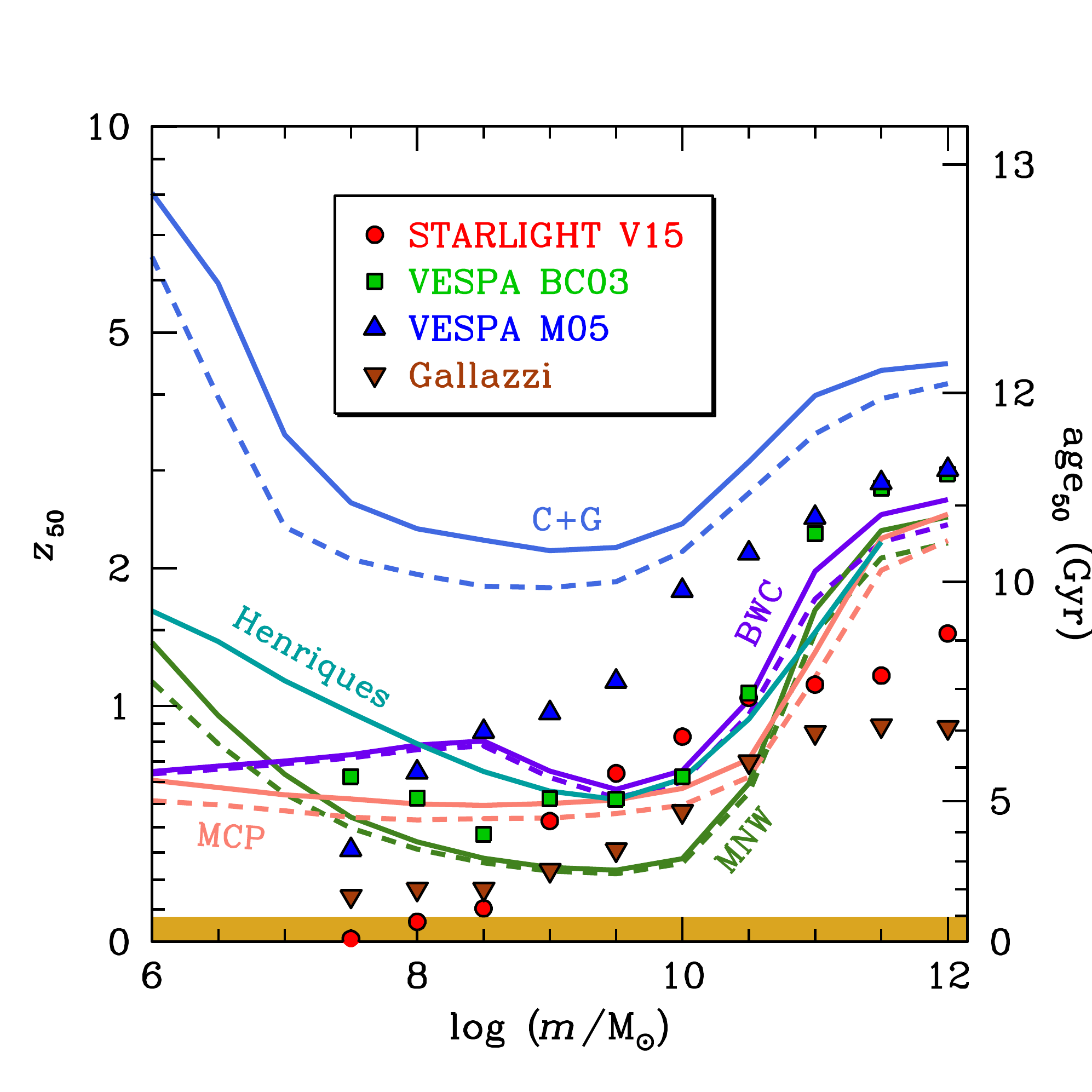} 
\caption{Median (over each mass bin) of median stellar ages versus galaxy stellar mass,  measured in the {\tt complete} sample of the SDSS Main Galaxy Sample with 3
 spectral models (\emph{symbols}) and predicted in
 galaxy formation models
  (\emph{lines}):
the physical model of Cattaneo et al. (2011) with  a more realistic truncation at low mass from Gnedin (2000) [\CG] (\emph{royal blue}), 
the empirical (abundance matching) models of Moster et al. (2013) [\MNW] (\emph{olive green}) and
Behroozi et al. (2013) [\BWC] (\emph{violet}), where all three model the stellar mass a a function of halo mass and redshift; the model 
 of Mutch et al. [\MCP] (\emph{salmon}), which has the stellar mass growth rate as the halo mass growth rate times a function of halo mass and redshift, and the Henriques et al. (2015) [\Hen]
 SAM (\emph{blue-green})  run on the
Millennium~II simulation (Boylan-Kolchin et al. 2009) for centrals only.
 The \emph{solid} and \emph{dashed lines} show the analytical galaxy formation models with the bursty and quiet halo merger schemes, respectively. 
The \emph{orange shaded region} highlights very young galaxies.
\label{fig:z50vsm_all}}
\end{figure}
\nocite{BoylanKolchin+09}

We now compare the spectral models in this paper with the galaxy formation models described in Paper~I \citep{Tweed+18}.
There, we had considered three analytical models where the galaxy stellar mass is a function of $z$=0 halo mass and redshift: one physical model \citep{Cattaneo+11}, improved with a more realistic cut off at low halo masses from the hydrodynamical simulations of \cite{Gnedin00} [\CG], and two empirical models (\citealp{Moster+13} [\MNW], \citealp{Behroozi+13} [\BWC])] that use abundance matching to link galaxy stellar masses with their halo masses.
We also considered a model where the galaxy stellar mass growth rate is equal to its halo
mass growth rate times a function of the $z$=0 halo mass and of redshift
(\citealp{Mutch+13} [\MCP]). Finally, we considered the SAM of
\cite{Henriques+15}, which has the advantage of having environmental effects
implicitly incorporated, but the disadvantage of worse mass resolution and
poorer statistics.
Note that the galaxy formation models were limited to central galaxies, either explicitly (for \Hen) or by construction of the Monte-Carlo halo merger trees (analytical models), while the spectral models were run on all galaxies. However, in Paper~I, we found that the fraction of VYGs  differs little between centrals and satellites in the \Hen\ SAM. 

Fig.~\ref{fig:z50vsm_all} compares the age-mass relations of the galaxy
formation models derived in Paper~I with those of the spectral models (also shown in Fig.~\ref{fig:agespecvsm}).
The spectral models exhibit a shallower decrease of ages with decreasing
stellar mass (i.e. shallower downsizing of mass) between $\lm = 10$ to  11.5.
While several galaxy formation models (\CG, \MNW, and \Hen)  predict \emph{upsizing} at masses below $10^9\,\msun$, this is not seen in the spectral models run on SDSS galaxies, except in the VESPA BC03 model.

\begin{figure}
\centering
\includegraphics[width=\hsize,viewport=60 40 550 730]{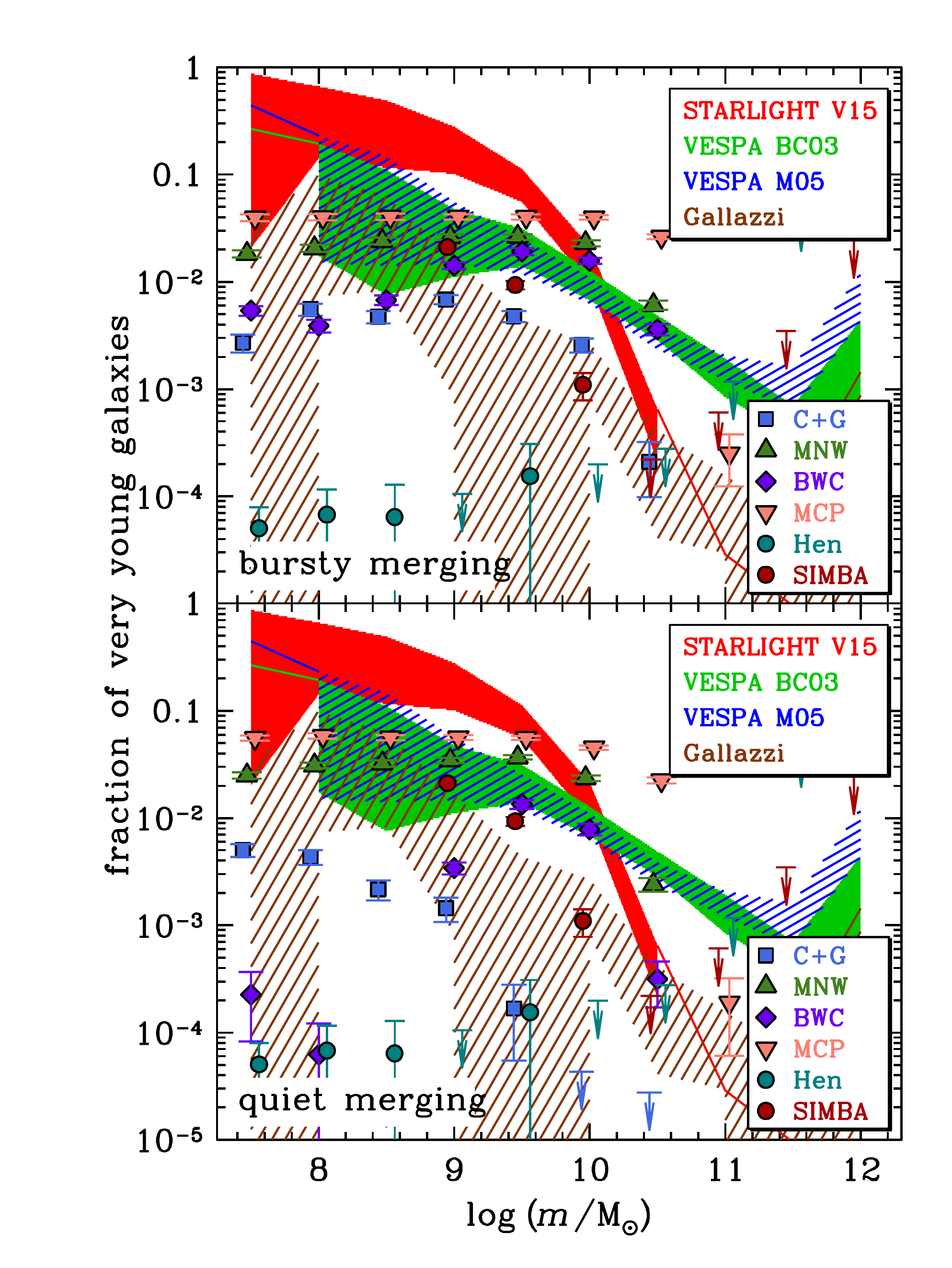} 
\caption{Fraction of very young galaxies versus stellar mass: constraints from the {\tt complete} sample of SDSS galaxy spectra (\emph{shaded regions} and \emph{lines}) confronted to the predictions from the galaxy formation models (\emph{symbols}, with same acronyms as in Fig.~\ref{fig:z50vsm_all}) of
Paper~I, for bursty (\emph{top}) and quiet (\emph{bottom}) halo merging. 
The \emph{upper} and \emph{lower envelopes} of the shaded regions are respectively the 84th percentiles of the {\tt liberal} VYG sample and the 16th percentiles of the {\tt conservative} VYG sample (no AGN and blue colour gradients). The \emph{lines} are the 84th percentiles of the {\tt liberal} VYG sample. 
The error bars for the  `Hen' (Henriques et al. 2015) SAM
and the SIMBA simulation (Dav\'e et al. 2019)
are binomial and the upper limits are $1\,\sigma$.
In the first 4 models, the galaxies are weighted, hence the error bars are estimated from 100 bootstraps.
\label{fig:modelsObs1}
}
\end{figure}

Can the differences in the VYG fractions of SDSS galaxies obtained with different spectral models and/or different ({\tt conservative}/{\tt liberal}) treatments of the selection effects be understood by comparing those results to the predictions of VYG fractions from models of galaxy formation? 
Fig.~\ref{fig:modelsObs1} compares the
VYG fractions derived from the SDSS spectra with 
the predicted fractions from galaxy formation models obtained in Paper~I, the
only study known to us that uses galaxy formation/evolution models to predict the fractions of VYGs versus stellar mass.
The two panels are for the two schemes of halo merging, one that comes with starbursts, and the other one that does not.

One first notices that the  different galaxy formation models display a much wider range of predicted VYG fractions than the spectral models do, although in the {\tt bursty} halo merging scheme the analytical models agree fairly well with one another. It is the SAM that disagrees the most by predicting orders of magnitude fewer VYGs.

The VYG fractions predicted by the galaxy formation models vary little with galaxy stellar mass, while the spectral models predict decreasing {\tt liberal} VYG fractions with increasing stellar mass.
One exception is the \BWC\ model in the {\tt quiet} merging scheme, which shows a peak in VYG fraction at $\lm=9.5$.

The match between the VYG fractions versus mass based on galaxy formation model predictions and on spectral modelling is generally not good. In particular, 
the STARLIGHT spectral model yields much higher {\tt conservative} VYG
fractions at $m<10^9\,\msun$ than any galaxy formation model predicts. Also,
the \Hen\ SAM predicts at least 100 times lower VYG fractions than even the {\tt conservative} spectral models. 
At $m = 10^{8.5} \,\msun$, the \Hen\ SAM predicts 1000 times fewer VYGs than the conservative estimate of the STARLIGHT V15 model.
However a closer look reveals that, in the {\tt bursty} halo merging scheme,
the \BWC\ model predicts VYG fractions that match well the VESPA VYG
fractions.

Fig.~\ref{fig:modelsObs1} also shows the first predictions of the fractions
of VYGs versus stellar mass at $z=0$ from a cosmological hydrodynamical
simulation, here the $100 \, h^{-1} \, \rm Mpc$ {\sc SIMBA}
simulation \citep{Dave+19}. This simulation has a more refined treatment of
AGN feedback for low and intermediate mass galaxies, and 
reproduces better observational properties of galaxies at $z=0$ than
analogous simulations of
the same resolution in space and mass ($1 \, h^{-1} \, \rm kpc$). 
Moreover, hydrodynamical simulations 
should be more realistic than the analytical models and the SAM, thanks to their
much better spatial resolution and their more refined subgrid physics (in
particular the AGN feedback in {\sc SIMBA}).
The VYG fractions found in {\sc SIMBA} are much higher  than in the SAM (up
to over 300 times at $\lm=9$), but
comparable to the analytical models, and quite similar to the upper envelope
of the Gallazzi spectral models.
However, the VYG fractions
are
overestimated at the lowest masses, because galaxies near the
lowest resolved mass
cannot form early and galaxy formation is delayed, as we found in Paper~I for
the \Hen\ SAM run on the lower resolution Millennium simulation.
In the SIMBA simulation, the SFHs appear to have converged for 256 star
particles, corresponding to $\lm = 9.3$. 

\subsection{Comparison to emission-line galaxies}
\label{sec:emline}
\begin{figure}
    \centering
    \includegraphics[width=\hsize,viewport=0 30 550 540]{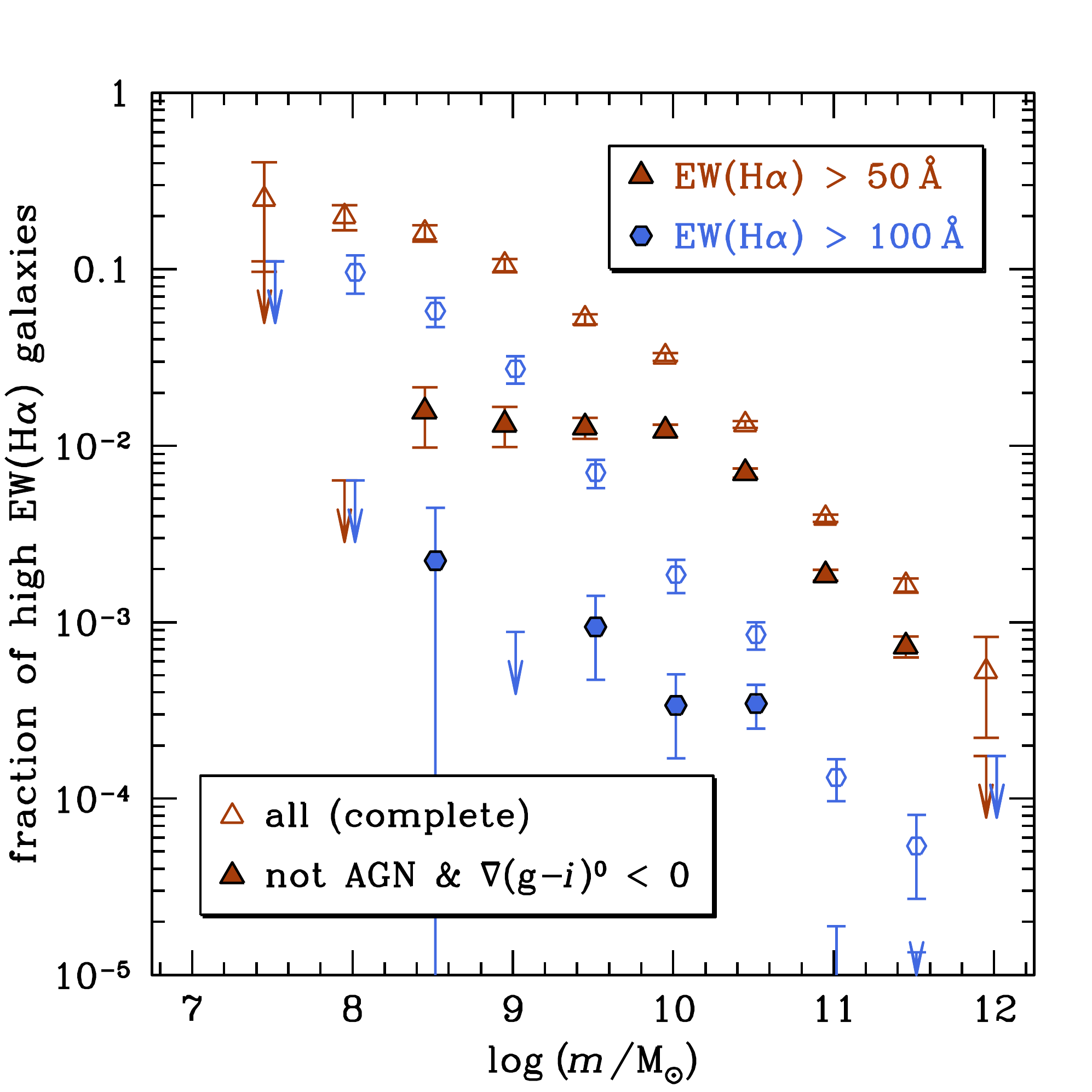}
    \caption{Fraction of galaxies in {\tt complete} sample (for all 3
      spectral models) that have their H$\alpha$ emission-line with
      equivalent width (determined from MPA/JHU, now using the sign
      convention of positive EW for emission lines) above a threshold, as a
      function of the mean of the log masses of the 3 spectral models (with
      weights 1/2, 1/4, and 1/4 for STARLIGHT V15, VESPA BC03, and VESPA M05,
      respectively. The \emph{open symbols} are the fraction of high EWs
      among all galaxies in the {\tt complete} sample, while the \emph{filled
        symbols} represent the fraction of galaxies in the {\tt complete}
      sample that have high EWs, are not AGN and have blue colour gradients.
      The \emph{error bars} are binomial with 90 per cent Wilson upper
      limits.
      The abscissa are slightly shifted for clarity.
    }
    \label{fig:fEWvsm}
\end{figure}

Strong emission lines in galaxy spectra are another indicator of very young stellar populations. 
As already mentioned in Sect.~\ref{sec:tests},
by measuring the ratio of the line flux over the continuum, the EW of an emission line essentially measures the ongoing SF rate over its integral over time, which to first order is the specific SF rate. 
According to \cite{Leitherer+99} using the {\sc starburst99} code, a single
burst of SF is characterised by an (H$\alpha$) EW that decreases
4 Myr after the burst as $t^{-4}$, becoming negligible ($<0.1\,$\AA) after 1 Gyr. On the
other hand, if the SF rate is uniform in time, the EW falls off much more slowly, roughly as $t^{-1/3}$, leading to values of 30 and 150\,\AA, depending on whether the initial mass function is truncated at 30 or $100\,\msun$ at the high end.

Fig.~\ref{fig:fEWvsm} displays the fraction of galaxies with high EWs; 1) those with EW $>$ 50 \AA\ and 2) those with EW $>$ 100 \AA. These fractions decrease with increasing mass, roughly in the same way as the fractions of VYGs. In particular, the conservative case of the fraction of galaxies in the {\tt complete} sample that have high H$\alpha$ EWs $>$ 50\,\AA, are not AGN, and have blue colour gradients (solid symbols), shows the same plateau at $\simeq 1.5$ per cent as does the fraction of VYGs with the same constraints in the two VESPA models (Fig.~\ref{fig:fyoungvsm_sfh}).

\begin{figure}
    \centering
    \includegraphics[width=\hsize,viewport=0 30 550 540]{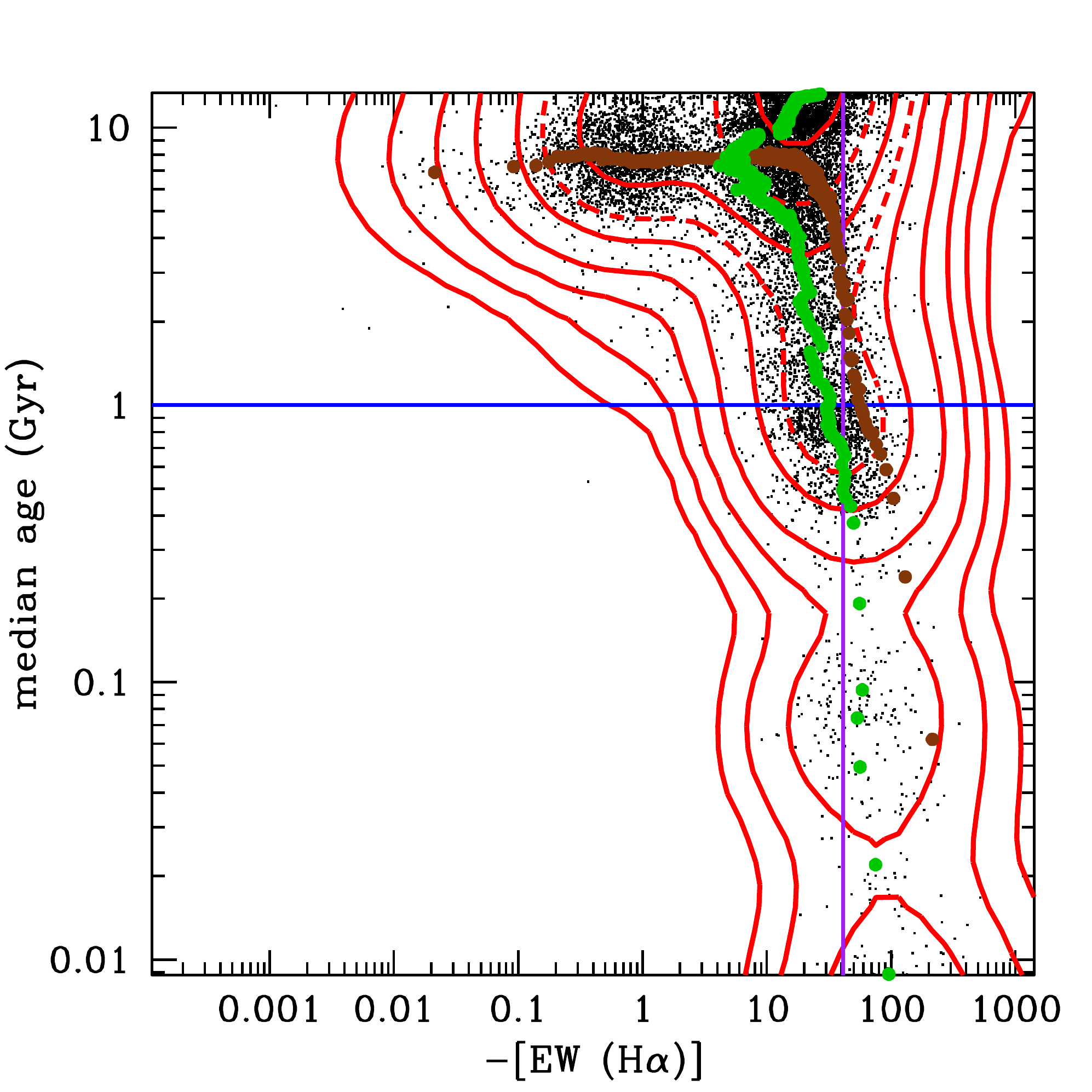}
    \caption{Median age (from the STARLIGHT V15 model fit to the continua of
      SDSS spectra) versus H$\alpha$
      equivalent width (from MPA/JHU) for the non-AGN galaxies
      in the {\tt clean} sample. Only 1 random galaxy among 20 is shown, while the
      contours (in steps of 0.5 dex in density -- and 2 intermediate ones
      shown by \emph{dashes} -- estimated in 40$\times$40
      cells) are for the full galaxy sample (the limits of the figure correspond to the limits of the data, after discarding systems with $\rm -EW(H\alpha) < 0$).
      The \emph{horizontal blue line} corresponds to a median age of 1 Gyr
      (7th percentile),
      while the  \emph{vertical purple line} corresponds to  $\rm
      -EW(H\alpha) = 41\,\AA$
      (93rd percentile).
      The \emph{brown circles} show the medians of the median ages in equal number  bins of
      EW (1000/bin), while the \emph{green circles} show the medians of EW in
      equal number bins of median age.}
    \label{fig:agevsEW}
\end{figure}
Fig.~\ref{fig:agevsEW} compares the ages and H$\alpha$ EWs of the {\tt
  clean} galaxy sample.
One notices 3 classes of galaxies in this figure:
a class of low EW with high
ages (upper left clump of points),
a dominant class of moderately high EW with high ages (upper right clump),
and a class of somewhat higher EW and low ages (lower clump).
Admittedly, the bimodal distribution of galaxy ages (even with STARLIGHT, see
Fig.~\ref{fig:pdfage}) may be an artifact of the
spectral modeling of the SFHs. 
Note that the dominant clump of high-EW old galaxies cannot be AGN, since we had filtered these out of the {\tt clean} sample for this figure.
Fig.~\ref{fig:agevsEW} displays
a clear anti-correlation  between ages and EWs ($r = 0.37$ from a Spearman rank correlation test, with a null $p$ value). However,
one might have hoped for an even stronger correlation, as can be seen by the
disagreement in the trajectories of median age vs EW (brown) and its
transpose (green).

\cite{Telles&Melnick18} analysed a sample of galaxies with EW(H$\alpha) >
50$\,\AA, EW(H$\beta) > 30$\,\AA\ and a rectangular selection of the upper
left portion of the BPT diagram. They fit parametric three-burst SFHs  to the
spectral energy distributions of these galaxies from the ultraviolet (UV) to the mid-IR. They found that the `young' stellar population contributes to typically less than 2 per cent of the stellar mass of their emission-line galaxies, and never more than 8 per cent. However, their `young' population is defined to be less than 10 Myr old, and unfortunately they do not provide the fraction of stellar mass in galaxies in their `intermediate' age population which spans 100 Myr to 1 Gyr in age.

\subsection{Number of very young galaxies in the SDSS Main Galaxy Sample}

\begin{figure}
\centering
\includegraphics[width=\hsize, viewport=0 30 550 540]{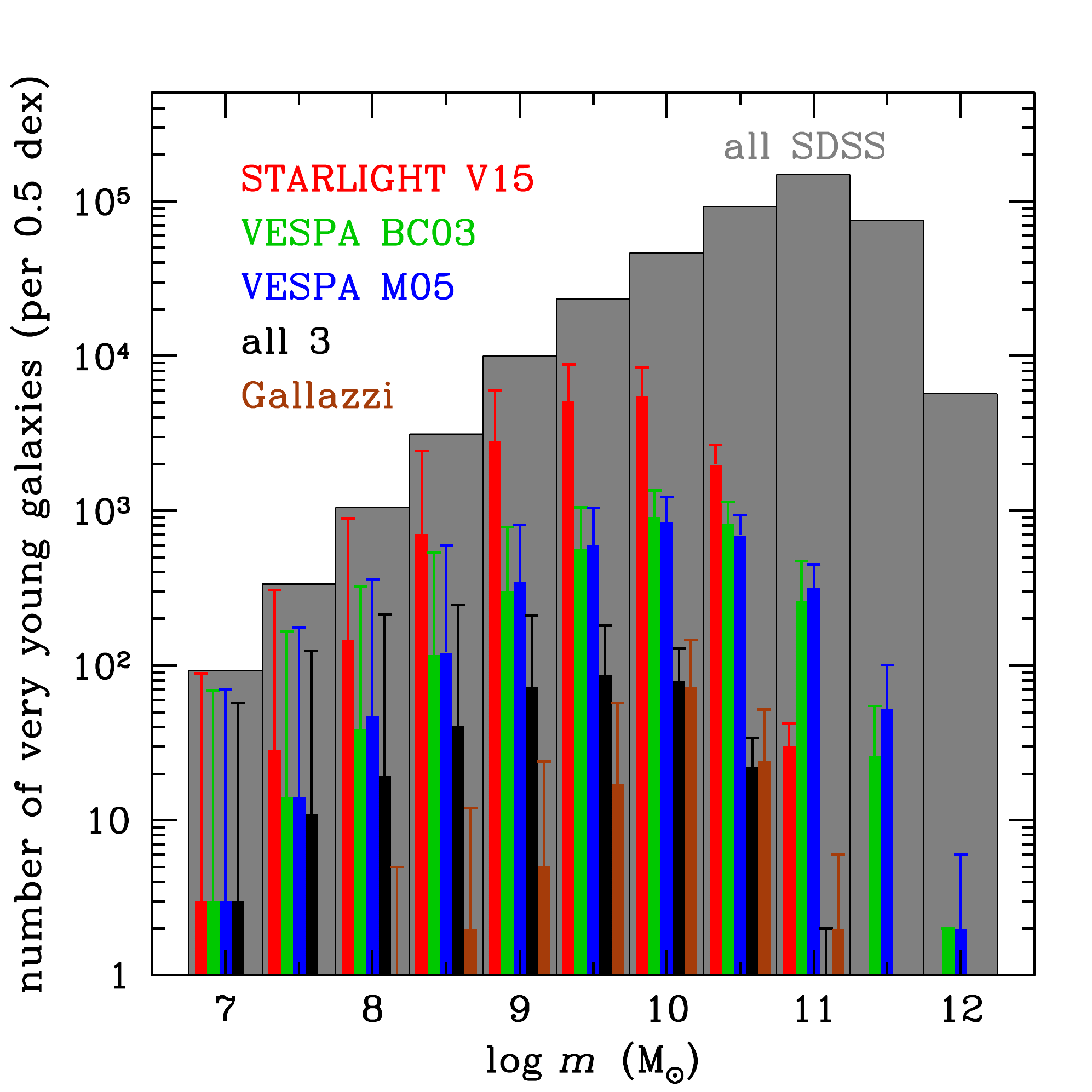}
\caption{Observed numbers of SDSS Main Galaxy Sample galaxies in {\tt clean}
  sample (\emph{grey bars}) and numbers 
  of  {\tt conservative} (\emph{thick bars}) and {\tt liberal}
  (\emph{error bars}) very young galaxies
  in the {\tt clean}
  sample,
  all as a function of
  stellar mass (averaged over the 3 models as in Fig.~\ref{fig:fyoungvsm_sfh}). The \emph{black bars} represent the subsample of these very young galaxies that are classified as very young in all three spectral models.
\label{fig:sdss}}
\end{figure}

\nocite{Tojeiro+09}

Given the non-negligible fractions of VYGs,
how many VYGs may lurk in the SDSS Main Galaxy Sample?
Fig.~\ref{fig:sdss} displays the numbers of VYGs in the {\tt clean} sample for the different spectral models, as well as the numbers of galaxies that are VYGs for all three spectral models (black bars).

The SDSS spectra analysed by STARLIGHT and VESPA  
lead to thousands of {\tt conservative} VYGs in the
{\tt clean} samples of the SDSS/MGS (16211, 3040, and 3010 for the STARLIGHT
V15, VESPA BC03, and VESPA M05 spectral models, respectively). These spectral
estimates lead to peaks in the number of VYG galaxies in 
the SDSS/MGS around $m = 10^{10} \,{\rm M}_{\odot}$.
This would mean that VYGs like I~Zw~18, but much more massive, are not so
rare, but rather ubiquitous.

There is also a large discrepancy in the predicted number of VYGs at large
stellar masses, between the VESPA models, according to which there are many VYGs with masses higher than $10^{11.25}\,\msun$ and the STARLIGHT V15 model, whose most massive VYG has $m=10^{10.82}\,\msun$.
The most massive {\tt conservative} VYG found by all three spectral models has 
$\lm = 10.35$ (STARLIGHT V15), 10.93 (VESPA BC03), 11.05 (VESPA M05), and corresponds to the same galaxy. Its weighted log mass is $\lm=10.67$.

\subsection{Can VYGs contain hidden old stellar populations?}
\label{sec:hide}
The higher fraction of VYGs found in our analysis of SFHs derived from SDSS
spectra, in comparison with the predictions from nearly all galaxy formation models (Fig.~\ref{fig:modelsObs1}), suggest that old stellar populations may be present in VYGs, but are hidden from our view and do not contribute strongly to the SED.
By selecting {\tt conservative} VYGs that have bluer global colours than
their fibre colours, we have attempted to eliminate galaxies that possess an old extended stellar population underlying the young stellar population. However, 
the old stellar population may have no effect on the global colour within the
extent of the galaxy in the SDSS image, and yet become increasingly important
with radius, dominating the integrated stellar mass at large radii. 

We examine this situation by considering the following simple model. Assume that the galaxy is composed of two stellar populations, one young and one old, 
which both follow S\'ersic surface brightness profiles,
\begin{equation}
  \Sigma(\theta) = \Sigma(0)\,\exp\left[-b(n)\left ({\theta\over \theta_{\rm e}}\right)^{1/n}
    \right] \ ,
  \label{Sersic}
\end{equation}
where $b(n) \simeq
  2\,n-1/3+0.009876/n$ \citep{Prugniel&Simien97}, 
with effective
(half-light) radii, $\theta_{\rm young}$ and $\theta_{\rm old}$, indices
$n_{\rm young}$ and $n_{\rm old}$, and with very different mass-to-light
ratios, $\Upsilon_{\rm young}$ and $\Upsilon_{\rm old}$, both assumed to be
independent of radius.
Our model includes the additional constraint that each
population contributes to half the total stellar mass within the fibre. This
is a conservative assumption, since our VYGs are selected to contain \emph{at
  least} half their stellar mass younger than 1 Gyr. We also assume that the
young population has a S\'ersic index $n_{\rm young}=1$ as expected for
exponential disks and choose $\theta_{\rm young}=3''$ (close to the median of the {\tt
  conservative} VYGs of 2.9~arcsec in the {\tt clean} sample).

Given that the integrated stellar mass (or luminosity), in cylindrical apertures, of the S\'ersic model follows (see \citealp{Graham&Colless97})
\begin{equation}
    m(\theta) = {2\pi\,n\over b^{2n}(n)}\,\gamma\left[2n,b(n)\left({\theta\over \theta_{\rm e}}\right)^{1/n}\right]\,\Sigma_0\,d^2\,\theta_{\rm e}^2 \ ,
    \label{L2}
\end{equation}
where $\Sigma_0$ is the central surface mass density, $d$ is the (cosmological angular) distance and $\gamma(a,x)$ is the lower incomplete gamma function, the ratio of total old mass over total young mass is
\begin{flalign}
    {m_{\rm old}\over m_{\rm young}} &= 
    {\left[m(\theta_{\rm fib})/m\right]_{\rm young}\over 
    \left[m(\theta_{\rm fib})/m\right]_{\rm old}}
    \label{mratio1} \\
    &=
    {
    P[2n_{\rm young},b(n_{\rm young})\left (\theta_{\rm fib}/\theta_{\rm young}\right)^{1/n_{\rm young}}]
    \over
     P[2n_{\rm old},b(n_{\rm old})\left (\theta_{\rm fib}/\theta_{\rm old}\right)^{1/n_{\rm old}}]
     } \ ,
     \label{mratio}
\end{flalign}
where $P(a,x) = \gamma(a,x)/\Gamma(a)$ is the lower regularised gamma function and where  equation~(\ref{mratio1}) comes from the equal contributions of the young and old populations to the mass within the fibre.

\begin{figure}
    \centering
    \includegraphics[width=\hsize,viewport=0 30 420 385]{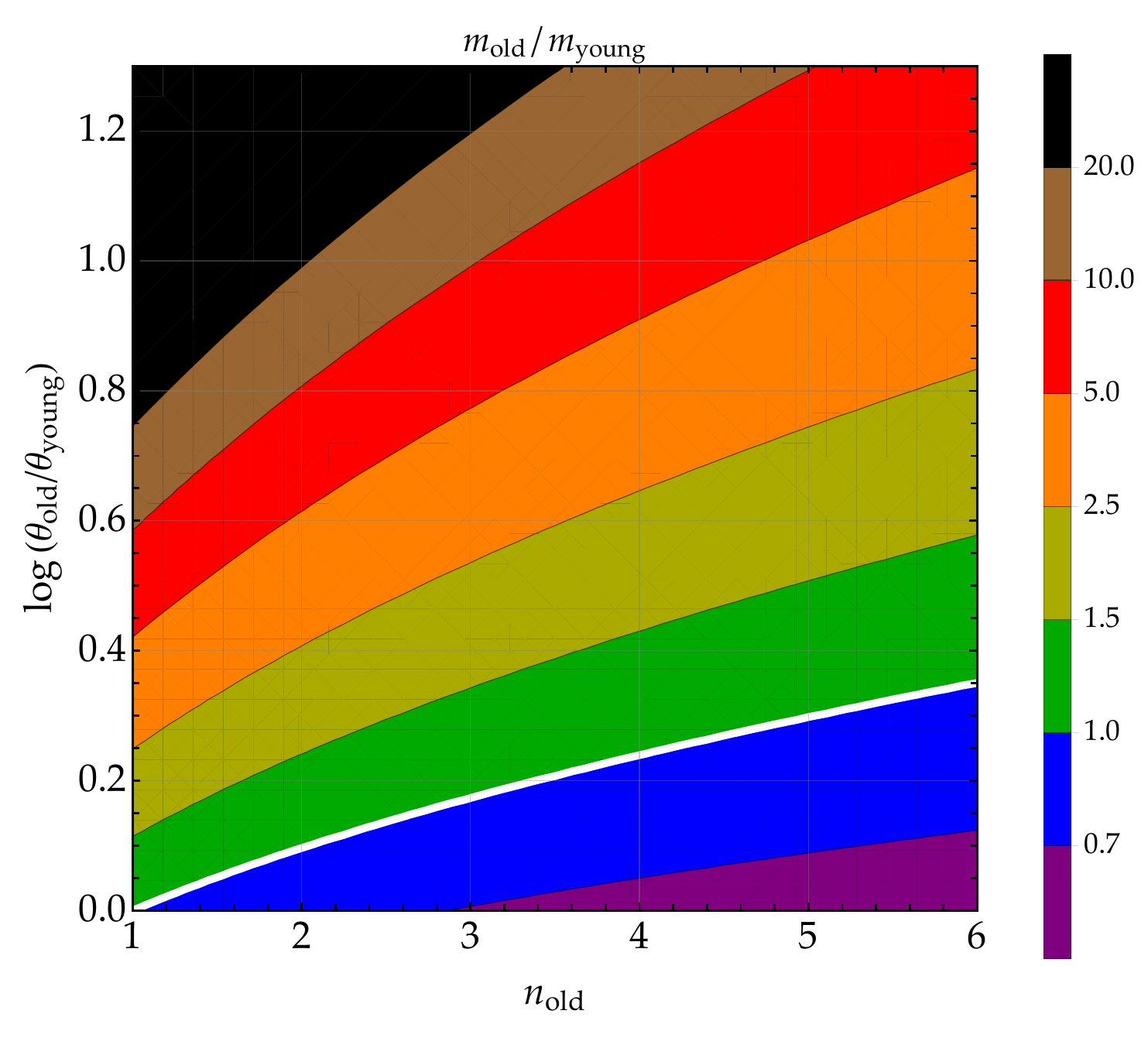}
    \caption{Ratios of old over young total masses, $m_{\rm old}/ m_{\rm young}$ (eq.~[\ref{mratio}]) as a function of the S\'ersic index of the old population, $n_{\rm old}$ and the ratio of effective radii $\theta_{\rm old}/\theta_{\rm young}$, assuming $n_{\rm young}=1$ and $\theta_{\rm young} = 3''$. The \emph{white curve}
    denotes a ratio of unity. }
    \label{fig:massratiocontours}
\end{figure}
Fig.~\ref{fig:massratiocontours} indicates that the  ratio
of total masses of the old versus young populations can be much larger than unity for sufficiently large effective radius of the old stellar population (the region above the white curve).

Could this old stellar population hide in our {\tt conservative} VYG sample whose galaxies have bluer global colours than their fibre colours? 
Since we assume that the mass-to-light ratio of each stellar population is independent of the aperture, their $(g-i)^0$ colours are also constants. 
Therefore, the integrated colour of the mix of the two populations at aperture $\theta$ is
\begin{flalign}
    (g\!-\!i)^0(\theta) &= -2.5\,\log \left[
    {L_g^{\rm young}(\theta) + L_g^{\rm old}(\theta)
    \over
     L_i^{\rm young}(\theta) + L_i^{\rm old}(\theta)
    } 
    \right]
    \nonumber \\
    &= -2.5\log\left [{\cal R}(\theta)\right] \ ,
    \label{gmieq}
\end{flalign}
where
\begin{equation}
    {\cal R}(\theta) = 
 {
    {\rm dex}\left[-0.4\,(g\!-\!i)_{\rm young}^0\right]+\lambda_i(\theta)\,
    {\rm dex}\left[-0.4\,(g\!-\!i)_{\rm old}^0\right]
    \over
    1+\lambda_i(\theta)
    } \ .
    \label{gmieq2}
\end{equation}
In equation~(\ref{gmieq2}), $\lambda_i(\theta)$ is the ratio of old to young $i$-band luminosities at the aperture of angle $\theta$:
\begin{flalign}
      \lambda_i(\theta) &= {L_{\rm old}(\theta) \over L_{\rm young}(\theta)} \nonumber \\
  & =   {\Upsilon_{i,\rm young}\over \Upsilon_{i,\rm old}}\,
      {
      \gamma\left(
      2n_{\rm old},b(n_{\rm old}) \,(\theta/\theta_{\rm old})^{1/n_{\rm old}}
      \right)
      \over 
        \gamma\left(
        2n_{\rm old},b(n_{\rm old})\,
        (\theta_{\rm fib}/\theta_{\rm old})^{1/n_{\rm old}}
        \right)} \nonumber \\
    & \quad \quad \times
  {
  \gamma\left(
  2n_{\rm young},b(n_{\rm young})\, (\theta/\theta_{\rm young})^{1/n_{\rm young}}
  \right)
      \over 
        \gamma\left(
        2n_{\rm young},b(n_{\rm young}) \,(\theta_{\rm fib}/\theta_{\rm young})^{1/n_{\rm young}}
        \right)}  \ ,
        \label{lambda}
\end{flalign}
for our adopted S\'ersic models and
our usual assumption of equal young and old contributions to the stellar mass within the fibre.

\begin{table}
\caption{Parameters of two-population model}
\begin{center}
  \tabcolsep 4pt
    \begin{tabular}{lcccccc}
    \hline
    \hline
    population & age & metallicity & $n$ & $\theta_{\rm e}$ & $\Upsilon_i$ & $(g-i)^0$ \\
    & (Gyr) & & (arcsec) & (solar) & \\
    \hline
    young & \ \,1 & solar &  1 & 3 & 0.56 & 0.52 \\
    old & 10 & solar & free & free & 2.52 & 1.16 \\
    \hline
    \end{tabular}
\end{center}
\parbox{\hsize}{Notes: The columns are:
1) population;
2) age;
3) S\'ersic index;
4) effective radius;
5) mass-to-light ratio in the $i$ band;
6) $(g-i)^0$ colour.}
    \label{tab:toy}
\end{table}
We illustrate our toy model by adopting
the STARLIGHT V15 spectral model, assuming that the young and old stellar
populations have respective ages of 1 and 10 Gyr, both with solar
metallicity.
Table~\ref{tab:toy} displays the $i$-band mass-to-light ratios and $(g-i)^0$
  colours of the young and old populations for the V15 single stellar
  population model,  determined
by cubic spline interpolation of both log $\Upsilon$ and $(g-i)^0$ vs. (log)
metallicity in the table for the BaSTI isochrones and the Kroupa Universal IMF (see Table~\ref{tab:sfhmodels}) in {\sf Magnitudes, colours and mass-to-light ratios for the SDSS filters (AB system)}.\footnote{http://www.iac.es/proyecto/miles/pages/photometric-predictions-based-on-e-miles-seds.php\label{fn:miles}} 

\begin{figure}
    \centering
    \includegraphics[width=0.85\hsize,viewport=0 90 370 310]{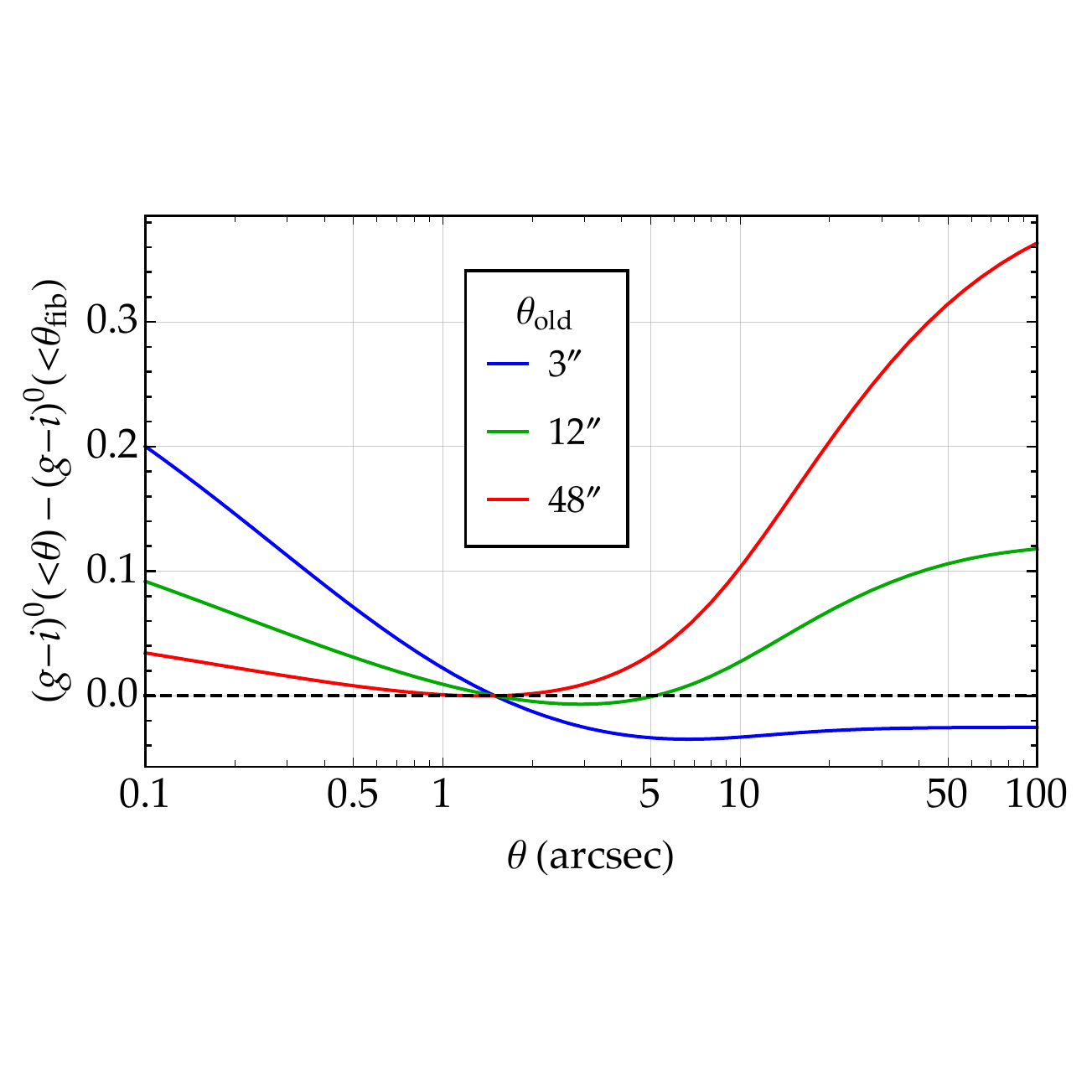}
    \caption{Colour profiles for a mixed population of  young and old stars, with $n_{\rm old}=3$ and other parameters from Table~\ref{tab:toy}.}
    \label{fig:gmiprofs}
\end{figure}
Fig.~\ref{fig:gmiprofs} shows that, contrary to intuition, the mix of a
young  stellar population with an old one does not necessarily lead to
a red colour gradient.
This can be understood by the differences in the
low and high index S\'ersic models: in a log-log plot the surface brightness
profile of an $n$=1 S\'ersic model is flat in the inner regions and falls
rapidly at large radii, while a higher index
has a much more
gradual decline at all radii. Thus, the high-$n$ S\'ersic surface brightness
profile dominates the $n$=1 one both at low and high radii. This is similar to
spiral galaxies, which are dominated by bulges in their inner regions,
spheroids in their outer regions, and with the disc dominating in between. Therefore, the colour of the mix of S\'ersic models
should have a blue gradient at small radii and a red gradient at large radii.
Conversely, the presence of a blue gradient does not necessarily
eliminate the possibility of an old, red stellar population at large
radii. 

\begin{figure}
    \centering
        \includegraphics[width=\hsize,viewport=0 20 430 385]{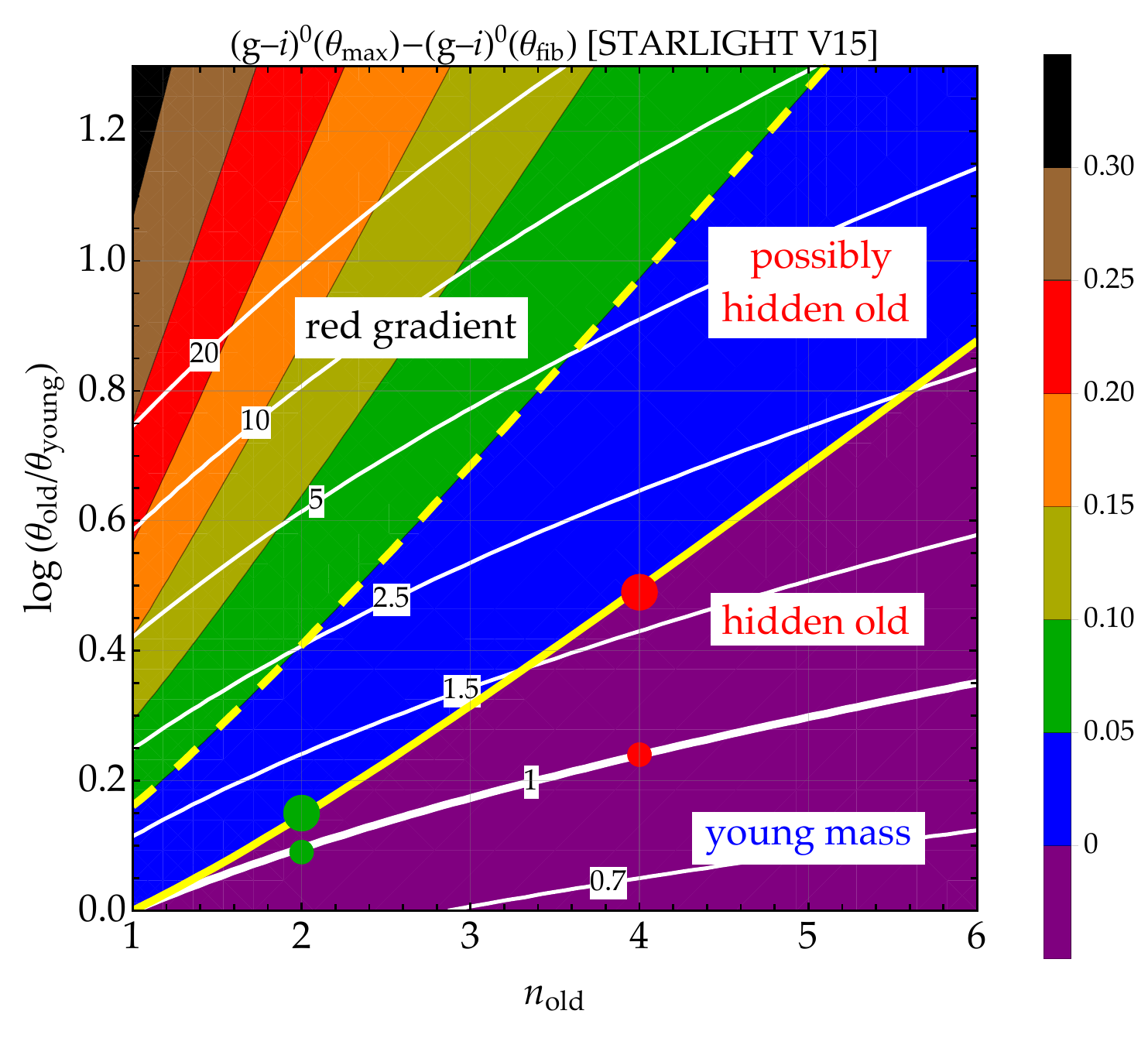}
    \caption{Difference in $(g\!-\!i)^0$ colours
      (eqs.~[\ref{gmieq}]-[\ref{lambda}]) between $\theta_{\rm max}=4$
      half-light radii of the young population (where SDSS measures the
      global photometry) and the fibre radius, as a function of the  S\'ersic index of the old population, $n_{\rm old}$ and the ratio of effective radii $\theta_{\rm old}/\theta_{\rm young}$, assuming the parameters given in Table~\ref{tab:toy}.
      The \emph{solid yellow curve} shows the strict upper limit for a blue
      gradient, while the \emph{dashed yellow curve} shows the upper limit
      allowing for a slight red difference of 0.05 magnitudes caused by
      photometric errors.
      The \emph{white curves} indicate the old over young mass ratios, with
      equality as a \emph{thick white curve}
      (as in Fig.~\ref{fig:massratiocontours}).
      The \emph{filled circles} represent four extreme cases that are
      discussed for future observability.
    }
    \label{fig:gmi}
\end{figure}

Fig.~\ref{fig:gmi} illustrates the difference between the global colour (difference of model magnitudes, measured at $\theta_{\rm max}=4\,\theta_{\rm young}$, because the SDSS photometric pipeline fits exponential profiles out to 4 effective radii\footnote{https://www.sdss.org/dr12/algorithms/magnitudes/}) and the fibre colour, again as a function of the old population S\'ersic index and of the ratio of old to young effective radii.
Fig.~\ref{fig:gmi} clearly shows that there is a range of effective radii of the old population that allows this old population to dominate the total mass (between the thick white curve and either of the yellow curves) without making the colour redder than $(g-i)_{\rm fibre}$.
The region between the white and solid yellow curves only allows bluer global
colours, while extending to the dashed yellow curve allows slightly redder global colours (by 0.05 mag), consistent with typical photometric errors.
For example, if $n_{\rm old}=3$, one can have an old stellar population if
$1.5 < \theta_{\rm old}/\theta_{\rm young} < 2.1$ (i.e. 0.18 to 0.32 in log)
for no red gradient, or up to 4.9 for $0 < (g-i)^0(\theta_{\rm
  max})-(g-i)^0(\theta_{\rm fib})<0.05$, if we allow for photometric errors.
The same model applied to the VESPA models, 
using the same online calculator to recompute the colours and mass-to-light
ratios,
yields  very similar curves as in Fig.~\ref{fig:gmi}.

We conclude that forcing a blue gradient does not rule out an underlying old stellar population of high S\'ersic index coupled with an effective radius that is greater but not very much greater than that of the young stellar population. Note that in high EW galaxies, the colours are strongly affected by the emission lines, so that bluer does not necessarily mean younger, as is clear in I~Zw~18 \citep{Papaderos&Ostlin12}. This reinforces the possibility of an old stellar population hiding in our {\tt conservative} sample of VYGs.

\begin{figure}
  \centering
  \includegraphics[width=0.95\hsize,viewport=0 15 375 375]{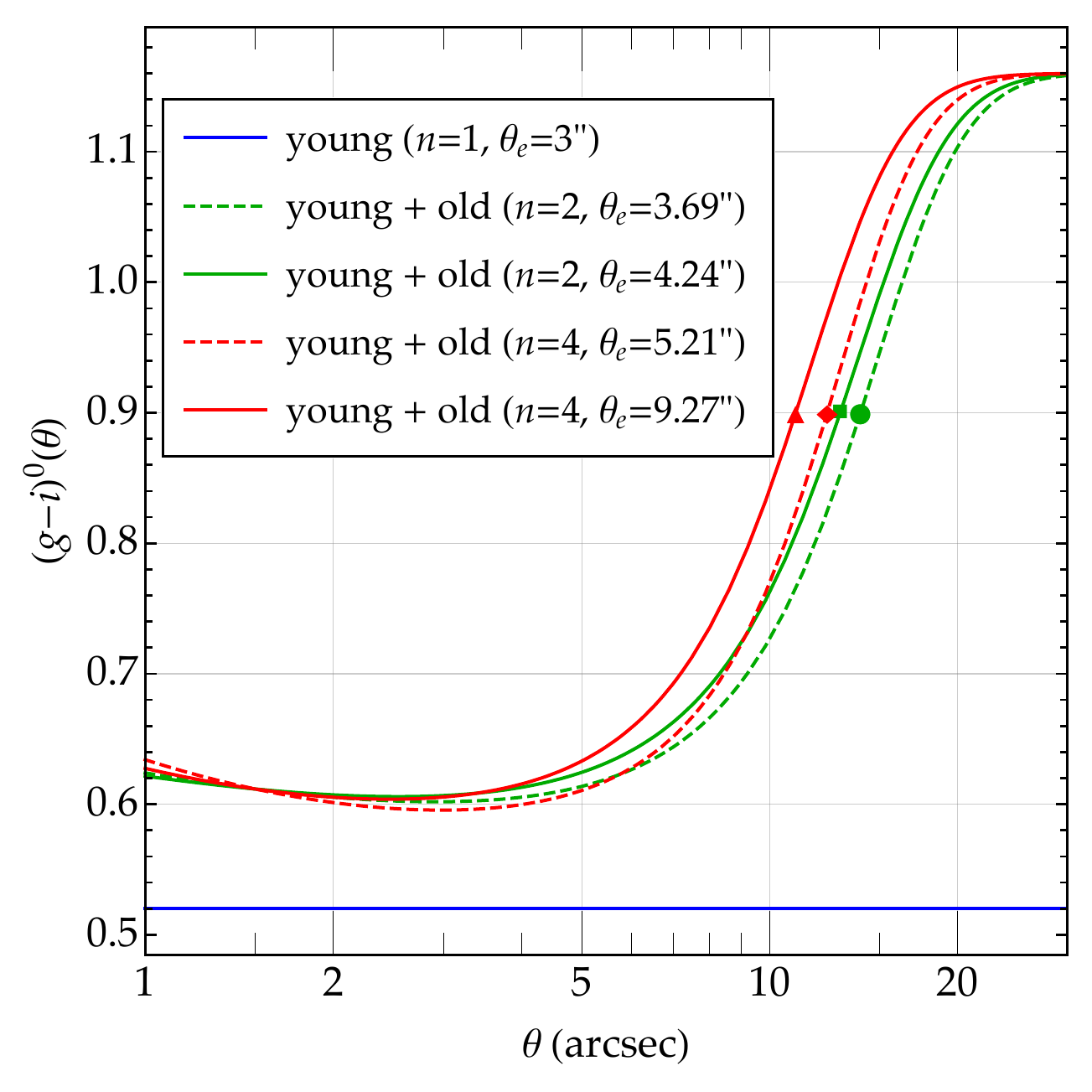} 
  \includegraphics[width=0.95\hsize,viewport=0 15 375 375]{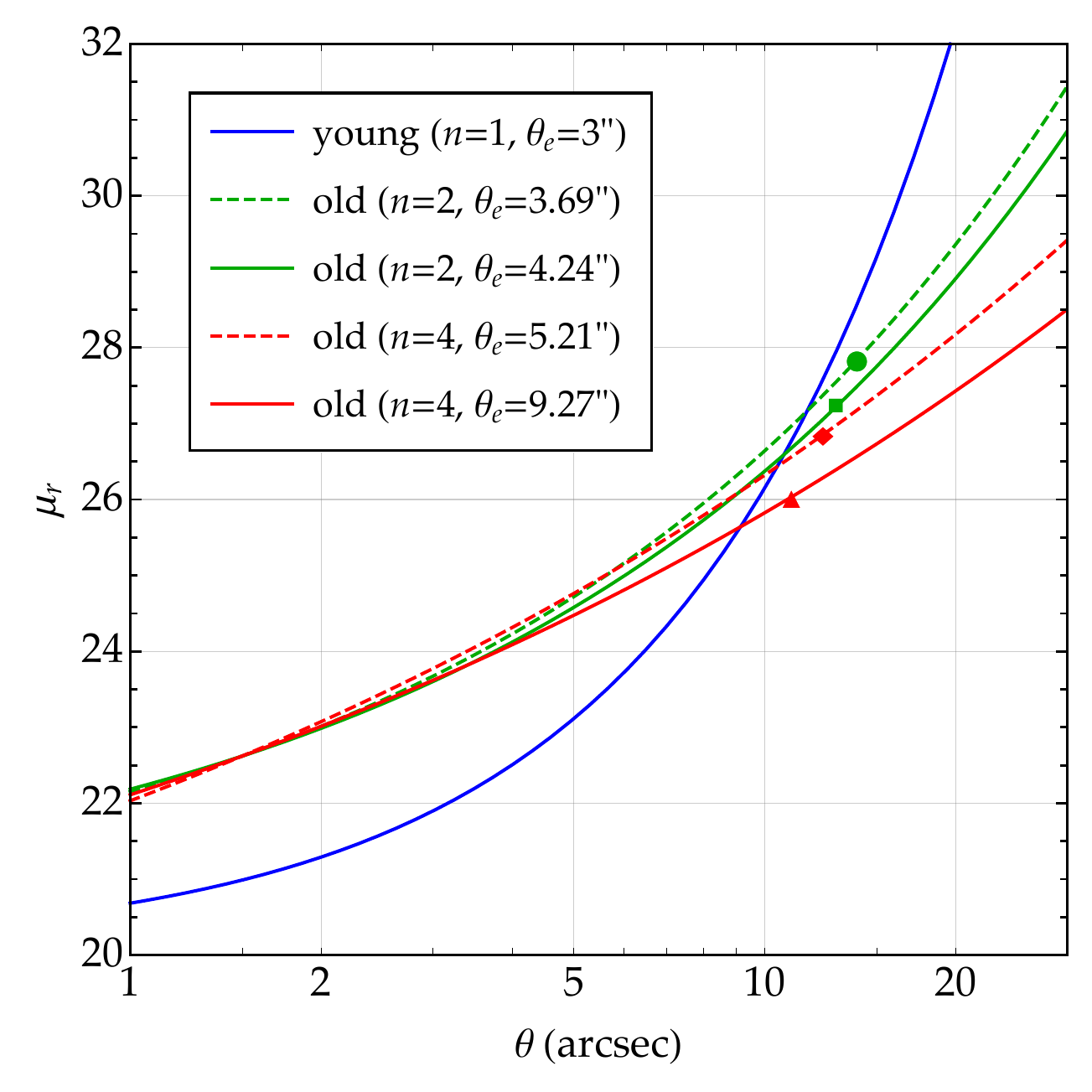}
  \caption{
        {\bf Top:} Colour profiles of the young stellar population as well as mixtures of
        the young with 4 extreme old
        populations (corresponding to the symbols shown in Fig.~\ref{fig:gmi})
        that 1) dominate the stellar mass, 2) account for half
    the stellar mass within the 3 arcsec SDSS fibre, and 3) have a blue
    gradient.
{\bf Bottom:} Surface brightness profiles for the same models as in the top panel.
In both panels, the \emph{symbols} correspond to
$(g-i)^0 = 0.9$, a typical 
colour that can be distinguished from the bluer inner colour.
  }
  \label{fig:mutheta}
\end{figure}

We finally ask how deep will observations be required to detect a possible underlying old
stellar population.
Fig.~\ref{fig:mutheta} shows the colour and surface brightness profiles for the
young stellar population and mixtures of this young population with
four extreme old
stellar populations, two with S\'ersic index  $n_{\rm old} = 2$ and two with $n_{\rm old} = 4$.
For each S\'ersic index, we adopt the extreme values for the effective
radii, $\theta_{\rm old}$, lying on the thick white (minimum effective
radius for old population dominating the mass) and yellow (maximum effective
radius to allow for a blue gradient up to $4\,\theta_{\rm young}$) curves of Fig.~\ref{fig:gmi}.
Assuming 0.1 mag photometric uncertainties for very deep observations, colours will
have 0.14 mag uncertainties, and a $2\,\sigma$ detection of a red
colour gradient will require a redward shift of 0.3 mag, i.e. an outer
 colour of $(g-i)^0 = 0.9$ given the typical inner colour of  $(g-i)^0 = 0.6$.
The top
panel of Fig.~\ref{fig:mutheta} indicates  $(g-i)^0 = 0.9$ is reached
between 12 and 15 arcsec. Reading off the
  corresponding values for the surface brightness in the bottom panel of
  Fig.~\ref{fig:mutheta}, we deduce that observations reaching a depth in the
  range from 26 mag$\,$arcsec$^{-2}$
  ($n_{\rm old}=4$, largest possible $\theta_{\rm old}$) to
  28 mag$\,$arcsec$^{-2}$ ($n_{\rm old}=2$, smallest possible $\theta_{\rm
    old}$)
  will be required to detect the possible hidden old
  stellar population. But S\'ersic indices of $n=4$ pertain to giant
  ellipticals, and a hidden old stellar population is likely to be less
  luminous and with a lower index, making it more difficult to detect (the
   surface brightness will need to reach 27 mag$\,$arcsec$^{-2}$).
  Repeating this analysis allowing for the $2\,\sigma$  detection of a 0.1
  mag redward shift (instead of 0.3 mag) brings the required
  surface brightness values to 25 to 27 mag$\,$arcsec$^{-2}$ (26 to 27 mag$\,$arcsec$^{-2}$ for $n_{\rm old} \leq 3$).

\subsection{The number of major starbursts and the reliability of the VYG classification}
\label{sec:csfh}
\begin{figure}
  \centering
   \includegraphics[width=0.9\hsize,viewport=0 30 700 585]{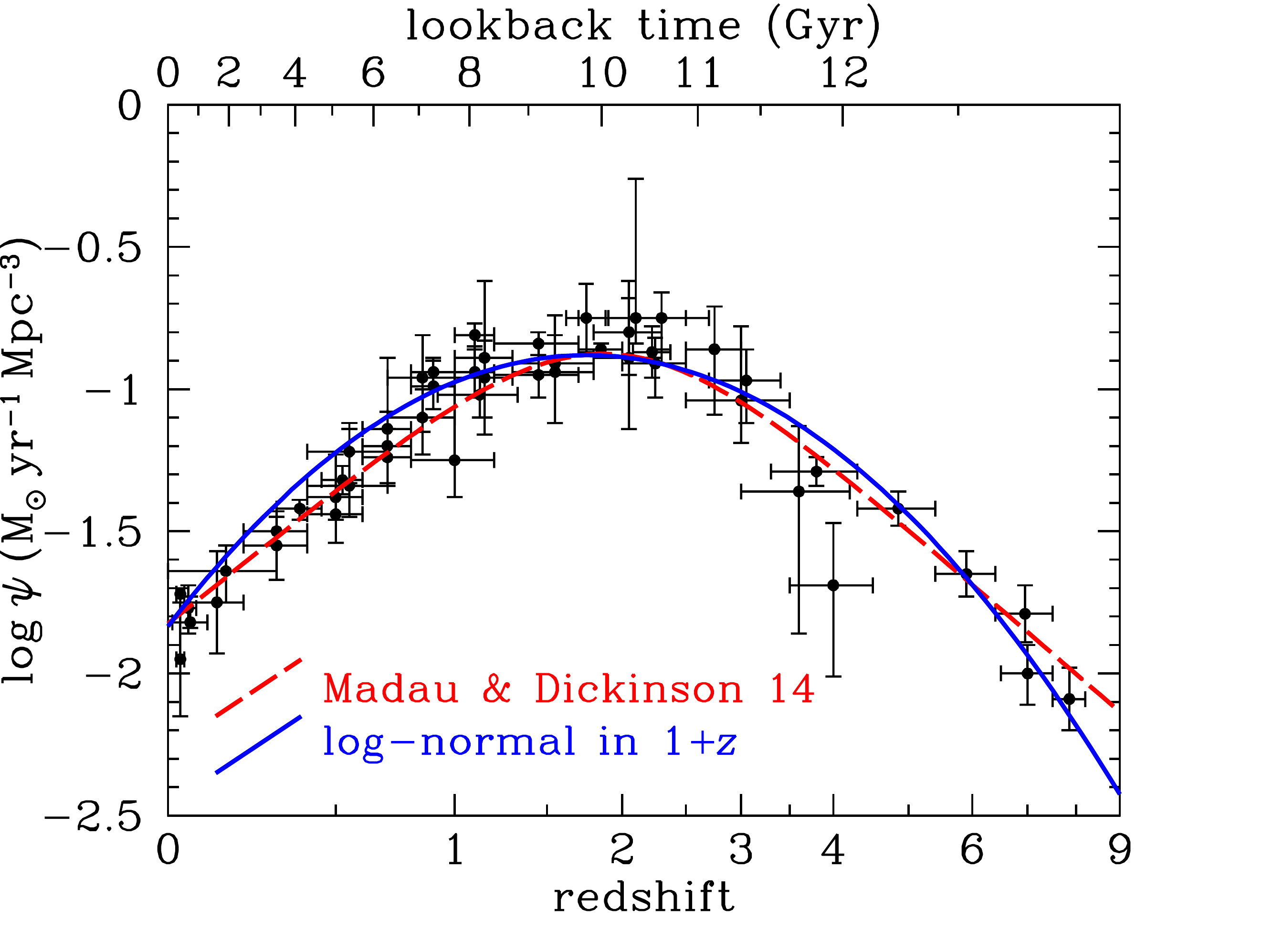}
  \caption{Evolution of the cosmic star formation density. The \emph{symbols} are
    the data compiled by Madau \& Dickinson (2014), while the \emph{dashed red} and
    \emph{solid blue curves} are two analytical fits from Madau \& Dickinson
    and from our lognormal fit to $\psi$ vs. $1+z$ (with mean of 0.44 and standard
    deviation 0.21), respectively.
  \label{fig:csfh}}
\end{figure}

Can the VYG fractions constrain the SFHs of galaxies?
The most realistic assumption is that the distribution of lookback times should follow the
cosmic star formation history (CSFH).
We first note the CSFH (denoted $\psi$) is very close to a log-normal in
$1+z$, i.e. $\psi(z) \simeq \psi_{\rm max}
\,\exp\{-[\log(1+z)-\mu]^2/(2\,\sigma^2)\}$.
This is shown in Figure~\ref{fig:csfh} for
our best-fit $\mu=0.44$ and $\sigma=0.21$
(with a peak CSFR $\psi=0.13\rm\,M_\odot\,yr^{-1}\,Mpc^{-3}$ at $z = 10^{0.44}-1 = 1.76$, with 84 per cent of the CSFR
occurring at redshifts $0.7 < z < 3.5$). The lognormal CSFH produces
a slightly smaller $\chi^2$ (in $\log \psi$) than the fit of \cite{Madau&Dickinson14}, despite
having one less parameter (leading to a reduced $\chi_\nu^2$ of 1.46 instead of
1.58).

The mean redshift of our
{\tt complete} sample is $z_{\rm obs}  = 0.07$, i.e. a
lookback time of 0.93 Gyr. This means
that the critical (maximum) redshift for our galaxies to be VYGs is the redshift corresponding to
lookback time of 1.93 Gyr, i.e. $z_{\rm crit} = 0.154$.
According to the lognormal CSFH, we expect the probability  $p_{\rm young}$
that a star in a galaxy
at $z = 0.07$ is younger than 1 Gyr will be
\begin{eqnarray} 
  p_{\rm young} &\!\!\!\!=\!\!\!\!& P(z_{\rm obs} < z < z_{\rm crit} \,|\, z > z_{\rm obs})
  \nonumber \\
  &\!\!\!\!=\!\!\!\!& 
            {
              \int_{z_{\rm obs}}^{z_{\rm crit}}
  \psi(z)\,/\,[(1+z)\,E(z)]\,{\rm d} z
\over
\int_{z_{\rm obs}}^\infty
\psi(z)\,/\,[(1+z)\,E(z)]\,{\rm d} z
}
=            0.026\ .
\label{fyoung1}
\end{eqnarray} 

\begin{figure}
  \centering
  \includegraphics[width=0.83\hsize,viewport=30 0 700 550]{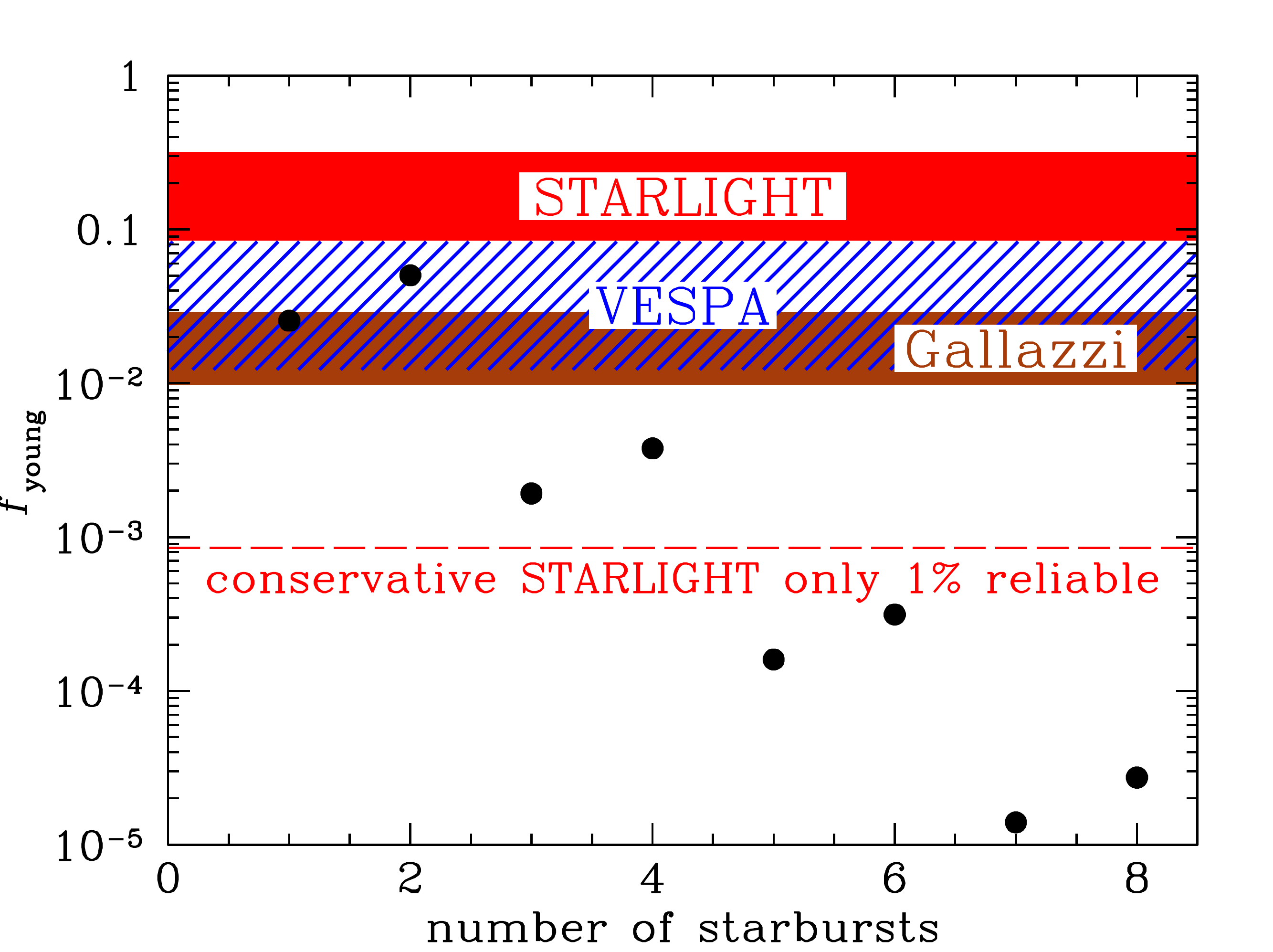}
  \caption{Comparison of prediction of mean fractions of very young galaxies
    of stellar mass $m>10^8\msun$: 
    predicted by the cosmic star formation history (from eq.~[\ref{cdfyoung}],
    \emph{filled circles})
and
obtained from SDSS (from eq.~[\ref{meanfyoung}],
\emph{shaded regions} extending from conservative to
    liberal).
  \label{fig:fracvsNb}}
\end{figure}

In one extreme situation, suppose that all galaxies have single burst SFHs. If we
draw random SF redshifts
from our lognormal CSFH,
we expect that VYGs will comprise 2.6 per cent of our sample of $z=0.07$
galaxies (eq.~[\ref{fyoung1}]).
At the other extreme, if the SFH of galaxies is continuous, the probability
of $>10^8$ (unrelated) stars to be younger than 1 Gyr, will be negligible.
More precisely, if each star has a
probability $p_{\rm young}$ of being younger than 1 Gyr, the probability  of $N$
unrelated stars or starbursts (of equal mass)
leading to at least half the stellar mass being younger than 1 Gyr
will be the cumulative distribution function of the binomial
distribution
\begin{equation}
  f_{\rm young} = \sum_{k=k_{\rm min}}^N{N \choose k} \,p_{\rm young}^k \left(1-p_{\rm
    young}\right)^{N-k} \ ,
  \label{cdfyoung}
\end{equation}
where $k_{\rm min} = N/2$ and $(N+1)/2$ for even and odd $N$, respectively. 
As shown in Figure~\ref{fig:fracvsNb} (symbols), Eq.~(\ref{cdfyoung})
yields  a VYG fraction of $2\,p_{\rm young}\,(1-p_{\rm young}/2) = 5$ per
cent for $N=2$ starbursts, but falls very rapidly to 
0.4 per cent
for $N=4$ starbursts,
$3\times 10^{-4}$ for $N=6$
starbursts (and below $10^{-10}$ for $N=20$ bursts of SF).

We compare these predicted VYG fractions
with the mean fractions predicted from a mass-limited sample:
\begin{equation}
  \left\langle f_{\rm young} \right\rangle = {\int_{m_{\rm min}}^\infty
    \phi(m)\,f_{\rm young}(m)\,{\rm d}m \over
 \int_{m_{\rm min}}^\infty
      \phi(m)\,{\rm d}m } \ ,
    \label{meanfyoung}
\end{equation}
where $f_{\rm young}(m)$ is the VYG fraction that we obtained for the {\tt
  complete} SDSS/MGS sample and $\phi(m)$ is the $z=0.07$ cosmic stellar mass function.
Adopting the double Schechter stellar mass function of \cite{Baldry+08} and
integrating down to minimum galaxy stellar mass $m_{\rm min} = 10^8\,\msun$,
we find {\tt liberal} mean VYG fractions of 32,
9
and 3 per cent for STARLIGHT
V15,
VESPA
and Gallazzi, respectively, and 
{\tt conservative}  fractions of 8 per cent for STARLIGHT V15 and 1~per cent
for
the
VESPA and Gallazzi models,
as shown in Figure~\ref{fig:fracvsNb} (shaded regions).
The geometric means of these {\tt liberal} and {\tt conservative} mean VYG
fractions are
16, 3, and 2 per cent for 
per cent with the STARLIGHT, VESPA, and Gallazzi models, respectively. 
The VYG classification with STARLIGHT  is, at best, 30 per cent reliable for the 2-burst model
($f_{\rm young}=0.05$), while with the other spectral models it is either incomplete and
fully reliable for the 2-burst model, or nearly complete and reliable for the
single-burst model. 
For 4 major starbursts, the VYG
classification would have reliability of only 2.5, 13 and 20 percent for STARLIGHT,
VESPA and Gallazzi, respectively.
We now focus on STARLIGHT, which predicts $\sim$\,8 times more frequent {\tt conservative} VYGs
than the other two models. Adopting the \emph{ansatz} that the {\tt conservative} VYG classification of
STARLIGHT is only 1~per~cent reliable, we deduce that $m > 10^8 \rm M_\odot$ galaxies
undergo on average at most 4 major starbursts.

This discussion is based on the full galaxy population. It cannot be applied
to specific populations.
It suggests that the decrease of the
fraction of VYGs with increasing galaxy
stellar
mass is a natural outcome of the more
bursty SFHs of low mass galaxies (as seen in cosmological hydrodynamical
simulations,
e.g. \citealp{Tollet+19}).


\subsection{Concluding remarks}
Our galaxy SFHs from SDSS spectra using different spectral models all indicate that the fraction of VYGs decrease with
stellar mass.
Nevertheless,  the fraction of VYGs remains an unsolved problem for two reasons:
1) the wide variation in estimates of VYG fractions, both among the spectral
models, and among the galaxy formation models;  2) the possibility of our missing
old stellar populations.
Turning this around, we believe that the very strong sensitivity of the
fraction of VYGs, to the galaxy formation modelling on one hand (and the
number of major starbursts per galaxy) and to the spectral
modelling of observed spectra on the other hand, make the study of the mass
variation of the fraction
of VYGs a strong constraint for both models of  galaxy formation
and evolution and for spectral models.

It will be worthwhile to confirm the existence of VYGs (some nearly as
massive  than our Milky Way), in several ways:
\vspace{-0.5\baselineskip}
\begin{enumerate}
    \item Analyze the spectra of the VYG candidates with Bayesian modelling (e.g. BEAGLE, \citealp{Chevallard&Charlot16}), for example to penalise against decreasing metal histories.
    \item Perform spectral fits to stacked spectra of VYG candidates.
    \item Analyze jointly SDSS spectra and photometry in the far- and near-UV
      from the Galaxy Evolution Explorer (GALEX) and in the near-IR from the
      Wide-field Infrared Survey Explorer (WISE), accounting for aperture
      effects (e.g. with BEAGLE).
    \item Obtain integral field spectroscopy of a
      large statistical sample of the VYG candidates to test whether they are
      globally young or if their youth is restricted to their inner regions
      probed by the SDSS fibres (instead of relying on the blue colour
      gradient as a way to avoid aperture effects, since the presence of such
      a gradient is insufficient to exclude old stellar populations).
    \item Obtain deep images of VYG candidates, e.g. with E-ELT, to search
      for low-surface brightness, extended old stellar populations that may
      dominate the stellar mass (as \citealp{Papaderos+02} have done for I~Zw~18).
    \item Stack hundreds or thousands of images of VYG candidates to achieve
      comparable depths to search for old stellar populations.. 
    \item Obtain deep near-IR images of nearby VYG candidates with the future James Web Space Telescope
      (JWST) to derive colour-magnitude diagrams of resolved stars deeper
      than previously done with HST.
\end{enumerate}

While we cannot be certain that the majority of the thousands of VYGs that we find in the SDSS
MGS truly have median ages below 1 Gyr, these VYGs may still hold clues to important
recent star formation. Indeed, in a
forthcoming article (Trevisan et al. in prep.), we will highlight the
specific properties that
distinguish VYGs from other star forming galaxies and
  discuss the
  impact of galaxy mergers in the formation of VYGs at low redshift.
Moreover, we
have begun radio observations at GMRT and VLA to understand the kinematics
and accretion
history of neutral gas around VYGs.

An electronic table of \textcolor{darkgreen}{the Clean sample of 404\,931} galaxies
\textcolor{red}{\sout{that are VYGs according to one of our 3
preferred 
spectral models}}
is provided in electronic form, with the top 10 lines shown
in Table~\ref{tab:elec}. 

\section*{Acknowledgments}
We thank Andrea Cattaneo, Daniel Kunth, Sophia Liannou, and Polychronis Papa\-deros
for enlightening discussions and
the
referee, Erik Tollerud, for
his
numerous constructive comments that strengthened this
article.  M.T. and T.X.T. are grateful to the hospitality of the Institut
d'Astrophysique de Paris, where a large part of this work was
performed. G.A.M.
acknowledges the Brazilian CNPq (grant \#451451/2019-8) and the Universidade
Federal do Rio 
Grande do Sul for its hospitality.  We
are grateful to Roberto Cid Fernandes for making his STARLIGHT code publicly
available  and
Rita Tojeiro for making the VESPA output publicly available.

This research has made use of the NASA/IPAC Extragalactic Database (NED),
which is operated by the Jet Propulsion Laboratory, California Institute of
Technology, under contract with the National Aeronautics and Space
Administration.

Funding for the Sloan Digital Sky Survey IV has been provided by the Alfred
P. Sloan Foundation, the U.S. Department of Energy Office of Science, and the
Participating Institutions.

\bibliography{}

\begin{thebibliography}{}
\makeatletter
\relax
\def\mn@urlcharsother{\let\do\@makeother \do\$\do\&\do\#\do\^\do\_\do\%\do\~}
\def\mn@doi{\begingroup\mn@urlcharsother \@ifnextchar [ {\mn@doi@}
  {\mn@doi@[]}}
\def\mn@doi@[#1]#2{\def\@tempa{#1}\ifx\@tempa\@empty \href
  {http://dx.doi.org/#2} {doi:#2}\else \href {http://dx.doi.org/#2} {#1}\fi
  \endgroup}
\def\mn@eprint#1#2{\mn@eprint@#1:#2::\@nil}
\def\mn@eprint@arXiv#1{\href {http://arxiv.org/abs/#1} {{\tt arXiv:#1}}}
\def\mn@eprint@dblp#1{\href {http://dblp.uni-trier.de/rec/bibtex/#1.xml}
  {dblp:#1}}
\def\mn@eprint@#1:#2:#3:#4\@nil{\def\@tempa {#1}\def\@tempb {#2}\def\@tempc
  {#3}\ifx \@tempc \@empty \let \@tempc \@tempb \let \@tempb \@tempa \fi \ifx
  \@tempb \@empty \def\@tempb {arXiv}\fi \@ifundefined
  {mn@eprint@\@tempb}{\@tempb:\@tempc}{\expandafter \expandafter \csname
  mn@eprint@\@tempb\endcsname \expandafter{\@tempc}}}

\bibitem[\protect\citeauthoryear{{Aloisi} et~al.,}{{Aloisi}
  et~al.}{2007}]{Aloisi+07}
{Aloisi} A.,  et~al., 2007, \mn@doi [\apjl] {10.1086/522368}, \href
  {http://adsabs.harvard.edu/abs/2007ApJ...667L.151A} {667, L151}

\bibitem[\protect\citeauthoryear{{Baldry}, {Glazebrook}  \& {Driver}}{{Baldry}
  et~al.}{2008}]{Baldry+08}
{Baldry} I.~K.,  {Glazebrook} K.,   {Driver} S.~P.,  2008, \mn@doi [\mnras]
  {10.1111/j.1365-2966.2008.13348.x}, \href
  {http://adsabs.harvard.edu/abs/2008MNRAS.388..945B} {388, 945}

\bibitem[\protect\citeauthoryear{{Baldwin}, {Phillips}  \&
  {Terlevich}}{{Baldwin} et~al.}{1981}]{BPT81}
{Baldwin} J.~A.,  {Phillips} M.~M.,   {Terlevich} R.,  1981, \mn@doi [\pasp]
  {10.1086/130766}, 93, 5

\bibitem[\protect\citeauthoryear{{Behroozi}, {Wechsler}  \&
  {Conroy}}{{Behroozi} et~al.}{2013}]{Behroozi+13}
{Behroozi} P.~S.,  {Wechsler} R.~H.,   {Conroy} C.,  2013, \mn@doi [\apj]
  {10.1088/0004-637X/770/1/57}, \href
  {http://adsabs.harvard.edu/abs/2013ApJ...770...57B} {770, 57}

\bibitem[\protect\citeauthoryear{{Boylan-Kolchin}, {Springel}, {White},
  {Jenkins}  \& {Lemson}}{{Boylan-Kolchin} et~al.}{2009}]{BoylanKolchin+09}
{Boylan-Kolchin} M.,  {Springel} V.,  {White} S.~D.~M.,  {Jenkins} A.,
  {Lemson} G.,  2009, \mn@doi [\mnras] {10.1111/j.1365-2966.2009.15191.x},
  \href {http://adsabs.harvard.edu/abs/2009MNRAS.398.1150B} {398, 1150}

\bibitem[\protect\citeauthoryear{{Bressan}, {Fagotto}, {Bertelli}  \&
  {Chiosi}}{{Bressan} et~al.}{1993}]{Bressan+93}
{Bressan} A.,  {Fagotto} F.,  {Bertelli} G.,   {Chiosi} C.,  1993, \aaps, \href
  {http://cdsads.u-strasbg.fr/abs/1993A%26AS..100..647B} {100, 647}

\bibitem[\protect\citeauthoryear{{Brinchmann}, {Charlot}, {White}, {Tremonti},
  {Kauffmann}, {Heckman}  \& {Brinkmann}}{{Brinchmann}
  et~al.}{2004}]{Brinchmann+04}
{Brinchmann} J.,  {Charlot} S.,  {White} S.~D.~M.,  {Tremonti} C.,  {Kauffmann}
  G.,  {Heckman} T.,   {Brinkmann} J.,  2004, \mn@doi [\mnras]
  {10.1111/j.1365-2966.2004.07881.x}, \href
  {http://adsabs.harvard.edu/abs/2004MNRAS.351.1151B} {351, 1151}

\bibitem[\protect\citeauthoryear{{Bruzual} \& {Charlot}}{{Bruzual} \&
  {Charlot}}{2003}]{Bruzual&Charlot03}
{Bruzual} G.,  {Charlot} S.,  2003, \mn@doi [\mnras]
  {10.1046/j.1365-8711.2003.06897.x}, \href
  {http://adsabs.harvard.edu/abs/2003MNRAS.344.1000B} {344, 1000}

\bibitem[\protect\citeauthoryear{{Cappellari} \& {Emsellem}}{{Cappellari} \&
  {Emsellem}}{2004}]{Cappellari&Emsellem04}
{Cappellari} M.,  {Emsellem} E.,  2004, \mn@doi [\pasp] {10.1086/381875}, \href
  {http://adsabs.harvard.edu/abs/2004PASP..116..138C} {116, 138}

\bibitem[\protect\citeauthoryear{{Cattaneo}, {Mamon}, {Warnick}  \&
  {Knebe}}{{Cattaneo} et~al.}{2011}]{Cattaneo+11}
{Cattaneo} A.,  {Mamon} G.~A.,  {Warnick} K.,   {Knebe} A.,  2011, \mn@doi
  [\aap] {10.1051/0004-6361/201015780}, \href
  {http://adsabs.harvard.edu/abs/2011A%26A...533A...5C} {533, A5}

\bibitem[\protect\citeauthoryear{{Chabrier}}{{Chabrier}}{2003}]{Chabrier03}
{Chabrier} G.,  2003, \mn@doi [\pasp] {10.1086/376392}, \href
  {http://adsabs.harvard.edu/abs/2003PASP..115..763C} {115, 763}

\bibitem[\protect\citeauthoryear{{Charlot} \& {Bruzual}}{{Charlot} \&
  {Bruzual}}{1991}]{Charlot&Bruzual91}
{Charlot} S.,  {Bruzual} G.,  1991, \mn@doi [\apj] {10.1086/169608}, \href
  {http://adsabs.harvard.edu/abs/1991ApJ...367..126C} {367, 126}

\bibitem[\protect\citeauthoryear{{Charlot} \& {Fall}}{{Charlot} \&
  {Fall}}{2000}]{Charlot&Fall00}
{Charlot} S.,  {Fall} S.~M.,  2000, \mn@doi [\apj] {10.1086/309250}, \href
  {http://adsabs.harvard.edu/abs/2000ApJ...539..718C} {539, 718}

\bibitem[\protect\citeauthoryear{{Chevallard} \& {Charlot}}{{Chevallard} \&
  {Charlot}}{2016}]{Chevallard&Charlot16}
{Chevallard} J.,  {Charlot} S.,  2016, \mn@doi [\mnras]
  {10.1093/mnras/stw1756}, \href
  {http://adsabs.harvard.edu/abs/2016MNRAS.462.1415C} {462, 1415}

\bibitem[\protect\citeauthoryear{{Cid Fernandes}, {Mateus}, {Sodr{\'e}},
  {Stasi{\'n}ska}  \& {Gomes}}{{Cid Fernandes} et~al.}{2005}]{CidFernandes+05}
{Cid Fernandes} R.,  {Mateus} A.,  {Sodr{\'e}} L.,  {Stasi{\'n}ska} G.,
  {Gomes} J.~M.,  2005, \mn@doi [\mnras] {10.1111/j.1365-2966.2005.08752.x},
  \href {http://cdsads.u-strasbg.fr/abs/2005MNRAS.358..363C} {358, 363}

\bibitem[\protect\citeauthoryear{{Conroy}, {Gunn}  \& {White}}{{Conroy}
  et~al.}{2009}]{Conroy+09}
{Conroy} C.,  {Gunn} J.~E.,   {White} M.,  2009, \mn@doi [\apj]
  {10.1088/0004-637X/699/1/486}, \href
  {http://adsabs.harvard.edu/abs/2009ApJ...699..486C} {699, 486}

\bibitem[\protect\citeauthoryear{{Contreras Ramos} et~al.,}{{Contreras Ramos}
  et~al.}{2011}]{ContrerasRamos+11}
{Contreras Ramos} R.,  et~al., 2011, \mn@doi [\apj]
  {10.1088/0004-637X/739/2/74}, \href
  {http://adsabs.harvard.edu/abs/2011ApJ...739...74C} {739, 74}

\bibitem[\protect\citeauthoryear{{Dav{\'e}}, {Angl{\'e}s-Alc{\'a}zar},
  {Narayanan}, {Li}, {Rafieferantsoa}  \& {Appleby}}{{Dav{\'e}}
  et~al.}{2019}]{Dave+19}
{Dav{\'e}} R.,  {Angl{\'e}s-Alc{\'a}zar} D.,  {Narayanan} D.,  {Li} Q.,
  {Rafieferantsoa} M.~H.,   {Appleby} S.,  2019, \mn@doi [\mnras]
  {10.1093/mnras/stz937}, \href
  {https://ui.adsabs.harvard.edu/abs/2019MNRAS.486.2827D} {486, 2827}

\bibitem[\protect\citeauthoryear{{Dressler} et~al.,}{{Dressler}
  et~al.}{2016}]{Dressler+16}
{Dressler} A.,  et~al., 2016, \mn@doi [\apj] {10.3847/1538-4357/833/2/251},
  \href {http://adsabs.harvard.edu/abs/2016ApJ...833..251D} {833, 251}

\bibitem[\protect\citeauthoryear{{Dressler}, {Kelson}  \&
  {Abramson}}{{Dressler} et~al.}{2018}]{Dressler+18}
{Dressler} A.,  {Kelson} D.~D.,   {Abramson} L.~E.,  2018, \mn@doi [\apj]
  {10.3847/1538-4357/aaedbe}, \href
  {http://adsabs.harvard.edu/abs/2018ApJ...869..152D} {869, 152}

\bibitem[\protect\citeauthoryear{{Fagotto}, {Bressan}, {Bertelli}  \&
  {Chiosi}}{{Fagotto} et~al.}{1994a}]{Fagotto+94a}
{Fagotto} F.,  {Bressan} A.,  {Bertelli} G.,   {Chiosi} C.,  1994a, \aaps,
  \href {http://cdsads.u-strasbg.fr/abs/1994A%26AS..104..365F} {104, 365}

\bibitem[\protect\citeauthoryear{{Fagotto}, {Bressan}, {Bertelli}  \&
  {Chiosi}}{{Fagotto} et~al.}{1994b}]{Fagotto+94b}
{Fagotto} F.,  {Bressan} A.,  {Bertelli} G.,   {Chiosi} C.,  1994b, \aaps,
  \href {http://cdsads.u-strasbg.fr/abs/1994A%26AS..105...29F} {105, 29}

\bibitem[\protect\citeauthoryear{{Gallazzi}, {Charlot}, {Brinchmann}, {White}
  \& {Tremonti}}{{Gallazzi} et~al.}{2005}]{Gallazzi+05}
{Gallazzi} A.,  {Charlot} S.,  {Brinchmann} J.,  {White} S.~D.~M.,   {Tremonti}
  C.~A.,  2005, \mn@doi [\mnras] {10.1111/j.1365-2966.2005.09321.x}, \href
  {http://adsabs.harvard.edu/abs/2005MNRAS.362...41G} {362, 41}

\bibitem[\protect\citeauthoryear{{Garilli}, {Maccagni}  \& {Andreon}}{{Garilli}
  et~al.}{1999}]{Garilli+99}
{Garilli} B.,  {Maccagni} D.,   {Andreon} S.,  1999, \aap, \href
  {http://adsabs.harvard.edu/abs/1999A%26A...342..408G} {342, 408}

\bibitem[\protect\citeauthoryear{{Girardi}, {Bressan}, {Chiosi}, {Bertelli}  \&
  {Nasi}}{{Girardi} et~al.}{1996}]{Girardi+96}
{Girardi} L.,  {Bressan} A.,  {Chiosi} C.,  {Bertelli} G.,   {Nasi} E.,  1996,
  \aaps, \href {http://adsabs.harvard.edu/abs/1996A%26AS..117..113G} {117, 113}

\bibitem[\protect\citeauthoryear{{Gnedin}}{{Gnedin}}{2000}]{Gnedin00}
{Gnedin} N.~Y.,  2000, \mn@doi [\apj] {10.1086/317042}, \href
  {http://adsabs.harvard.edu/abs/2000ApJ...542..535G} {542, 535}

\bibitem[\protect\citeauthoryear{{Graham} \& {Colless}}{{Graham} \&
  {Colless}}{1997}]{Graham&Colless97}
{Graham} A.,  {Colless} M.,  1997, \mnras, 287, 221

\bibitem[\protect\citeauthoryear{{Henriques}, {White}, {Thomas}, {Angulo},
  {Guo}, {Lemson}, {Springel}  \& {Overzier}}{{Henriques}
  et~al.}{2015}]{Henriques+15}
{Henriques} B.~M.~B.,  {White} S.~D.~M.,  {Thomas} P.~A.,  {Angulo} R.,  {Guo}
  Q.,  {Lemson} G.,  {Springel} V.,   {Overzier} R.,  2015, \mn@doi [\mnras]
  {10.1093/mnras/stv705}, \href
  {http://adsabs.harvard.edu/abs/2015MNRAS.451.2663H} {451, 2663}

\bibitem[\protect\citeauthoryear{{Izotov} \& {Thuan}}{{Izotov} \&
  {Thuan}}{1998}]{Izotov&Thuan98}
{Izotov} Y.~I.,  {Thuan} T.~X.,  1998, \mn@doi [\apj] {10.1086/305440}, \href
  {https://ui.adsabs.harvard.edu/abs/1998ApJ...497..227I} {497, 227}

\bibitem[\protect\citeauthoryear{{Izotov} \& {Thuan}}{{Izotov} \&
  {Thuan}}{2004}]{Izotov&Thuan04}
{Izotov} Y.~I.,  {Thuan} T.~X.,  2004, \mn@doi [\apj] {10.1086/424990}, \href
  {http://cdsads.u-strasbg.fr/abs/2004ApJ...616..768I} {616, 768}

\bibitem[\protect\citeauthoryear{{Izotov}, {Guseva}  \& {Thuan}}{{Izotov}
  et~al.}{2011}]{Izotov+11}
{Izotov} Y.~I.,  {Guseva} N.~G.,   {Thuan} T.~X.,  2011, \mn@doi [\apj]
  {10.1088/0004-637X/728/2/161}, \href
  {https://ui.adsabs.harvard.edu/abs/2011ApJ...728..161I} {728, 161}

\bibitem[\protect\citeauthoryear{{Izotov}, {Thuan}, {Guseva}  \&
  {Liss}}{{Izotov} et~al.}{2018}]{Izotov+18}
{Izotov} Y.~I.,  {Thuan} T.~X.,  {Guseva} N.~G.,   {Liss} S.~E.,  2018, \mn@doi
  [\mnras] {10.1093/mnras/stx2478}, \href
  {http://adsabs.harvard.edu/abs/2018MNRAS.473.1956I} {473, 1956}

\bibitem[\protect\citeauthoryear{{Kauffmann} et~al.,}{{Kauffmann}
  et~al.}{2003}]{Kauffmann+03}
{Kauffmann} G.,  et~al., 2003, \mnras, 341, 33

\bibitem[\protect\citeauthoryear{{Kewley}, {Dopita}, {Sutherland}, {Heisler}
  \& {Trevena}}{{Kewley} et~al.}{2001}]{Kewley+01}
{Kewley} L.~J.,  {Dopita} M.~A.,  {Sutherland} R.~S.,  {Heisler} C.~A.,
  {Trevena} J.,  2001, \mn@doi [\apj] {10.1086/321545}, \href
  {https://ui.adsabs.harvard.edu/abs/2001ApJ...556..121K} {556, 121}

\bibitem[\protect\citeauthoryear{{Kroupa}}{{Kroupa}}{2001}]{Kroupa01}
{Kroupa} P.,  2001, \mn@doi [\mnras] {10.1046/j.1365-8711.2001.04022.x}, \href
  {http://cdsads.u-strasbg.fr/abs/2001MNRAS.322..231K} {322, 231}

\bibitem[\protect\citeauthoryear{{La Barbera}, {de Carvalho}, {de La Rosa},
  {Lopes}, {Kohl-Moreira}  \& {Capelato}}{{La Barbera}
  et~al.}{2010}]{LaBarbera+10}
{La Barbera} F.,  {de Carvalho} R.~R.,  {de La Rosa} I.~G.,  {Lopes} P.~A.~A.,
  {Kohl-Moreira} J.~L.,   {Capelato} H.~V.,  2010, \mn@doi [\mnras]
  {10.1111/j.1365-2966.2010.16850.x}, \href
  {http://adsabs.harvard.edu/abs/2010MNRAS.408.1313L} {408, 1313}

\bibitem[\protect\citeauthoryear{{Le Borgne} et~al.,}{{Le Borgne}
  et~al.}{2003}]{LeBorgne+03}
{Le Borgne} J.-F.,  et~al., 2003, \mn@doi [\aap] {10.1051/0004-6361:20030243},
  \href {http://cdsads.u-strasbg.fr/abs/2003A%26A...402..433L} {402, 433}

\bibitem[\protect\citeauthoryear{{Leitherer} et~al.,}{{Leitherer}
  et~al.}{1999}]{Leitherer+99}
{Leitherer} C.,  et~al., 1999, \mn@doi [\apjs] {10.1086/313233}, \href
  {http://adsabs.harvard.edu/abs/1999ApJS..123....3L} {123, 3}

\bibitem[\protect\citeauthoryear{{Lejeune}, {Cuisinier}  \& {Buser}}{{Lejeune}
  et~al.}{1998}]{Lejeune+98}
{Lejeune} T.,  {Cuisinier} F.,   {Buser} R.,  1998, \mn@doi [\aaps]
  {10.1051/aas:1998405}, \href
  {http://adsabs.harvard.edu/abs/1998A%26AS..130...65L} {130, 65}

\bibitem[\protect\citeauthoryear{{Madau} \& {Dickinson}}{{Madau} \&
  {Dickinson}}{2014}]{Madau&Dickinson14}
{Madau} P.,  {Dickinson} M.,  2014, \araa, 52, 415

\bibitem[\protect\citeauthoryear{{Maraston}}{{Maraston}}{2005}]{Maraston05}
{Maraston} C.,  2005, \mn@doi [\mnras] {10.1111/j.1365-2966.2005.09270.x},
  \href {http://cdsads.u-strasbg.fr/abs/2005MNRAS.362..799M} {362, 799}

\bibitem[\protect\citeauthoryear{{Maraston} \& {Str{\"o}mb{\"a}ck}}{{Maraston}
  \& {Str{\"o}mb{\"a}ck}}{2011}]{Maraston&Stromback11}
{Maraston} C.,  {Str{\"o}mb{\"a}ck} G.,  2011, \mn@doi [\mnras]
  {10.1111/j.1365-2966.2011.19738.x}, \href
  {http://adsabs.harvard.edu/abs/2011MNRAS.418.2785M} {418, 2785}

\bibitem[\protect\citeauthoryear{{Marigo} \& {Girardi}}{{Marigo} \&
  {Girardi}}{2007}]{Marigo&Girardi07}
{Marigo} P.,  {Girardi} L.,  2007, \mn@doi [\aap] {10.1051/0004-6361:20066772},
  \href {http://adsabs.harvard.edu/abs/2007A%26A...469..239M} {469, 239}

\bibitem[\protect\citeauthoryear{{Marigo}, {Girardi}, {Bressan}, {Groenewegen},
  {Silva}  \& {Granato}}{{Marigo} et~al.}{2008}]{Marigo+08}
{Marigo} P.,  {Girardi} L.,  {Bressan} A.,  {Groenewegen} M.~A.~T.,  {Silva}
  L.,   {Granato} G.~L.,  2008, \mn@doi [\aap] {10.1051/0004-6361:20078467},
  \href {http://adsabs.harvard.edu/abs/2008A%26A...482..883M} {482, 883}

\bibitem[\protect\citeauthoryear{{Moster}, {Naab}  \& {White}}{{Moster}
  et~al.}{2013}]{Moster+13}
{Moster} B.~P.,  {Naab} T.,   {White} S.~D.~M.,  2013, \mn@doi [\mnras]
  {10.1093/mnras/sts261}, \href
  {http://adsabs.harvard.edu/abs/2013MNRAS.428.3121M} {428, 3121}

\bibitem[\protect\citeauthoryear{{Mutch}, {Croton}  \& {Poole}}{{Mutch}
  et~al.}{2013}]{Mutch+13}
{Mutch} S.~J.,  {Croton} D.~J.,   {Poole} G.~B.,  2013, \mn@doi [\mnras]
  {10.1093/mnras/stt1453}, \href
  {http://cdsads.u-strasbg.fr/abs/2013MNRAS.435.2445M} {435, 2445}

\bibitem[\protect\citeauthoryear{{O'Sullivan} et~al.,}{{O'Sullivan}
  et~al.}{2014}]{O'Sullivan+14a}
{O'Sullivan} E.,  et~al., 2014, \mn@doi [\apj] {10.1088/0004-637X/793/2/73},
  \href {http://adsabs.harvard.edu/abs/2014ApJ...793...73O} {793, 73}

\bibitem[\protect\citeauthoryear{{Papaderos} \& {{\"O}stlin}}{{Papaderos} \&
  {{\"O}stlin}}{2012}]{Papaderos&Ostlin12}
{Papaderos} P.,  {{\"O}stlin} G.,  2012, \mn@doi [\aap]
  {10.1051/0004-6361/201117551}, \href
  {http://adsabs.harvard.edu/abs/2012A%26A...537A.126P} {537, A126}

\bibitem[\protect\citeauthoryear{{Papaderos}, {Izotov}, {Thuan}, {Noeske},
  {Fricke}, {Guseva}  \& {Green}}{{Papaderos} et~al.}{2002}]{Papaderos+02}
{Papaderos} P.,  {Izotov} Y.~I.,  {Thuan} T.~X.,  {Noeske} K.~G.,  {Fricke}
  K.~J.,  {Guseva} N.~G.,   {Green} R.~F.,  2002, \mn@doi [\aap]
  {10.1051/0004-6361:20021023}, \href
  {http://adsabs.harvard.edu/abs/2002A%26A...393..461P} {393, 461}

\bibitem[\protect\citeauthoryear{{Pietrinferni}, {Cassisi}, {Salaris}  \&
  {Castelli}}{{Pietrinferni} et~al.}{2004}]{Pietrinferni+04}
{Pietrinferni} A.,  {Cassisi} S.,  {Salaris} M.,   {Castelli} F.,  2004,
  \mn@doi [\apj] {10.1086/422498}, \href
  {http://cdsads.u-strasbg.fr/abs/2004ApJ...612..168P} {612, 168}

\bibitem[\protect\citeauthoryear{{Pietrinferni}, {Cassisi}, {Salaris}  \&
  {Castelli}}{{Pietrinferni} et~al.}{2006}]{Pietrinferni+06}
{Pietrinferni} A.,  {Cassisi} S.,  {Salaris} M.,   {Castelli} F.,  2006,
  \mn@doi [\apj] {10.1086/501344}, \href
  {http://cdsads.u-strasbg.fr/abs/2006ApJ...642..797P} {642, 797}

\bibitem[\protect\citeauthoryear{{Prugniel} \& {Simien}}{{Prugniel} \&
  {Simien}}{1997}]{Prugniel&Simien97}
{Prugniel} P.,  {Simien} F.,  1997, \aap, 321, 111

\bibitem[\protect\citeauthoryear{{Salpeter}}{{Salpeter}}{1955}]{Salpeter55}
{Salpeter} E.~E.,  1955, \apj, 121, 161

\bibitem[\protect\citeauthoryear{{S{\'a}nchez-Bl{\'a}zquez}
  et~al.,}{{S{\'a}nchez-Bl{\'a}zquez} et~al.}{2006}]{Sanchez-Blazquez+06}
{S{\'a}nchez-Bl{\'a}zquez} P.,  et~al., 2006, \mn@doi [\mnras]
  {10.1111/j.1365-2966.2006.10699.x}, \href
  {http://cdsads.u-strasbg.fr/abs/2006MNRAS.371..703S} {371, 703}

\bibitem[\protect\citeauthoryear{{Sarzi} et~al.,}{{Sarzi}
  et~al.}{2006}]{Sarzi+06}
{Sarzi} M.,  et~al., 2006, \mn@doi [\mnras] {10.1111/j.1365-2966.2005.09839.x},
  \href {http://adsabs.harvard.edu/abs/2006MNRAS.366.1151S} {366, 1151}

\bibitem[\protect\citeauthoryear{{Searle} \& {Sargent}}{{Searle} \&
  {Sargent}}{1972}]{Searle&Sargent72}
{Searle} L.,  {Sargent} W.~L.~W.,  1972, \mn@doi [\apj] {10.1086/151398}, \href
  {http://adsabs.harvard.edu/abs/1972ApJ...173...25S} {173, 25}

\bibitem[\protect\citeauthoryear{{Skillman} \& {Kennicutt}}{{Skillman} \&
  {Kennicutt}}{1993}]{Skillman&Kennicutt93}
{Skillman} E.~D.,  {Kennicutt} Robert~C. J.,  1993, \mn@doi [\apj]
  {10.1086/172868}, \href
  {https://ui.adsabs.harvard.edu/abs/1993ApJ...411..655S} {411, 655}

\bibitem[\protect\citeauthoryear{{Strauss} et~al.,}{{Strauss}
  et~al.}{2002}]{Strauss+02}
{Strauss} M.~A.,  et~al., 2002, \aj, 124, 1810

\bibitem[\protect\citeauthoryear{{Telles} \& {Melnick}}{{Telles} \&
  {Melnick}}{2018}]{Telles&Melnick18}
{Telles} E.,  {Melnick} J.,  2018, \mn@doi [\aap]
  {10.1051/0004-6361/201732275}, \href
  {http://adsabs.harvard.edu/abs/2018A%26A...615A..55T} {615, A55}

\bibitem[\protect\citeauthoryear{{Thomas} et~al.,}{{Thomas}
  et~al.}{2013}]{Thomas+13}
{Thomas} D.,  et~al., 2013, \mn@doi [\mnras] {10.1093/mnras/stt261}, \href
  {http://adsabs.harvard.edu/abs/2013MNRAS.431.1383T} {431, 1383}

\bibitem[\protect\citeauthoryear{{Tojeiro}, {Wilkins}, {Heavens}, {Panter}  \&
  {Jimenez}}{{Tojeiro} et~al.}{2009}]{Tojeiro+09}
{Tojeiro} R.,  {Wilkins} S.,  {Heavens} A.~F.,  {Panter} B.,   {Jimenez} R.,
  2009, \mn@doi [\apjs] {10.1088/0067-0049/185/1/1}, \href
  {http://cdsads.u-strasbg.fr/abs/2009ApJS..185....1T} {185, 1}

\bibitem[\protect\citeauthoryear{{Tollet}, {Cattaneo}, {Macci{\`o}}, {Dutton}
  \& {Kang}}{{Tollet} et~al.}{2019}]{Tollet+19}
{Tollet} {\'E}.,  {Cattaneo} A.,  {Macci{\`o}} A.~V.,  {Dutton} A.~A.,   {Kang}
  X.,  2019, \mn@doi [\mnras] {10.1093/mnras/stz545}, \href
  {https://ui.adsabs.harvard.edu/abs/2019MNRAS.485.2511T} {485, 2511}

\bibitem[\protect\citeauthoryear{{Tremonti} et~al.,}{{Tremonti}
  et~al.}{2004}]{Tremonti+04}
{Tremonti} C.~A.,  et~al., 2004, \mn@doi [\apj] {10.1086/423264}, \href
  {http://adsabs.harvard.edu/abs/2004ApJ...613..898T} {613, 898}

\bibitem[\protect\citeauthoryear{{Trevisan}, {Mamon}  \&
  {Khosroshahi}}{{Trevisan} et~al.}{2017}]{Trevisan+17}
{Trevisan} M.,  {Mamon} G.~A.,   {Khosroshahi} H.~G.,  2017, \mn@doi [\mnras]
  {10.1093/mnras/stw2588}, \href
  {http://adsabs.harvard.edu/abs/2017MNRAS.464.4593T} {464, 4593}

\bibitem[\protect\citeauthoryear{{Tweed}, {Mamon}, {Thuan}, {Cattaneo},
  {Dekel}, {Menci}, {Calura}  \& {Silk}}{{Tweed} et~al.}{2018}]{Tweed+18}
{Tweed} D.~P.,  {Mamon} G.~A.,  {Thuan} T.~X.,  {Cattaneo} A.,  {Dekel} A.,
  {Menci} N.,  {Calura} F.,   {Silk} J.,  2018, \mn@doi [\mnras]
  {10.1093/mnras/sty507}, \href
  {http://adsabs.harvard.edu/abs/2018MNRAS.477.1427T} {477, 1427} (Paper~I)

\bibitem[\protect\citeauthoryear{{Vazdekis}, {S{\'a}nchez-Bl{\'a}zquez},
  {Falc{\'o}n-Barroso}, {Cenarro}, {Beasley}, {Cardiel}, {Gorgas}  \&
  {Peletier}}{{Vazdekis} et~al.}{2010}]{Vazdekis+10}
{Vazdekis} A.,  {S{\'a}nchez-Bl{\'a}zquez} P.,  {Falc{\'o}n-Barroso} J.,
  {Cenarro} A.~J.,  {Beasley} M.~A.,  {Cardiel} N.,  {Gorgas} J.,   {Peletier}
  R.~F.,  2010, \mn@doi [\mnras] {10.1111/j.1365-2966.2010.16407.x}, \href
  {http://cdsads.u-strasbg.fr/abs/2010MNRAS.404.1639V} {404, 1639}

\bibitem[\protect\citeauthoryear{{Vazdekis} et~al.,}{{Vazdekis}
  et~al.}{2015}]{Vazdekis+15}
{Vazdekis} A.,  et~al., 2015, \mn@doi [\mnras] {10.1093/mnras/stv151}, \href
  {http://cdsads.u-strasbg.fr/abs/2015MNRAS.449.1177V} {449, 1177}

\bibitem[\protect\citeauthoryear{{Vogt}, {Dopita}  \& {Kewley}}{{Vogt}
  et~al.}{2013}]{Vogt+13}
{Vogt} F.~P.~A.,  {Dopita} M.~A.,   {Kewley} L.~J.,  2013, \mn@doi [\apj]
  {10.1088/0004-637X/768/2/151}, \href
  {http://cdsads.u-strasbg.fr/abs/2013ApJ...768..151V} {768, 151}

\makeatother
\end{thebibliography}


\appendix

\onecolumn

\begin{landscape}
\section{Table of very young galaxies}

\begin{table}
  \caption{Table of \textcolor{darkgreen}{Clean sample} of \textcolor{darkgreen}{404\,931} galaxies
    \textcolor{red}{\sout{that are
      very young according to at least one of our three preferred spectral models}}}
  \begin{center}
      \tabcolsep=2pt
    \begin{tabular}{rcrrrrccccccccccccccccccc}
      \hline
      \hline
      ID & RA & \multicolumn{1}{c}{Dec} & \multicolumn{1}{c}{plate} & \multicolumn{1}{c}{MJD} & \multicolumn{1}{c}{fiber} & Compl. & blue & AGN &
      $\log(m/\msun)$ & age$_{50}$$\,$(Gyr) &  &
      $\log(m/\msun)$ & age$_{50}$$\,$(Gyr) &  &
      $\log(m/\msun)$ & age$_{50}$$\,$(Gyr) & &
      \multicolumn{3}{c}{SFH (SL V15 - no evol)}
      & & \multicolumn{3}{c}{SFH (SL V15 - evol)}\\
      \cline{2-3}
      \cline{10-11}
      \cline{13-14}
      \cline{16-17}
      \cline{19-21}
      \cline{23-25}
      & \multicolumn{2}{c}{deg J2000} & & & & \multicolumn{3}{c}{\ \ \ gradient}   &
      \multicolumn{2}{c}{SL V15} & &
      \multicolumn{2}{c}{VESPA BC03} & &
      \multicolumn{2}{c}{VESPA M05} & &
      0--1$\,$Gyr & 1--6.1$\,$Gyr & $>$6.1$\,$Gyr
      && 0--1$\,$Gyr & 1--6.1$\,$Gyr & $>$6.1$\,$Gyr \\
      (1) & (2) & \multicolumn{1}{c}{(3)} & \multicolumn{1}{c}{(4)} & \multicolumn{1}{c}{(5)} & (6) & (7) & (8) & (9) & (10) & (11)
      & & (12) & (13) & & (14) & (15) & & (16) & (17) & (18) & & (19) & (20)
      & (21) \\ 
      \hline
     1 & 0.0065 & $ -0.0926$ & 387 & 51791 & 186 & Y & N & T & 10.96 & \ \, 5.84 &  & 10.67 & 10.33 &  & 10.70 & 11.43 &  & 0.00 & 0.56 & 0.44 &  & 0.00 & 0.57 & 0.43\\ 
     2 & 0.0083 & $ 15.6972$ & 750 & 52235 & 456 & Y & Y & T & 11.04 & \ \, 6.58 &  & 10.91 & \ \, 9.44 &  & 10.63 & \ \, 7.06 &  & 0.01 & 0.43 & 0.56 &  & 0.01 & 0.45 & 0.54\\ 
     3 & 0.0087 & $ 15.8817$ & 750 & 52235 & 459 & Y & Y & N & 10.05 & 12.93 &  & 10.20 & \ \,3.28 &  & 10.25 & 11.33 &  & 0.08 & 0.04 & 0.89 &  & 0.12 & 0.04 & 0.83\\ 
     4 & 0.0118 & $ 14.7155$ & 750 & 52235 & 174 & N & N & N & \ \, 9.62 & \ \,1.36 &  & \ \, 9.63 & \ \,3.31 &  & \ \, 9.66 & \ \,2.74 &  & 0.42 & 0.36 & 0.21 &  & 0.51 & 0.34 & 0.15\\ 
     5 & 0.0134 & $ -1.1130$ & 685 & 52203 & 232 & N & Y & N & \ \, 8.75 & \ \,0.79 &  & \ \, 8.92 & \ \, 7.23 &  & \ \,8.74 & \ \,2.67 &  & 0.95 & 0.04 & 0.01 &  & 0.96 & 0.03 & 0.01\\ 
     6 & 0.0139 & $-10.7211$ & 650 & 52143 & 211 & Y & Y & T & 11.78 & 12.00 &  & 10.99 & \ \,5.12 &  & 10.91 & \ \,8.19 &  & 0.01 & 0.18 & 0.81 &  & 0.01 & 0.20 & 0.79\\ 
     7 & 0.0144 & $ 14.1982$ & 750 & 52235 & 92 & Y & Y & T & 11.61 & \ \,7.43 &  & 11.23 & \ \,     9.61 &  & 11.17 & \ \, 9.07 &  & 0.01 & 0.38 & 0.61 &  & 0.01 & 0.42 & 0.57\\ 
     8 & 0.0174 & $ -8.7342$ & 650 & 52143 & 446 & Y & Y & T & 11.27 & \ \,9.31 &  &    10.95 & 10.89 &  & 10.83 & \ \, 7.05 &  & 0.00 & 0.21 & 0.79 &  & 0.00 & 0.22 & 0.78\\ 
     9 & 0.0192 & $ -8.9438$ & 650 & 52143 & 447 & N & Y & N & \ \, 8.70 & \ \,0.47 &  & \ \,9.31 & \ \,5.35 &  & \ \,8.82 & \ \,3.89 &  & 0.74 & 0.26 & 0.00 &  & 0.79 & 0.21 & 0.00\\ 
    10 & 0.0198 & $  0.7817$ & 387 & 51791 & 446 & Y & Y & T & 11.92 & \ \,8.64 &  & 11.53 & 11.24 &  & 11.57 & 11.09 &  & 0.00 & 0.21 & 0.79 &  & 0.00 & 0.22 & 0.77\\ 
      \end{tabular}
  \end{center}

  \noindent Notes: the columns are:
  (1): unique identifier;
  (2) and (3): equatorial coordinates;
  (4), (5) and (6): SDSS plate, MJD and fiber;
  (7): is galaxy in {\tt complete} sample?
  (8): does galaxy have a blue colour gradient? ($(g-i)_{\rm model}^0 <
  (g-i)_{\rm fiber}^0$);
  (9): is galaxy an AGN: `Y' (yes, above \citealp{Kewley+01} curve), `T' (transition between curves of
  \citeauthor{Kewley+01} and \citealp{Kauffmann+03}), `N' (no, below
  \citeauthor{Kauffmann+03} curve);
  (10): stellar mass according to STARLIGHT with V15 model;
  (11): median age  according to STARLIGHT with V15 model;
  (12): stellar mass according to VESPA with BC03 2-component dust model;
  (13): median age  according to VESPA with BC03 2-component model;
  (14): stellar mass according to VESPA with M05 2-component dust model;
  (15): median age  according to VESPA with M05 2-component model;
  (16) to (18): star formation history in respective bins of 0-1 Gyr, 1-6.12
  Gyr, and over 6.12 Gyr, using STARLIGHT with the V15 model, with no
  correction for stellar evolution;
  (19) to (21): star formation history in respective bins of 0-1 Gyr, 1-6.12
  Gyr, and over 6.12 Gyr, using STARLIGHT with the V15 model, now
  correcting for stellar evolution.
  The full table for the 404\,931 galaxies of the {\tt clean} sample is available online.
  \label{tab:elec}
\end{table}
\label{lastpage}

\bsp
\end{landscape}


\end{document}